\documentclass[preprint,10pt]{elsarticle}

\usepackage[a4paper,margin=1in]{geometry}
\usepackage{placeins}
\usepackage{booktabs}
\usepackage[colorlinks,citecolor=blue,urlcolor=black,bookmarks=false,hypertexnames=true]{hyperref}
\usepackage{graphicx}
\usepackage{mathtools}

\usepackage{multirow}
\usepackage{bigdelim} 
\usepackage{xcolor}

\usepackage{siunitx}      
\usepackage{tikz}
\usepackage{subcaption} 
\usepackage{hyperref}
\usepackage{nicematrix} 
\usepackage{float} 
\usetikzlibrary{decorations.pathreplacing}
\usepackage{color,soul}

\usepackage{array}
\usepackage{ragged2e}
\usepackage{caption}
\usepackage[most]{tcolorbox}
\newcommand{\underlabel}[2]{\underset{\text{\normalsize #2}}{#1}}

\usepackage{xcolor}
\usepackage{empheq}

\usepackage{amssymb}

\usepackage{amsmath}

\usepackage{amsthm}

\newtheoremstyle{myremark}
  {3pt}
  {3pt}
  {}
  {}
  {\bfseries}
  {.}
  { }
  {}

\theoremstyle{myremark}

\journal{}

\begin{document}

\begin{frontmatter}



\title{A constitutive framework for distortional-mode-dependent failure in soft materials: Tension–compression asymmetry and beyond}
\author{Yogesh C. Chandrashekar}
\author{Kshitiz Upadhyay\corref{cor1}}
\cortext[cor1]{Corresponding author. Email: kshitizu@umn.edu}
\address{Department of Aerospace Engineering and Mechanics, University of Minnesota, Minneapolis, MN 55455, USA}

\begin{abstract}
Soft materials often exhibit pronounced tension--compression asymmetry (TCA) in their softening and failure behavior, a feature that conventional hyperelastic and continuum-damage formulations generally fail to capture within a unified framework. This work presents a Lode-invariant-based hyperelastic softening model for distortional-mode-dependent failure in soft materials, in which mode dependence is introduced through a bi-failure construction with distinct tensile and compressive energy limiters. The proposed model extends Volokh’s classical energy-limiting approach by embedding a Lode-angle-dependent weighting function, thereby ensuring a smooth and physically consistent transition of failure behavior across distortion modes directly within the constitutive description of the bulk response, without introducing internal damage variables. Agarose hydrogels (1, 2, and 3\,\%\,w/v) serve as the model system for validation. The framework accurately reproduces experimental stress--stretch responses in uniaxial tension and compression, capturing concentration-dependent stiffness and failure energetics. Using parameters calibrated solely from combined uniaxial data, the model successfully predicts pure shear behavior---including softening and failure---thereby demonstrating strong cross-mode predictive capability. To further assess thermodynamic consistency and distortion-mode sensitivity, the model’s free-energy landscape is analyzed across the full Lode-invariant space, confirming a smooth and physically consistent response under diverse loading conditions. Parameter evolution with concentration follows power-law scaling, enabling interpolation and predictive validation at intermediate concentrations (evaluated at 2.5\,\%\,w/v). Overall, the proposed formulation provides a physically interpretable constitutive framework for tension--compression-asymmetric softening and distortional-mode-dependent failure, and establishes a foundation for three-dimensional failure mapping in soft materials.

\end{abstract}


\end{frontmatter}



\section{Introduction}
\label{sec:Intro}
 
Soft materials, such as biological tissues, gels, and foams, exhibit complex mechanical behavior characterized by large deformations, nonlinear stress–strain response, gradual damage accumulation, and a pronounced loading-mode based asymmetry between tension and compression, known as tension–compression asymmetry (TCA) \cite{avril2026state, goriely2015mechanics, bohringer2025compression, anssari2022modelling, martonova2025generalized, carrillo2013nonlinear, holzapfel2025biomechanics, Moerman2016, myers2015continuous, drozdov2020tension, cortes2012extra, sun2010review, reddipaga2026construction}. Beyond their elastic characteristics, the inelastic mechanisms governing softening and failure differ fundamentally between deformation modes: under compression, these materials are prone to local buckling and wrinkling \cite{huang2016effects, du2014variational, du2020tension, tang2019phase, stewart2016wrinkling}, whereas tensile loading promotes microcrack formation driven by the collective rupture of molecular bonds at weak cross-links \cite{ konale2025modeling, volokh2017fracture, notbohm2015microbuckling, linka2018fatigue}. Despite this well-documented mechanical asymmetry, continuum frameworks capable of accurately capturing such failure-related irreversible phenomena in soft materials remain underdeveloped \cite{anssari2024unified, pena2019failure}. This gap is particularly consequential for biomedical applications, where accurate modeling of tissue softening and rupture—such as in brain, vascular, and organ injury—is essential for understanding, diagnosing, and mitigating conditions including stroke, cerebral hemorrhage, and traumatic brain injury \cite{liu2025tube, rausch2017modeling, zhu2019visco,Upadhyay2022_headinjurymodel, Bayly2021}.

Hyperelasticity provides the fundamental constitutive framework for describing the nonlinear elastic response of soft materials \cite{franceschini2006brain, mendis1995constitutive, Lohr2022, wang20243d}. Classical formulations such as neo-Hookean, Mooney–Rivlin \cite{mooney1940theory, rivlin1948large, rivlin1949largeb, rivlin1951large}, Ogden \cite{ogden1972large}, along with their variants \cite{anssari2024generalisation, anssari2022modelling}, remain widely adopted because of their tractable strain energy functions, conveniently expressed in terms of principal strain invariants or principal stretches, and their ability to represent large deformations. Nevertheless, their predictive capability is often limited to moderate loading regimes and lacks the ability to capture failure or rupture behavior \cite{martins1998numerical, pena2009mullins, volokh2007hyperelasticity}. Consequently, recent efforts have focused on integrating failure theories into hyperelastic frameworks, yielding constitutive descriptions that capture damage and rupture while retaining computational efficiency \cite{anssari2023continuous, volokh2007hyperelasticity, volokh2010modeling, pena2014computational}. Such extensions are pivotal for advancing the design and analysis of soft materials in both engineering and biomedical applications. 

To place these developments in context, it is useful to distinguish between the two primary forms through which damage manifests in soft materials: continuous and discontinuous softening \cite{anssari2024unified, pena2011prediction, pena2019failure}. Continuous softening occurs under large stretches as a gradual reduction in stiffness along the primary loading path, ultimately leading to failure through micro-crack nucleation and growth \cite{volokh2008prediction, pena2011prediction, anssari2024unified, pena2010constitutive, calvo2009modelling}. In contrast, discontinuous softening appears upon unloading and is associated with the Mullins effect \cite{mullins1969softening, harwood1965stress, pena2011mechanical}, hysteresis, and permanent set—manifested as residual strains after unloading \cite{anssari2024continuous, munoz2008experimental, alastrue2008experimental, pena2014computational}.

Among the various damage formulations proposed, the Ogden–Roxburgh pseudo-elastic framework \cite{ogden1999pseudo} remains widely used to model discontinuous softening via a stress-softening function defined in terms of the maximum strain energy attained. However, this function is activated only upon unloading \cite{chagnon2004relevance} and thus neglects softening associated with the primary loading path and the ensuing failure \cite{holzapfel2017modeling, pena2011prediction, chagnon2004relevance}. Continuous softening, on the other hand, is most naturally addressed within the continuum damage mechanics (CDM) paradigm, originally rooted in the phenomenological frameworks developed by Kachanov \cite{kachanov1958time} and Rabotnov \cite{rabotnov1980elements}. CDM employs internal damage variables to progressively degrade the intact strain-energy function—an approach first devised for small-strain isotropic damage \cite{lemaitre1984use} and later extended to finite-strain kinematics \cite{simo1987fully, gurtin1981simple}.

While CDM effectively describes softening behavior, the damage variable lacks a direct physical interpretation, as it is an internal thermodynamic state variable \cite{voyiadjis2015investigation, keller2011thermodynamic, volokh2007hyperelasticity}. Moreover, as noted by Weisbecker et al. \cite{weisbecker2012layer} and Anssari-Benam et al. \cite{anssari2024continuous, anssari2023continuous}, CDM formulations intrinsically couple elastic parameters with the damage variable, hindering independent characterization of the intrinsic material response and complicating experimental calibration. These limitations, along with difficulties in reproducing large-strain responses, have motivated the development of alternative, more physically grounded frameworks for modeling damage in soft solids.

Alternative approaches embed material-dependent damage mechanisms at the microstructural or bond level within continuum formulations. Representative examples from the fibrous tissue literature include: (i) fibril-sliding models—capturing irreversible proteoglycan-bridge sliding across collagen fibrils (Schmidt et al. \cite{schmidt2014statistical}; Gasser and Miller \cite{gasser2011irreversible, miller2021microstructurally}); (ii) fibril chain-breakage models incorporating stochastic-structurally based damage \cite{rodriguez2006stochastic, rodriguez2008finite, hamedzadeh2018constitutive}; and (iii) formulations that incorporate the effects of collagen cross-links in addition to sliding \cite{holzapfel2020damage}, providing a qualitative description of rupture in collagenous soft tissues. While these models are rooted in physical mechanisms and yield valuable mechanistic insight, they typically require fibril-scale information that is difficult to obtain and become computationally demanding when applied to macroscopic problems \cite{anssari2024unified, lu2025waviness}. Moreover, their material- and architecture-specific assumptions restrict general applicability \cite{xiao2021micromechanical, mai2019damage, miller2022bottom, dargazany2009network}. 

To address these limitations, Gao and Klein~\cite{gao1998numerical, gao1997elastic} proposed the more general Virtual Internal Bond (VIB) model, which bridges atomistic and continuum scales by idealizing materials as networks of atoms interconnected by cohesive bonds. Unlike conventional cohesive zone models, which treat fracture and elasticity as separate phenomena through distinct cohesive surface and bulk elements, 
the VIB framework unifies both elastic and fracture behavior within a single continuum description. This is achieved by incorporating atomic bond potentials into the constitutive response via the Cauchy–Born rule \cite{born1940stability, klein2001physics}, thereby establishing a rigorous atomistic–continuum connection~\cite{park2009potential, park2009unified, volokh2005modified, liu2018predicting, ji2004mechanical, sun2025damage}. The VIB framework has been widely employed to capture the nucleation and propagation of cracks in various classes of materials ~\cite{gao1998numerical, klein1998crack, gao2003modeling, volokh2006approach, thiagarajan2004finite}, and implemented in finite element (FE) frameworks (e.g., via UMAT in ABAQUS) to reproduce material softening and failure~\cite{zhang2002numerical}. However, the inherent softening behavior in the VIB formulation often induces numerical instabilities due to strain localization and loss of ellipticity, leading to convergence issues and singular stiffness matrices~\cite{zhang2002numerical, thiagarajan2004finite}. Ensuring stability typically requires restrictive measures--such as locally stiffened elements or predefined crack paths~\cite{zhang2002numerical}--which limit its predictive flexibility. Thus, despite its strong physical foundation, the VIB approach remains computationally demanding and less flexible, motivating the pursuit of more efficient continuum-level, energy-based formulations~\cite{volokh2005modified, thiagarajan2004finite}.

As a macroscopic alternative to continuum–atomistic approaches, Volokh et al. \cite{volokh2004nonlinear, volokh2007hyperelasticity, volokh2010modeling} introduced a physically motivated phenomenological framework that describes damage-induced softening through energy limiters embedded directly within the hyperelastic constitutive formulation. When coupled with the material sink concept \cite{volokh2017fracture, faye2019effect, bui2021localized} to account for failure propagation, this framework offers a physically interpretable and computationally efficient alternative to conventional CDM. It eliminates the need for ad hoc damage thresholds or evolution laws typically invoked in CDM, thereby establishing a new paradigm for modeling damage and failure in soft materials. The energy limiters framework has since been integrated with various hyperelastic formulations to capture bulk failure in a wide range of soft materials and biological tissues, as extensively reported in the literature \cite{volokh2022modeling, volokh2019constitutive, volokh2015cavitation, volokh2011cavitation, volokh2011modeling, volokh2008prediction, lev2019thermoelastic, goswami2025mechanics, li2016invariant, li2018constitutive}.

Despite the advances achieved in prior studies, conventional softening hyperelasticity and related frameworks fail to capture a critical aspect of soft solids: distortion-mode-dependent asymmetry, particularly TCA. In the context of hyperelasticity with energy limiters, this limitation arises from the use of a single set of energy limiting parameters ($\Phi$, $m$) within the stress-softening formulation, which constrains the model’s ability to differentiate between tensile and compressive responses. Consequently, most existing models are calibrated exclusively using uniaxial tension data, and their predictive capability under different distortion modes or complex multi-loading conditions remains limited \cite{anssari2023continuous, li2016invariant}. This restricted treatment of TCA is not unique to damage models—it also persists in purely hyperelastic formulations— highlighting a fundamental gap in current constitutive descriptions of soft materials. 

The significance of TCA is well established through numerous experimental studies. Normand et~al.~\cite{normand2000new} demonstrated that agarose hydrogels exhibit substantially different elastic moduli, failure stresses, and failure strains in tension and compression. Similar asymmetric responses were reported in ballistic gelatin by Pasumarthy and Tippur~\cite{pasumarthy2016mechanical}, and in gellan gels by Tang et~al.~\cite{tang1997stress}. Notbohm et~al.~\cite{notbohm2015microbuckling} further emphasized the pivotal role of TCA in biological networks, linking compressive stiffness loss and fibrin microbuckling to cellular mechanosensing in fibrous matrices. More recently, Voyiadjis et~al.~\cite{voyiadjis2018hyperelastic} and Zhu et~al.~\cite{zhu2019visco} showed that variations in compressibility between tension and compression significantly contribute to TCA, highlighting the necessity of simultaneously accounting for both asymmetry and volumetric effects in constitutive descriptions of brain-tissue mechanics. Together, these studies reveal that both the intrinsic hyperelastic response and the damage tolerance of soft materials are fundamentally asymmetric between tension and compression~\cite{zhang2019fracture, tang2019phase, anssari2022modelling}. Building on this foundation, Latorre et~al.~\cite{latorre2017understanding} theoretically demonstrated that, within the Valanis–Landel~\cite{valanis1967strain, valanis2022valanis} framework, complete characterization of incompressible hyperelastic materials requires incorporating both tensile and compressive branches from uniaxial tests. Consistent with this, subsequent investigations showed that material parameters calibrated using only one deformation mode lead to poor predictive capability under multiaxial or combined loading conditions~\cite{upadhyay2019thermodynamics, Upadhyay2020, ogden2004fitting, destrade2017methodical}. Collectively, this body of evidence establishes TCA as a defining characteristic of soft solids and exposes a central limitation of current constitutive frameworks---their inability to represent distortion-mode-dependent asymmetry---thereby restricting predictive accuracy under complex loading states.

Progress has been made toward extending hyperelastic formulations to capture TCA in the elastic regime. Among these, the Ogden model can implicitly represent asymmetry through the coefficients~$\alpha_i$~\cite{ogden1972large, bohringer2025compression, Moerman2016, voyiadjis2018hyperelastic, Lohr2022, budday2017mechanical, budday2017rheological}, which simultaneously govern the degree of nonlinearity in the stress–strain response and the extent of tension--compression disparity.  While this offers some flexibility, independent control over these two features is not possible \cite{Moerman2016}. To overcome this limitation, Du et~al.~\cite{du2020tension, du2014variational} extended Ambartsumyan’s classical bi-modulus concept~\cite{ambartsumyan1986elasticity} to a hyperelastic framework by introducing distinct shear moduli for tension and compression based on principal stretches. This approach provides a physically intuitive representation of TCA in the purely elastic regime. However, its applicability remains confined to intact (i.e., undamaged) hyperelasticity—specifically, to separable Valanis–Landel-type formulations (e.g., neo-Hookean, Ogden)~\cite{valanis1967strain, zhang2019fracture}—and therefore cannot account for asymmetry arising from irreversible phenomena such as stress softening or failure.  

This limitation highlights a broader gap: while progress has been made in representing elastic asymmetry, extending these principles to the inelastic regime has proven challenging. TCA accompanying damage and failure remains poorly captured within existing hyperelastic-softening or failure frameworks. Moreover, the absence of an explicit energy potential that distinguishes between different distortion modes hinders the formulation of bounding relations capable of rigorously separating tensile and compressive energy contributions at finite strains. ~\autoref{tab:damage_models} presents a concise overview of existing damage models and their limitations in the context of the present study. While these available approaches successfully reproduce various aspects of softening and fracture, none explicitly encode distortion-mode sensitivity or TCA within a unified, thermodynamically consistent potential. Collectively, these unresolved gaps motivate the generalized framework proposed herein, developed to capture distortion-mode-dependent softening and failure.

Building upon this motivation, we extend Volokh’s classical energy limiters framework \cite{volokh2007hyperelasticity, volokh2010modeling, volokh2013review} by introducing distortional mode–dependent energy limiters formulated in terms of the Lode invariants of the Hencky strain. These invariants—particularly the third invariant \(K_3\) (Lode angle)—uniquely characterizes distinct distortion modes \cite{criscione2000invariant, chen2012general, neff2016geometry, landauer2019experimental, prasad2020analysis, martin2018non, kulwant2023semi, falope2024energetic}, thereby providing a natural basis for defining a bi-failure criterion that captures TCA in softening behavior. The proposed extension is integrated with the strain energy density function of Prasad and Kannan \cite{prasad2020analysis}, which employs the same invariant framework to introduce distortion-mode-dependent shear moduli and thereby reproduce TCA in the elastic response \cite{kulwant2023semi, prasad2020analysis}. Together, these developments establish a unified Lode invariant-based hyperelastic–damage formulation capable of representing asymmetry in both elastic and inelastic regimes.

To evaluate the predictive capability of the proposed framework, we selected agarose hydrogels as a model material due to their widespread use as tissue phantoms in soft-tissue injury studies \cite{bremer2024ballistic, tutwiler2020rupture, sun2025damage} and their elastomeric behavior under short-time loading \cite{lu2020pseudo, Upadhyay2020}. Experimental characterization involved calibrating the model to combined tension--compression responses of agarose gels at 1, 2, and 3\% w/v concentrations using full-field Digital Image Correlation (DIC) for enhanced accuracy. The calibrated model was then validated under pure shear loading to assess its predictive capability beyond the training modes. To further examine the framework’s thermodynamic consistency and generalizability, we analyzed its energy landscape across the full Lode-invariant space, providing a comprehensive characterization of the mode-dependent response. Additionally, the model was applied to an intermediate (2.5\%) agarose concentration not used in calibration to explore generalizability across polymer network densities and potential connections to the micromechanical origins of softening and failure \cite{jones1990rigid, ramzi1998structure, guenet2006agarose}.

The remainder of the article is organized as follows: Section~\ref{sec:preliminaries} discusses the concept of energy limiters to model softening-hyperelasticity-based material failure and the continuum framework of Lode-invariants-based hyperelasticity; Section~\ref{sec:classical-energy-limiters} reviews the microstructural origins of asymmetry in deformation-induced softening; Section~\ref{sec:prop_formulation} introduces a novel Lode-invariants-based constitutive damage model and bi-failure criterion; Section~\ref{sec:methodology} details specimen preparation, testing, and calibration procedures; Section~\ref{sec:results_discussion} validates and benchmarks the model across deformation modes and gel concentrations; and Section~\ref{sec:summary_conclusion} concludes with key findings and future directions.

\begin{table}[t]
\centering
\caption{Summary of representative damage‐modeling frameworks for soft materials. Each approach is briefly evaluated in terms of its principal strengths, key limitations, and relevance to tension–compression asymmetry (TCA).}
\setlength{\tabcolsep}{3.8pt}
\renewcommand{\arraystretch}{1.1}
\footnotesize
\begin{tabular}{
>{\RaggedRight\arraybackslash}p{3.2cm}
>{\RaggedRight\arraybackslash}p{5.9cm}
>{\RaggedRight\arraybackslash}p{6.1cm}}
\toprule
\textbf{Framework} & \textbf{Key Merits} & \textbf{Principal Limitations} \\
\midrule
\textbf{Pseudo‐elasticity} 
&
Captures unloading hysteresis and stress‐softening with simple closed forms; computationally efficient for cyclic tests. &
Describes discontinuous softening only; lacks representation of progressive bond‐rupture–driven degradation. \\

\addlinespace[0.8mm]
\textbf{Continuum Damage Mechanics (CDM)} 
&
General framework for continuous softening; versatile and FE‐ready with numerous evolution laws. &
Internal damage variable lacks direct physical meaning; coupling between elastic constants and damage complicates calibration; TCA not explicitly encoded. \\

\addlinespace[0.8mm]
\textbf{Microstructural models} 
&
Physically interpretable; captures molecular-to-network degradation mechanisms. &
Computationally intensive and material-specific; difficult to generalize or upscale. \\

\addlinespace[0.8mm]
\textbf{Virtual Internal Bond (VIB)} 
&
Bridges continuum and atomistic scales through cohesive bond potentials; embeds strength and softening. &
Calibration complexity; significant computational overhead; \\

\addlinespace[0.8mm]
\textbf{Energy limiters (i.e., energy-limited hyperelasticity)} 
&
Physically interpretable via bounded strain energy; avoids ad hoc thresholds; efficiently couples with standard hyperelastic forms. &
Conventional forms assume symmetric softening and failure; no explicit distortion‐mode sensitivity or tension–compression distinction. \\
\bottomrule
\end{tabular}
\label{tab:damage_models}
\end{table}

\section{Preliminaries: Kinematics, the classical energy limiters approach to modeling soft material failure, and Lode-invariants-based hyperelasticity}
\label{sec:preliminaries}
\subsection{Kinematics}
\label{sec:kinematics}
In the framework of finite strain theory, the motion of a continuum body is described by a bijective mapping $\boldsymbol{\chi}$, which assigns to each material point $\mathbf{X}$ in the reference configuration its spatial position $\mathbf{x}$ in the current configuration at time $t$ as follows \cite{holzapfel2002nonlinear}: 

\begin{equation}
\mathbf{x} = \boldsymbol{\chi}(\mathbf{X}, t).
\end{equation}

The displacement vector field $\mathbf{u}$ characterizing the motion is consequently defined as
\begin{equation}
\mathbf{u}(\mathbf{X}, t) = \mathbf{x}(\mathbf{X}, t) - \mathbf{X}.
\label{eq:displacement}
\end{equation}

A fundamental kinematic quantity is the deformation gradient tensor $\mathbf{F}$, a two-point tensor that describes the local transformation of material line elements from the reference configuration to the current configuration, given as \cite{holzapfel2002nonlinear}

\begin{equation}
\mathbf{F}(\mathbf{X}, t) = \frac{\partial \boldsymbol{\chi}(\mathbf{X}, t)}{\partial \mathbf{X}} = \nabla_{\!\mathbf{X}}\, \mathbf{x}.
\label{eq:Fdef}
\end{equation}

Consistent with experimental observations in soft materials~\cite{bergstrom2015mechanics, li2019multiscale, mythravaruni2019incompressibility, doi2009gel}, we adopt the incompressibility constraint $J=\det\mathbf{F}(\mathbf{X},t)=1$.

    \begin{figure}[hbt!]
    \centering
    \includegraphics[width=5.186in, height=2.367in]{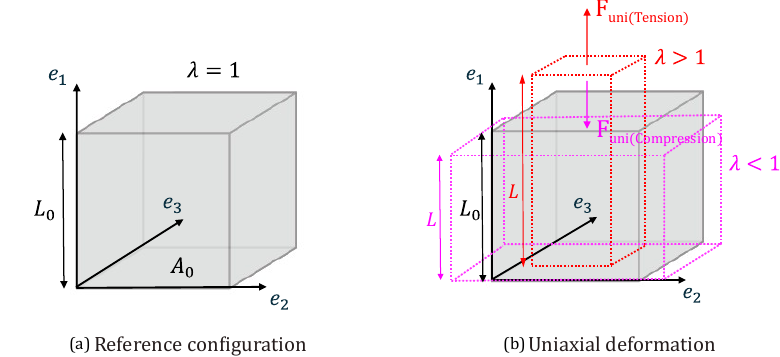}
    \caption{Unit cube in the (a) reference configuration and (b) under uniaxial deformation. Tension (\(\lambda>1\)) is shown in red; compression (\(0<\lambda<1\)) in magenta.}
    \label{fig:uniaxial_cube}
\end{figure}

Throughout this paper, we use uniaxial and pure shear deformation modes for calibration and/or validation.
\begin{itemize}
\item \textbf{Uniaxial deformation (compression and tension)}: Figure \ref{fig:uniaxial_cube} illustrates the uniaxial deformation of a unit cube. \(\{\mathbf{e}_1,\mathbf{e}_2,\mathbf{e}_3\}\) denote the orthonormal basis and \(\{\lambda_1,\lambda_2,\lambda_3\}\) the corresponding principal stretches. Owing to the incompressibility constraint,
\begin{equation}
\det\mathbf{F}=\lambda_1\lambda_2\lambda_3=1.
\label{eq:incompressible}
\end{equation}

For axial loading along \(\mathbf{e}_1\), the stretch $\lambda_1$ becomes
\begin{equation}
\lambda_1 = \lambda=\frac{L}{L_0},
\label{eq:stretch_def}
\end{equation}
with initial and current gauge lengths \(L_0\) and \(L\), respectively. Under Eq. \eqref{eq:incompressible}, the associated principal stretches are \(\{\lambda,\lambda^{-1/2},\lambda^{-1/2}\}\), giving the point-wise mapping
\begin{equation}
x_1=\lambda X_1,\qquad x_2=\lambda^{-1/2}X_2,\qquad x_3=\lambda^{-1/2}X_3,
\label{eq:uniaxial_map}
\end{equation}
and the deformation gradient
\begin{equation}
\mathbf{F}_{\text{uni}}=
\begin{bmatrix}
\lambda & 0 & 0\\[2pt]
0 & \lambda^{-1/2} & 0\\[2pt]
0 & 0 & \lambda^{-1/2}
\end{bmatrix}.
\label{eq:F_uni}
\end{equation}

\begin{figure}[hbt!]
    \centering
    \includegraphics[width=5.698in, height=2.117in]{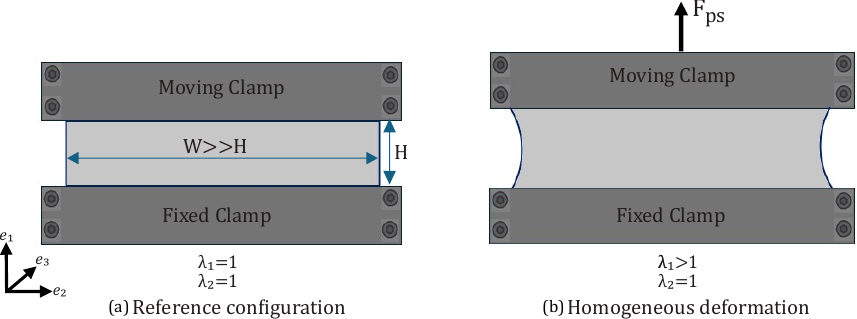} 
    \caption{Schematic illustration of a pure shear loading setup in the (a) reference configuration and (b) under homogeneous deformation.}
    \label{fig:pure_shear_setup} 
\end{figure}

\item \textbf{Pure shear deformation}: In the classical pure shear geometry (Rivlin–Thomas \cite{rivlin1953rupture}), illustrated in Fig. \ref{fig:pure_shear_setup}, a nearly homogeneous state is obtained when the specimen width \(W\) far exceeds the gauge height \(H\) (\(W\gg H\)) and the loading is applied parallel to \(H\). Considering a strictly incompressible material, the deformation gradient in pure shear is then expressed using principal stretches \(\{\lambda,1,\lambda^{-1}\}\) with $\lambda$ stretch along the loading direction \(\mathbf{e}_1\), yielding
\begin{equation}
x_1=\lambda X_1,\qquad x_2=X_2,\qquad x_3=\lambda^{-1}X_3
\label{eq:pure_shear_map}
\end{equation}
and the deformation gradient
\begin{equation}
\mathbf{F}_{\text{ps}}=
\begin{bmatrix}
\lambda & 0 & 0\\[2pt]
0 & 1 & 0\\[2pt]
0 & 0 & \lambda^{-1}
\end{bmatrix}.
\label{eq:F_ps}
\end{equation}
\end{itemize}

With the uniaxial and pure shear mappings established in Eqs. \eqref{eq:F_uni} and \eqref{eq:F_ps}, we next describe the classical energy limiters approach to modeling damage and failure in soft materials, whose inability to capture TCA formed a major motivation for this work.

\subsection{Classical energy limiters approach for modeling continuous softening}
\label{subsec:Classical_energy_limiters}

The constitutive response of a hyperelastic solid is specified through its strain energy density function \(W(\mathbf{F})\), which represents the stored energy per unit reference volume. A fundamental property of conventional hyperelasticity is that \( W\) grows indefinitely with increasing deformation, i.e.,
\begin{equation}
\mathbf{F} \to \infty \Rightarrow W \to \infty.
\label{eq:hyperelasticity_assumptions}
\end{equation}

While mathematically convenient, this result is physically unrealistic, as it implies that materials can sustain infinite deformation without failure, contradicting the empirical reality of finite energy storage and damage accumulation in any material.

To overcome this limitation and capture the onset of damage and bulk failure in soft materials, Volokh et al.~\cite{volokh2022modeling, volokh2007hyperelasticity, volokh2004nonlinear, volokh2008multiscale, Trapper2010} introduced the concept of energy limiters. This approach imposes a physically motivated upper bound on the strain energy density by introducing material-specific failure parameters (i.e., energy limiters) that capture the maximum energy an infinitesimal material volume can hold before failing. To this end, a bounded, modified strain energy density function $\psi$ was introduced ~\cite{volokh2021new} as given below:
\begin{flalign}
  \text{Volokh damage model:} && \psi(W(\mathbf{F})) = \psi_{f} - \psi_{e}(W(\mathbf{F}))\; =\frac{\Phi}{m} \left\{ \Gamma\left(\frac{1}{m}, 0\right) - \Gamma\left(\frac{1}{m}, \left(\frac{W(\mathbf{F})}{\Phi}\right)^m\right) \right\},&&
  \label{eq:volokh}
\end{flalign}
where
\begin{equation}
\Gamma(s, x) = \int_{x}^{\infty} t^{s-1} \exp(-t)  dt
\end{equation}
is the upper incomplete gamma function. Here, $\psi_f$ is the saturation energy and $\psi_e$ is the recoverable elastic energy. $W$ is the strain energy density in the undamaged material. The modified strain energy density function $\psi$ captures continuous softening and failure via the essential failure condition:

\begin{equation}
    \mathbf{F} \to \infty \Rightarrow \psi(W(\mathbf{F})) \to \psi_{f}=\frac{\Phi}{m}\Gamma\left(\frac{1}{m}, 0\right).
    \label{eq:volokh_classic_failure}
\end{equation}

The softening response is thus governed by two physically interpretable parameters: the pseudo--failure energy $\Phi$ that captures the onset of failure in the material (for $m=1$, $\psi_{f}=\Phi$), and the softening sharpness parameter $m$, which controls the sharpness of the transition to instability in the stress–strain response as softening progresses. Note, the model reduces to the conventional intact hyperelastic formulation (i.e., $W$) as $\Phi \rightarrow \infty$.

By frame indifference and assuming an isotropic response, \(W\) of an intact, undamaged material can be written as a function of invariants of any coaxial/isomorphic strain measure built from \(\mathbf F\). Common examples include the neo-Hookean and Mooney–Rivlin models \cite{mooney1940theory, rivlin1948large} that are based on the principal invariants $\left(I_1, I_2, I_3\right)$ of the right Cauchy–Green tensor $\mathbf{C}=\mathbf{F}^\mathrm{T}\mathbf{F}$, and the Ogden model \cite{ogden1972large} that is based on the principal stretches $\left(\lambda_1, \lambda_2, \lambda_3\right)$.

\begin{figure}[hbt!]
    \centering
    \includegraphics[width=3.49in, height=1.92in]{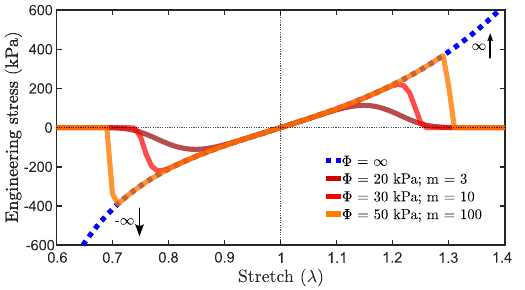}
    \caption{Typical uniaxial engineering stress–stretch responses of the Volokh damage model, considering the one-term Ogden strain energy density to represent intact material behavior. The dashed blue curve shows the undamaged response, while the solid curves demonstrate the softening response for different parameter pairs $(\Phi, m)$, as indicated in the legend. All curves use the following Ogden parameters: $\mu = 300~\mathrm{kPa}$ and $\alpha = 7$.}
    \label{fig:sym}
\end{figure}

Regardless of the choice of \(W\), the Volokh model always predicts a symmetric uniaxial stress–strain response. To demonstrate this, let's consider the Ogden hyperelastic model to represent intact behavior in Eq. (\ref{eq:volokh}):
\begin{equation}
W_{\mathrm{Ogden}}=\frac{2\mu}{\alpha^{2}}\bigl(\lambda_1^{\alpha}+\lambda_2^{\alpha}+\lambda_3^{\alpha}-3\bigr),
\label{eq:ogden}
\end{equation}
where \(\mu\) is the instantaneous shear modulus and \(\alpha\neq 0\) is the nonlinearity parameter. The corresponding principal Cauchy stress components can be obtained as
\begin{equation}
{T_i}=-p+\lambda_i\,\frac{\partial \psi}{\partial \lambda_i}\quad(\text{no sum}),
\label{eq:Cauchy-stretch-Volokh}
\end{equation}
where the Lagrange multiplier $p$ imposes incompressibility constraint (Eq. (\ref{eq:incompressible})). Under homogeneous uniaxial loading in $\mathbf{e}_1$ direction (Eq. (\ref{eq:F_uni})), we have \(\lambda_1=\lambda\), \(\lambda_2=\lambda_3=\lambda^{-1/2}\). Accordingly, the partial derivatives in Eq.~\eqref{eq:Cauchy-stretch-Volokh} can be expressed as:

\begin{equation}
\frac{\partial \psi}{\partial \lambda_1}= \frac{\partial \psi}{\partial W} \frac{\partial W}{\partial \lambda_1}= \frac{\partial \psi}{\partial W}\frac{2\mu}{\alpha}\,\lambda^{\alpha-1},\qquad
\frac{\partial \psi}{\partial \lambda_2}= \frac{\partial \psi}{\partial W} \frac{\partial W}{\partial \lambda_2}= \frac{\partial \psi}{\partial W}\frac{2\mu}{\alpha}\,\lambda^{-\tfrac{\alpha}{2}-1}
\label{eq:Volokh-Ogden-derivatives}
\end{equation}

Applying the chain rule to Eq.~\eqref{eq:volokh} then yields the stress reduction factor, $\mathrm{Red}(\lambda)$, which governs the onset and evolution of softening.

\begin{equation} 
\frac{\partial \psi}{\partial W} = \mathrm{Red}(\lambda) = \exp\!\left[-\!\left(\frac{W}{\Phi}\right)^{m}\right].
\label{eq:Volokh-derivative-wrt-W}
\end{equation}

Using Eqs.~(\ref{eq:Volokh-Ogden-derivatives})–(\ref{eq:Volokh-derivative-wrt-W}) in Eq.~(\ref{eq:Cauchy-stretch-Volokh}), and enforcing the traction boundary condition of a nonzero axial stress \(T_{1}\) with \(T_{2}=T_{3}=0\), allows determination of the Lagrange multiplier \(p\). Substituting this expression back into Eq.~(\ref{eq:Cauchy-stretch-Volokh}) yields the uniaxial Cauchy stress with the energy-limiting (failure) description:
\begin{flalign}
&\textrm{Volokh damage model:} &&
T_{\mathrm{Volokh}}^{(\mathrm{uniaxial})}
= \exp\!\left[-\!\left(\frac{2\mu}{\Phi\alpha^{2}}\bigl(\lambda^{\alpha}+2\lambda^{-\alpha/2}-3\bigr)\right)^{m}\right]
\frac{2\mu}{\alpha}\bigl(\lambda^{\alpha}-\lambda^{-\tfrac{\alpha}{2}}\bigr). &&
\label{eq:T11_uniax_volokh}
\end{flalign}

The corresponding first Piola–Kirchhoff (engineering) stress component, noting that \(\mathbf{P}=\mathbf{T}\mathbf{F}^{-\mathrm{T}}\), follows:
\begin{flalign}
&\textrm{Volokh damage model:} &&
P_{\mathrm{Volokh}}^{(\mathrm{uniaxial})}
=\frac{T_{\mathrm{Volokh}}^{(\mathrm{uniaxial})}}{\lambda}
=\exp\!\left[-\!\left(\frac{2\mu}{\Phi\alpha^{2}}\bigl(\lambda^{\alpha}+2\lambda^{-\alpha/2}-3\bigr)\right)^{m}\right]
\frac{2\mu}{\alpha}\bigl(\lambda^{\alpha-1}-\lambda^{-(1+\tfrac{\alpha}{2})}\bigr). &&
\label{eq:P11_uniax_volokh}
\end{flalign}

Figure \ref{fig:sym} shows the uniaxial engineering stress–stretch responses predicted using Eq. (\ref{eq:P11_uniax_volokh}) with Ogden model parameters $\mu = 300~\mathrm{kPa}$ and $\alpha = 7$ but varying failure parameters $\Phi$ and $m$. The dashed blue curve represents the intact hyperelastic response, while the solid curves illustrate the model's sensitivity to parameters $\Phi$ and $m$. 

A decrease in $\Phi$ causes an earlier deviation from the intact curve, signaling premature failure initiation. Conversely, a lower $m$ value reduces the steepness of the post-peak stress decay, leading to a more gradual softening response. A spectrum of failure behaviors—from gradual to abrupt, brittle-like failure (when $m \gg 1$)—can be captured through this model.

From Fig. \ref{fig:sym}, it is clear that the Volokh damage model enforces a perfectly symmetric tension–compression damage response. This inherent limitation prevents it from capturing the distinct, experimentally observed TCA (and deformation-mode-dependence in general) in continuous softening and failure in soft materials. This limitation forms a key motivation of the present work. 

We now describe the concept of Lode invariants and their recent application in deriving strain energy density functions that can efficiently capture TCA in elastic, intact material response.

\subsection{Lode-invariants-based hyperelasticity and the Prasad–Kannan model}
\label{subsec:lode_invariants_hyperelasticity}

Capturing the intrinsic TCA in soft materials requires a constitutive framework that can distinguish between different modes of distortion while upholding objectivity and compliance with thermodynamic principles. Classical strain-invariant formulations often fall short in this regard as the conventional invariants of the Cauchy–Green tensor cannot uniquely identify the deformation mode. To address this limitation, we leverage the advantages of a set of orthonormal invariants derived from the logarithmic (Hencky) strain tensor \cite{freed1995natural, hencky1931law}, originally introduced by Criscione et~al.~\cite{criscione2000invariant, criscione2002direct}. These invariants provide a complete and physically interpretable description of isotropic deformation states with mutually orthogonal stress components, wherein the contributions from dilatation and distortion are \textit{a priori} additively separable~\cite{flory1961thermodynamic, neff2015exponentiated, chen2012general, diani2005combining}.

Adopting the Hencky strain measure, $\mathbf{E} = \ln \mathbf{V}$, where $\mathbf{V}$ is the left stretch tensor obtained from the polar decomposition $\mathbf{F} = \mathbf{V}\mathbf{R}$ \cite{criscione2000invariant, neff2016geometry, prasad2020analysis}, the isotropic Lode invariants are defined as
\begin{equation}
\label{eq:K_invariant_formulas}
\underlabel{%
K_{1}=\frac{1}{\sqrt{3}}\operatorname{tr}(\ln\mathbf V)%
}{``amount of dilatation"}
\qquad
\underlabel{%
K_{2}=\bigl\|\operatorname{dev}(\ln\mathbf V)\bigr\|%
}{``magnitude of distortion"}
\qquad
\underlabel{%
K_{3}=\frac{1}{3}\sin^{-1}\!\left(
\frac{\sqrt{6}\,\operatorname{tr}\!\left[(\operatorname{dev}\ln\mathbf V)^{3}\right]}
     {\bigl\|\operatorname{dev}(\ln\mathbf V)\bigr\|^{3}}\right)%
}{``mode of distortion"}
\end{equation}

Each Lode invariant in Eq. (\ref{eq:K_invariant_formulas}) possesses distinct physical interpretations \cite{criscione2000invariant, chen2012general, prasad2020analysis, kulwant2023semi, landauer2019experimental} as given below: 
\begin{itemize}
    \item \(K_{1}\in(-\infty,\infty)\) quantifies the extent of volumetric change (dilatation) during deformation. Under the incompressibility constraint, \(K_{1} = 0\).
    \item \(K_{2}\in[0,\infty)\) quantifies the magnitude of distortion.
    \item \(K_{3}\in[-\pi/6,\pi/6]\), called the Lode angle, uniquely captures the mode (or type) of distortion. Note, \(K_{3}=-\pi/6\) for uniaxial compression, \(K_{3}=0\) for pure shear, and \(K_{3}=\pi/6\) for uniaxial tension.
\end{itemize}

For an isotropic hyperelastic solid, the strain energy density can thus be expressed as a function of the Lode invariants,
\begin{equation}
W = W(K_1, K_2, K_3),
\end{equation}
which, under incompressibility ($K_1=0$), simplifies to
\begin{equation}
W = W(K_2, K_3).
\end{equation}

Following Criscione et~al.~\cite{criscione2000invariant, falope2024energetic}, the Cauchy stress has an isochoric component associated with the Lagrange multiplier $p$ and a deviatoric component associated with $\mathbf{E}$, namely

\begin{equation}
\mathbf{T} =  - p\mathbf{I} + \frac{\partial W}{\partial \mathbf{E}}.
\label{eq:general_cauchy}
\end{equation}

In terms of the Lode invariants of $\mathbf{E}$, we have

\begin{equation}
\mathbf{T} = -p\,\mathbf{I} + \frac{\partial W}{\partial K_2}\,\mathbf{N}_1 + \frac{1}{K_2}\frac{\partial W}{\partial K_3}\,\mathbf{N}_2,
\label{eq:stress_rep_compact_PK}
\end{equation}
where $\mathbf{N}_1$ and $\mathbf{N}_2$ are mutually orthogonal, deviatoric unit tensors associated with distortional magnitude and mode, respectively (explicit forms are provided in~\ref{sec:appendix_A}). The scalar response functions are

\begin{equation}
\gamma_1 = \frac{\partial W}{\partial K_2} = \mathbf{T} : \mathbf{N}_1, 
\qquad
\gamma_2 = \frac{1}{K_2}\frac{\partial W}{\partial K_3} = \mathbf{T} : \mathbf{N}_2,
\label{eq:response_functions}
\end{equation}
where \(:\) denotes the double contraction operator, such that 
\( \mathbf{A} : \mathbf{B} = A_{ij}B_{ij} \) for any two second-order tensors 
\(\mathbf{A}\) and \(\mathbf{B}\).

To demonstrate the utility of this Lode-invariants-based framework, we adopt the Prasad–Kannan formulation ~\cite{prasad2020analysis, kulwant2023semi} as the intact hyperelastic potential, originally constructed from brain tissue data to define the first response function. This potential was constructed by imposing universal compatibility condition under uniaxial deformation, namely that all Cauchy stress components vanish except along the loading direction. In the Lode-invariant basis, this requirement translates to the orthogonality condition of the Cauchy stress tensor with respect to the second kinematic tensor, $\gamma_2 = \mathbf{T} : \mathbf{N}_2^{(\text{uniaxial})} = 0$, where the admissible functional form of the potential was determined through an \textit{a priori} analysis under the requirement of satisfying the Baker–Ericksen (B--E) inequalities \cite{baker1954inequalities}. This formulation further leverages the unique advantages of Lode invariants \cite{chen2012general, criscione2000invariant} to ensure thermodynamic consistency and smooth mode-dependence. The strain energy density function of the Prasad–Kannan model is expressed as
\begin{equation}
\begin{aligned}
W_{\mathrm{PK}}(K_2, K_3) &= \frac{\mu}{2} K_2^2
+ \frac{a \left[ \exp\!\left(  K_2 \mathcal{G}(K_3) \right) - 1 \right]}{ \mathcal{G}(K_3)}
- \frac{1}{2} a  K_2^2 \mathcal{G}(K_3)
- a K_2,
\end{aligned}
\label{eq:W_K2K3}
\end{equation}
where $\mu>0$ is the shear modulus\footnote{We retain the notation of Prasad and Kannan \cite{prasad2020analysis} in referring to the parameter \(\mu\) as the shear modulus. Under the present normalization of the quadratic small-strain term, however, the infinitesimal shear modulus in the strict linearized sense is \(G=\mu/2\).}and $a>0$ is the nonlinear strain-stiffening parameter. The Lode angle-dependent modulus function $\mathcal{G}(K_3)$, derived directly from the B–E inequalities, is given by
\begin{equation}
\mathcal{G}(K_3) =b_0 \left(
\frac{\exp\!\left( \frac{b_1}{2} - b_1 \cos\!\left(K_3 + \frac{\pi}{6}\right) \right)}{b_1}
+ \cos\!\left(K_3 + \frac{\pi}{6}\right)
+ \frac{1}{6}(\sqrt{7} - 2)\right),
\end{equation}
with model constants $b_0,b_1>0$. This function is positive, smooth, and strictly decreasing from uniaxial compression to tension, ensuring $\gamma_2(K_2,K_3=\pm\pi/6)=0$ and thereby maintaining universal uniaxial compatibility.

Substituting Eq.~(\ref{eq:W_K2K3}) into Eq.~(\ref{eq:response_functions}) yields the first and second response functions of the Prasad–Kannan model as given below \cite{prasad2020analysis}:

\begin{equation}
\gamma_{1,\mathrm{PK}}
    = \frac{\partial W_{\mathrm{PK}}}{\partial K_2}
    = a\!\left( \exp\!\big[K_2\,\mathcal{G}(K_3)\big] - 1 \right)
      - a\,K_2\,\mathcal{G}(K_3)
      + \mu\,K_2 ,
\label{eq:gamma1}
\end{equation}

\begin{equation}
\gamma_{2,\mathrm{PK}}
    = \frac{1}{K_2}\frac{\partial W_{\mathrm{PK}}}{\partial K_3}
    = -\,\frac{
        a\!\left(
            K_2^{2}\,\mathcal{G}(K_3)^{2}
            - 2K_2\,\mathcal{G}(K_3)\,\exp\!\big[K_2\,\mathcal{G}(K_3)\big]
            + 2\,\exp\!\big[K_2\,\mathcal{G}(K_3)\big]
            - 2
        \right)\mathcal{G}'(K_3)
        }{
        2K_2\,\mathcal{G}(K_3)^{2}
        }.
\label{eq:gamma2}
\end{equation}

The scalar functions $\gamma_{1,\mathrm{PK}}(K_2,K_3)$ and $\gamma_{2,\mathrm{PK}}(K_2,K_3)$ fully characterize the distortional stress response of the intact material and naturally embed mode dependence through the modulus function $\mathcal{G}(K_3)$. The function $\mathcal{G}(K_3)$ is constructed to monotonically decrease from compression to tension, consistent with experimental observations of human brain tissue reported by Budday et al.~\cite{budday2017mechanical, prasad2023new}. Thus, the Prasad–Kannan model has been shown to capture both bulk and nonlinear shear responses of soft tissues with high fidelity~\cite{prasad2020analysis}, making it an effective baseline model upon which the present damage and failure extension is built.

In summary, the Lode-invariants-based formulation provides a physically interpretable framework that decouples volumetric, distortional, and mode-dependent contributions to deformation. By employing the Prasad–Kannan potential as the intact hyperelastic response, the present study establishes a foundation for extending the energy limiters approach to account for TCA and distortion-mode-dependent softening in a thermodynamically consistent framework. The following section builds upon this formulation to introduce the proposed mode-sensitive failure extension incorporating mode-specific energy limiters. Before presenting this proposed model, we first examine the microstructural origins of TCA in soft material softening and failure.

\section{Microstructural Origins of Tension–Compression Asymmetry (TCA) in Soft Material Failure}
\label{sec:classical-energy-limiters}

From a mechanical standpoint, soft materials such as elastomers, hydrogels and biological tissues can be viewed as three-dimensional crosslinked polymer networks capable of sustaining large, finite deformations \cite{holzapfel2025biomechanics, treloar1975physics, treloar1943elasticity, treloar1943elasticity2, flory1943statistical, flory1944network, steck2023multiscale, sun2012highly}. Building on this microstructural perspective, the mechanical deformation and progression of damage in fibrous networks under tension can be delineated into three characteristic regimes, each reflecting a distinct stage of hierarchical structural reorganization \cite{islam2018effect}, as outlined below:

\begin{enumerate}
    \item \textbf{Regime I (Linear Elastic Response)}\\
    At small strains, geometric nonlinearity is minimal, load-bearing fibers respond elastically, and no major network reorientation occurs.

    \item \textbf{Regime II (Microstructural Realignment and Strain Stiffening)}\\
    With increasing deformation, fibers progressively align with the loading direction; transversely oriented fibers bend to facilitate the alignment of others. This reorganization introduces pronounced geometric nonlinearity, produces strain stiffening, and is accompanied by notable Poisson contraction.

    \item \textbf{Regime III (Stress-Path Establishment and Failure Onset)}\\
    Beyond a critical stretch, nearly straight, aligned chains form continuous load-bearing pathways across the material. As stretch increases, these chains carry growing loads until the weakest crosslink ruptures, marking the transition from distributed microdamage to macroscopic failure.
\end{enumerate}

An analogous three-stage evolution emerges under compression, but with fundamentally different governing processes \cite{filla2023multiscale, liang2017phase}. The response transitions from an initially linear regime to buckling-dominated softening and ultimately to a densification regime characterized by contact-driven stiffening and energy dissipation.

\begin{enumerate}
    \item \textbf{Regime I (Linear Elastic Response)}\\
    At small compressive strains, the material exhibits an effective stiffness comparable to that in tension, with negligible network reorganization and minimal fiber deformation.

    \item \textbf{Regime II (Buckling and Softening)}\\
    As compressive strain increases, fibers buckle and rearrange, producing a distributed reduction in global load-carrying capacity and a softening response that remains spatially delocalized in random networks.

    \item \textbf{Regime III (Densification and Rapid Stiffening)}\\
    At large compressive strains, extensive fiber–fiber contacts form, leading to localized densification, the emergence of crisscrossed bundles, and the activation of frictional and adhesive interactions. These effects collectively produce a sharp increase in stiffness and substantial energy dissipation, thereby governing the material’s capacity for energy absorption under compressive loading.

\end{enumerate}

These processes highlight how the interplay between distortional loading, microstructural evolution, and energy dissipation governs softening and ultimate failure in soft solids. TCA is a defining manifestation of this behavior, reflected in systematically distinct tensile and compressive responses. This asymmetry is particularly pronounced in biological and polymeric soft materials—such as brain tissue, fibrous gels, and elastomeric networks \cite{budday2017mechanical,budday2017rheological,drozdov2020tension}. The distinct mechanisms activated under tension—such as cavitation and microcrack initiation—and under compression—such as fibrillation, crack buckling, and network densification—demonstrate that deformation mode serves as an intrinsic regulator of the mechanical response, including material failure \cite{du2014variational, du2020tension, tang2019phase, volokh2017fracture, notbohm2015microbuckling,long2012crack,sun2022rheology}. In biopolymer networks, compressive instabilities commonly originate in weakly connected regions near free surfaces, initially manifesting as surface wrinkles and creases that subsequently propagate into the interior.  Such behavior reflects the loss of structural stability in the underlying crosslinked microstructure when subjected to increasing compressive distortion~\cite{ed2021poroviscoelasticity, yang2020inelasticity, doi2009gel, huang2016effects}.
 Collectively, these observations underscore the need for constitutive formulations that explicitly relate failure energetics to the mode of distortion, forming the basis for the generalized Lode-invariant energy-limiting model developed in the subsequent section.

\section{A Generalized Framework for Mode-Dependent Failure Using Lode Invariants}
\label{sec:prop_formulation}

The preceding section demonstrated that experimental trends motivate refining the classical  energy-limiting form $\psi(W\left(\mathbf{F}\right)) = \psi_f - \psi_e(W\left(\mathbf{F}\right))$
(see Section~\ref{subsec:Classical_energy_limiters},
Eqs.~\eqref{eq:volokh}–\eqref{eq:volokh_classic_failure}) to account for the TCA observed in softening and rupture.

\subsection{Bi-Failure construction with Lode–angle–dependent interpolation}

\begin{figure}[hbt!]
    \centering
    \includegraphics[width=5.925in, height=2.971in]{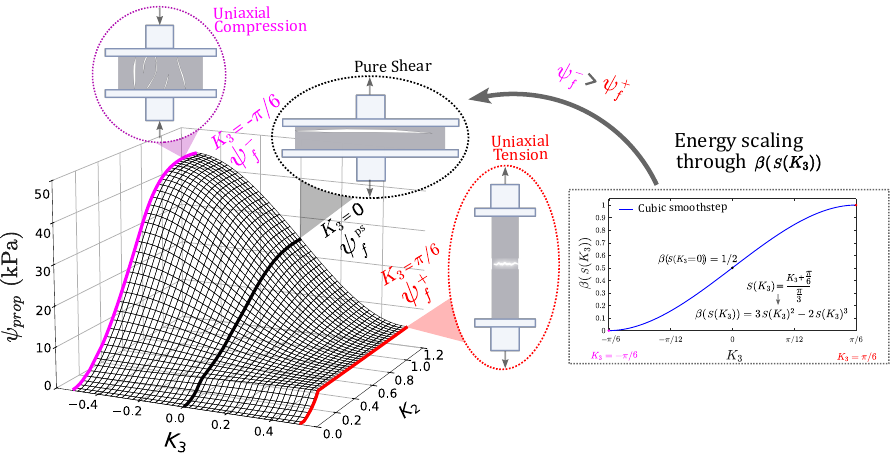}
    \caption{%
    A schematic of the proposed Lode-invariants-based bi-failure framework, with 
    $\psi_{\mathrm{prop}}(W(K_2, K_3))$ as the mode-dependent strain energy density. 
    The colored boundary curves correspond to uniaxial compression $(K_3=-\pi/6)$ (magenta), pure shear $(K_3=0)$, 
    and uniaxial tension $(K_3=\pi/6)$ (red), with distinct compressive and tensile failure energies 
    $\psi_f^{-}$ and $\psi_f^{+}$ indicated. 
    The inset shows the Lode-weighting function $\beta(s(K_3))$ that provides a smooth cubic smoothstep interpolation between the compressive $(\beta=0)$ and tensile $(\beta=1)$ branches, enabling a continuous, 
    mode-dependent description of softening and failure across the $(K_2,K_3)$ space and capturing 
    TCA.
    }
    \label{fig:asymm_prop}
\end{figure}

Extending the classical formulation necessitates a mechanism that separates tensile and compressive failure responses while ensuring consistency across all distortional modes. As depicted in Fig.~\ref{fig:asymm_prop}, we accomplish this through a \emph{bi-failure construction} that assigns distinct tensile $(+)$ and compressive $(-)$ limiters and introduces mode dependence via the Lode invariants $K_2$ and $K_3$ (defined in Section~\ref{subsec:lode_invariants_hyperelasticity}). The proposed strain energy density $\psi_{\text{prop}}$ is given by:

\begin{flalign}
&\textbf{Proposed asymmetric failure model:} &&
\psi_{\text{prop}}(W(K_2, K_3), K_3)
= \psi_{\mathrm{failure}}^{(\mathrm{prop})}(K_3)
- \psi_{\mathrm{elastic}}^{(\mathrm{prop})}(W(K_2, K_3), K_3), &&
\label{eq:Psi_prop_compact}
\end{flalign}

where $\psi_{\mathrm{failure}}^{(\mathrm{prop})}$ and $\psi_{\mathrm{elastic}}^{(\mathrm{prop})}$ denote the \textit{mode-dependent failure energy} and \textit{recoverable elastic energy}, respectively, defined as:

\begin{subequations}
\begin{align}
&\textbf{Proposed failure energy:} &&
\psi_{\mathrm{failure}}^{(\mathrm{prop})}(K_3)
= [1 - \beta(s(K_3))]\,\psi_f^{-} + \beta(s(K_3))\,\psi_f^{+}, && \\[3pt]
&\textbf{Proposed elastic energy:} &&
\psi_{\mathrm{elastic}}^{(\mathrm{prop})}(W(K_2, K_3), K_3)
= [1 - \beta(s(K_3))]\,\psi_e^{-}(W(K_2, K_3))
+ \beta(s(K_3))\,\psi_e^{+}(W(K_2, K_3)). &&
\end{align}
\label{eq:Psi_prop_components}
\end{subequations}

\noindent

Here $\psi_f^{-}$ and $\psi_f^{+}$ denote the \emph{compressive} and \emph{tensile} failure energies, respectively.  
These quantities are \emph{constants} for a given material; in the proposed model, they are defined as (cf. Eq. \eqref{eq:volokh_classic_failure}):

\begin{equation}
\left\{
\begin{aligned}
\psi_{f}^{+} &= \frac{\Phi^{+}}{m^{+}}\, 
  \Gamma\!\left(\frac{1}{m^{+}},\, 0\right),\\
\psi_{f}^{-} &= \frac{\Phi^{-}}{m^{-}}\, 
  \Gamma\!\left(\frac{1}{m^{-}},\, 0\right),
\end{aligned}
\right.
\label{eq:Psi_f_pm}
\end{equation}
where $\Phi^{\pm}$ and $m^{\pm}$ are the pseudo–failure energies and softening sharpness parameters, respectively, governing the failure energies under tensile $(+; K_3=\pi/6)$ and compressive $(-; K_3=-\pi/6)$ loading.

In contrast, the quantities $\psi_e^{+}(W(K_2,K_3))$ and $\psi_e^{-}(W(K_2,K_3))$ represent the \emph{tension-dominated} $(+)$ and \emph{compression-dominated} $(-)$ recoverable elastic energies, respectively, defined as (cf. Eq. (\ref{eq:volokh})):

\begin{equation}
\left\{
\begin{aligned}
\psi_{e}^{+}(W(K_2, K_3)) &=
\frac{\Phi^{+}}{m^{+}}\,
\Gamma\!\left(
  \frac{1}{m^{+}},\,
  \Bigl[\tfrac{W(K_2, K_3)}{\Phi^{+}}\Bigr]^{m^{+}}
\right), \\[4pt]
\psi_{e}^{-}(W(K_2, K_3)) &=
\frac{\Phi^{-}}{m^{-}}\,
\Gamma\!\left(
  \frac{1}{m^{-}},\,
  \Bigl[\tfrac{W(K_2, K_3)}{\Phi^{-}}\Bigr]^{m^{-}}
\right),
\end{aligned}
\right.
\label{eq:Psi_e_pm_1}
\end{equation}


Alternatively, the proposed strain energy density (Eq.~\eqref{eq:Psi_prop_compact}) may be recast in a form that explicitly connects it to branch-specific softening (cf. Eq.~\eqref{eq:volokh}) within the bi-failure framework as follows:

\begin{equation}
\begin{aligned}
\psi_{\text{prop}}(W(K_2, K_3), K_3) =[1 - \beta(s(K_3))]\,\psi^-(W(K_2, K_3)) + \beta(s(K_3))\,\psi^+(W(K_2, K_3))   \\[4pt]
\end{aligned}
\label{eq:psi_pm}
\end{equation}

Here, $\psi^+(W(K_2, K_3))$ and $\psi^-(W(K_2, K_3))$ represent the bounded (Volokh-type) potentials associated with tensile- and compressive-dominated distortion modes, respectively. These branch energies are defined according to Eqs.~\eqref{eq:Psi_f_pm} and \eqref{eq:Psi_e_pm_1} as

\begin{equation}
\left\{
\begin{aligned}
\psi^+(W(K_2, K_3)) &=  \psi_{f}^{+}
- \psi_{e}^{+}(W(K_2, K_3)) ,
 \\[4pt]
\psi^-(W(K_2, K_3))  &=
\psi_{f}^{-}
- \psi_{e}^{-}(W(K_2, K_3)),
\end{aligned}
\right.
\label{eq:psi_pm_exp}
\end{equation}

Although each branch is parameterized by its respective uniaxial energy limiters (i.e., $(\Phi^{+}, m^{+})$ or $(\Phi^{-}, m^{-})$), the functions $\psi_e^{\pm}$ depend on the intact elastic potential $W(K_2,K_3)$ and thus inherit explicit dependence on both Lode invariants $(K_2,K_3)$. Consequently, $\psi_e^{+}$ and $\psi_e^{-}$ vary smoothly with the distortion mode and quantify the recoverable portion of stored energy available prior to failure for any state characterized by $(K_2,K_3)$, while $\psi_f^{\pm}$ defines the fixed energetic thresholds identified in the canonical uniaxial modes. For the special cases of uniaxial tension ($K_3=\pi/6$) and uniaxial compression ($K_3=-\pi/6$), the corresponding recoverable elastic energies are reduced to:

\begin{equation}
\left\{
\begin{aligned}
\psi_{e}^{+}(W(K_2, K_3 ))\bigg|_{K_3=\tfrac{\pi}{6}} &=
\frac{\Phi^{+}}{m^{+}}\,
\Gamma\!\left(
  \frac{1}{m^{+}},\,
  \Bigl[\tfrac{W(K_2, K_{3}=\frac{\pi}{6})}{\Phi^{+}}\Bigr]^{m^{+}}
\right), \\[4pt]
\psi_{e}^{-}(W(K_2, K_3 ))\bigg|_{K_3=-\tfrac{\pi}{6}} &=
\frac{\Phi^{-}}{m^{-}}\,
\Gamma\!\left(
  \frac{1}{m^{-}},\,
  \Bigl[\tfrac{W(K_2, K_{3}=-\frac{\pi}{6}}{\Phi^{-}}\Bigr]^{m^{-}}
\right),
\end{aligned}
\right.
\label{eq:Psi_e_pm}
\end{equation}

\vspace{4pt}
\noindent

The weighting function \(\beta(s(K_3))\in[0,1]\) is introduced to provide a smooth interpolation between the compressive branch, for which \(\beta(s(-\pi/6))=0\), and the tensile branch, for which \(\beta(s(\pi/6))=1\).

The exact equation for \(\beta(s(K_3))\) is motivated by the zero end-derivative condition \cite{prasad2020analysis}
\begin{equation}
\left.\frac{\partial \psi_{\mathrm{prop}}}{\partial K_3}\right|_{K_3=\pm \frac{\pi}{6}} = 0,
\end{equation}
which is necessary to preserve compatibility with uniaxial deformation \cite{reddipaga2026construction}. In the present framework, this requirement implies that the weighting function must satisfy the following endpoint conditions (see Eq.~\eqref{eq:psi_K3}):

\begin{equation}
\left.\frac{d\beta}{dK_3}\right|_{K_3=\pm \frac{\pi}{6}}=0,
\end{equation}

so that
\begin{equation}
\left.\gamma_{2}^{(\mathrm{prop})}\right|_{K_3=\pm \frac{\pi}{6}}=0,
\qquad
\left.\mathbf{T}_{\mathrm{prop}}:\mathbf{N}_2\right|_{K_3=\pm \frac{\pi}{6}}=0.
\end{equation}

To enforce these conditions, we introduce the normalized Lode coordinate
\begin{equation}
s(K_3)=\frac{K_3+\pi/6}{\pi/3},
\qquad
s(K_3)\in[0,1],
\label{eq:normalized_lode_coordinate}
\end{equation}
and adopt the cubic smoothstep interpolation
\begin{align}
\beta(s(K_3))
&=
3\,s(K_3)^2-2\,s(K_3)^3,
\qquad
K_3\in\left[-\frac{\pi}{6},\frac{\pi}{6}\right].
\label{eq:Lode_weightage}
\end{align}
Its derivative with respect to \(K_3\) is
\begin{equation}
\frac{d\beta}{dK_3}
=
\beta'(s(K_3))
=
\frac{18}{\pi}s(K_3)\bigl(1-s(K_3)\bigr)
=
\frac{9}{2\pi}\left(1-\frac{36K_3^2}{\pi^2}\right),
\label{eq:Lode_weightage_derivative}
\end{equation}
which vanishes automatically at both endpoints. This construction therefore ensures exact recovery of the tensile and compressive branches at the canonical uniaxial states; see Section~\ref{subsubsec:uniaxial_deformation} for details.


Substituting the proposed uniaxial (compression and tension) failure energies and elastic energies (Eqs. \eqref{eq:Psi_f_pm}--\eqref{eq:Psi_e_pm_1}) in Eq. (\ref{eq:Psi_prop_components}), the general expression for the proposed strain energy density with distortion-mode softening $\psi_{\text{prop}}(W(K_2, K_3), K_3)$ (Eq. (\ref{eq:Psi_prop_compact})) becomes:

\begin{equation}
\begin{aligned}
\psi_{\text{prop}}(W(K_2, K_3), K_3) = & 
\underbrace{\frac{\Phi^{-}}{m^{-}}\Gamma\left(\frac{1}{m^{-}},0\right) + \beta(s(K_3))\left[\frac{\Phi^{+}}{m^{+}}\Gamma\left(\frac{1}{m^{+}},0\right) - \frac{\Phi^{-}}{m^{-}}\Gamma\left(\frac{1}{m^{-}},0\right)\right]}_{\psi_{failure}(K_3)} \\
& \hspace{-12em} - \underbrace{\left\{\frac{\Phi^{-}}{m^{-}}\Gamma\left(\frac{1}{m^{-}},\left[\frac{W(K_2,K_3)}{\Phi^{-}}\right]^{m^{-}}\right) + \beta(s(K_3))\left[\frac{\Phi^{+}}{m^{+}}\Gamma\left(\frac{1}{m^{+}},\left[\frac{W(K_2,K_3)}{\Phi^{+}}\right]^{m^{+}}\right) - \frac{\Phi^{-}}{m^{-}}\Gamma\left(\frac{1}{m^{-}},\left[\frac{W(K_2,K_3)}{\Phi^{-}}\right]^{m^{-}}\right)\right]\right\}}_{\psi_{elastic}(W(K_2,K_3),K_3)}
\end{aligned}
\label{eq:Psi_prop_expanded}
\end{equation}

The proposed formulation captures deformation-mode-dependent softening and damage evolution as an intrinsic function of the Lode invariants, thereby unifying tension, compression, and intermediate distortional modes within a single thermodynamically consistent framework. Note that following earlier extensions to energy limiters formulations by Volokh~et~al.~\cite{volokh2014irreversibility}, irreversibility can be incorporated through a switch parameter~$\zeta$ implemented via a Heaviside function~$H(\zeta)$ to enforce path dependence once the failure threshold is reached. In the present study, however, unloading and healing are not considered; hence, a reversible form is adopted by setting~$\zeta=0$, yielding~$H(\zeta)=1$ (see~\ref{appendix:irreversibility}).

In what follows, we derive the Cauchy and engineering stress--stretch relationships for three primary distortion states, establishing a predictive framework for calibration and validation.

\subsection{Stress--Stretch relationships for key homogeneous deformation modes}
\label{subsec:stress_stretch}

For the proposed softening model $\psi_{\mathrm{prop}}(W(K_2, K_3))$, the Cauchy stress is derived via the chain rule applied to the constitutive relation defined in Eq.\eqref{eq:stress_rep_compact_PK}:

\begin{equation}
\mathbf{T}_{\mathrm{prop}}(W({K_2,K_3}), K_3)
= -p \mathbf{I} 
+ \frac{\partial \psi_{\mathrm{prop}}}{\partial K_2} \mathbf{N}_1
+ \frac{1}{K_2}\frac{\partial \psi_{\mathrm{prop}}}{\partial K_3} \mathbf{N}_2 .
\label{eq:general_stress_rule_proposed}
\end{equation}

where the partial derivatives of the proposed strain energy density with respect to the invariants $K_i$ follow directly from the chain rule using Eqs.~\eqref{eq:psi_pm} and ~\eqref{eq:psi_pm_exp} and can be written as

\begin{equation}
\frac{\partial \psi_{\mathrm{prop}}}{\partial K_2}=  \gamma_{1}^{(\mathrm{prop})}
=
\left[(1-\beta(s(K_3)))\frac{\partial \psi^-}{\partial W} + \beta(s(K_3)) \frac{\partial \psi^+}{\partial W}\right]
\frac{\partial W}{\partial K_2},
\label{eq:psi_K2}
\end{equation}

\begin{equation}
\frac{1}{K_2}\frac{\partial \psi_{\mathrm{prop}}}{\partial K_3} = \gamma_{2}^{(\mathrm{prop})}
=
\frac{1}{K_2}\beta'(s(K_3))\left(\psi^{+}(W)-\psi^{-}(W)\right)
+
\left[(1-\beta(s(K_3)))\frac{\partial \psi^-}{\partial W} + \beta(s(K_3)) \frac{\partial \psi^+}{\partial W}\right]
\frac{1}{K_2}\frac{\partial W}{\partial K_3}.
\label{eq:psi_K3}
\end{equation}

Here, $\gamma_{1}^{(\mathrm{prop})}$ and $\gamma_{2}^{(\mathrm{prop})}$ denote the first and second scalar response functions of the proposed model, respectively. Substituting Eqs.~\eqref{eq:psi_K2}–\eqref{eq:psi_K3} into Eq.~\eqref{eq:general_stress_rule_proposed}, the general expression for the proposed Cauchy stress tensor becomes:
\begin{equation}
\begin{aligned}
\mathbf{T}_{\mathrm{prop}}(W({K_2,K_3}), K_3) 
= -p \mathbf{I}
&+ \left[(1-\beta(s(K_3)))\frac{\partial \psi^-}{\partial W} + \beta(s(K_3)) \frac{\partial \psi^+}{\partial W}\right]
\left(
\frac{\partial W}{\partial K_2}\mathbf{N}_1
+
\frac{1}{K_2}\frac{\partial W}{\partial K_3}\mathbf{N}_2
\right) \\
&+ \frac{\beta'(s(K_3))}{K_2}
\left(\psi^{+}(W)-\psi^{-}(W)\right)\mathbf{N}_2.
\end{aligned}
\label{eq:stress_proposed_final}
\end{equation}

The additional contribution proportional to $\beta'(s(K_3))$ (Eq.~\eqref{eq:Lode_weightage_derivative}) originates from the explicit dependence of the bi-failure interpolation on the Lode angle $K_3$. Similarly, the terms $\dfrac{\partial \psi^{\pm}}{\partial W} = \exp\!\left[-\!\left(\dfrac{W}{\Phi^{\pm}}\right)^{m^{\pm}}\right]$ correspond to the branch-specific stress reduction factors introduced in Eq.~\eqref{eq:Volokh-derivative-wrt-W}, which govern the softening behavior within each branch.

Recall that \(W(K_2, K_3)\) denotes the strain energy density of the intact/undamaged material. In this study, we adopt the Prasad–Kannan model $W_{\mathrm{PK}}$ (Eq.~\eqref{eq:W_K2K3}) to represent the intact response. Thus, the corresponding intact scalar response functions,  
\(\tfrac{\partial W_\mathrm{PK}}{\partial K_2} = \gamma_{1,\mathrm{PK}}\) and \(\tfrac{1}{K_2}\tfrac{\partial W_\mathrm{PK}}{\partial K_3} = \gamma_{2,\mathrm{PK}}\) (Eqs.~\eqref{eq:gamma1}–\eqref{eq:gamma2}), capture the intact stress-stretch responses which are subsequently modulated within the proposed asymmetric energy-limiting framework.

\subsubsection{Uniaxial deformation}
\label{subsubsec:uniaxial_deformation}


Uniaxial deformation is defined kinematically by the principal stretches $\lambda_1 = \lambda, \quad \lambda_2 = \lambda_3 = \lambda^{-1/2}$ (Eq. (\ref{eq:F_uni})). The corresponding Lode invariants are $K_2 = \tfrac{\sqrt{3}}{2}|\ln \lambda|$, and $K_3 = \pi/6$ for tension and $K_3 = -\pi/6$ for compression. At these limiting distortion modes, the Lode-weighting function $\beta(s(K_3))$ attains its boundary values (Eq.~\eqref{eq:Lode_weightage}), reducing the general potential $\psi_{\text{prop}}(W(K_2,K_3), K_3)$ to its mode-specific forms:
\begin{equation}
\psi_{\mathrm{prop}}^{(\mathrm{tension})}(W({K_2,K_3}),K_3)\bigg|_{K_3=\tfrac{\pi}{6}}
= \frac{\Phi^{+}}{m^{+}}
\!\left[
\Gamma\!\left(\frac{1}{m^{+}},0\right)
- \Gamma\!\left(
\frac{1}{m^{+}},
\Bigl(\frac{W(K_2,K_3=\frac{\pi}{6})}{\Phi^{+}}\Bigr)^{m^{+}}
\right)
\right],
\label{eq:Psi_prop_ten_simplified}
\end{equation}
and
\begin{equation}
\psi_{\mathrm{prop}}^{(\mathrm{comp})}(W({K_2,K_3}),K_3)\bigg|_{K_3=-\tfrac{\pi}{6}}
= \frac{\Phi^{-}}{m^{-}}
\!\left[
\Gamma\!\left(\frac{1}{m^{-}},0\right)
- \Gamma\!\left(
\frac{1}{m^{-}},
\Bigl(\frac{W(K_2,K_3=-\frac{\pi}{6})}{\Phi^{-}}\Bigr)^{m^{-}}
\right)
\right].
\label{eq:Psi_prop_comp_simplified}
\end{equation}

Remember, by invoking the universal relation for uniaxial deformation and the orthogonality of kinematic tensors, the second response function vanishes \cite{prasad2020analysis}, $\gamma_2 = \mathbf{T} : \mathbf{N}_2^{(\text{uniaxial})} = 0$ (see Eq.~\eqref{eq:response_functions}), implying the uniaxial stress is governed solely by $\mathbf{N}_1^{(\text{uniaxial})}$. Further, using the spectral decomposition of the Hencky strain $\mathbf{E} = \ln \mathbf{V} $, we have (see \cite{criscione2000invariant})
\begin{subequations}
\label{eq:N1_uniaxial_forms}
\begin{align}
{\mathbf{N}}_{1}^{(\text{tension})}
&= \sqrt{\tfrac{2}{3}}\!\left(
\mathbf{q}_1\!\otimes\!\mathbf{q}_1
- \tfrac{1}{2}\mathbf{q}_2\!\otimes\!\mathbf{q}_2
- \tfrac{1}{2}\mathbf{q}_3\!\otimes\!\mathbf{q}_3
\right), \\[3pt]
{\mathbf{N}}_{1}^{(\text{comp})}
&= \sqrt{\tfrac{2}{3}}\!\left(
-\mathbf{q}_1\!\otimes\!\mathbf{q}_1
+ \tfrac{1}{2}\mathbf{q}_2\!\otimes\!\mathbf{q}_2
+ \tfrac{1}{2}\mathbf{q}_3\!\otimes\!\mathbf{q}_3
\right),
\end{align}
\end{subequations}
wherein $\mathbf{q}_i$ ($i = 1,2,3$) represents the respective principal directions of $\mathbf{V}$.

With $\mathbf{T} = {T} _1^{(\mathrm{uniaxial})}\,\mathbf{q}_1 \otimes \mathbf{q}_1$ under uniaxial loading, enforcing the traction-free lateral boundary conditions ($T_{2}^{(\mathrm{uniaxial})}=T_{3}^{(\mathrm{uniaxial})}=0$) allows determination of the Lagrange multiplier $p$. For the respective cases of uniaxial tension and compression, this yields:
\begin{subequations}
\begin{align}
p^{(\text{tension})}(W({K_2,K_3}),K_3)\bigg|_{K_3=\tfrac{\pi}{6}}
&= -\frac{1}{\sqrt{6}}\,
    \frac{\partial W}{\partial K_2}\,
    \frac{\partial \psi_{\text{prop}}}{\partial W}
    \bigg|_{K_3=\tfrac{\pi}{6}},\\
p^{(\text{comp})}(W({K_2,K_3}),K_3)\bigg|_{K_3=-\tfrac{\pi}{6}}
&= \frac{1}{\sqrt{6}}\,
    \frac{\partial W}{\partial K_2}\,
    \frac{\partial \psi_{\text{prop}}}{\partial W}
    \bigg|_{K_3=-\tfrac{\pi}{6}}.
\end{align}
\label{eq:T_uniaxial}
\end{subequations}

Substituting Eqs. (\ref{eq:Psi_prop_ten_simplified})--(\ref{eq:T_uniaxial}) into Eq. (\ref{eq:general_stress_rule_proposed}) yields the Cauchy stress tensors for the corresponding uniaxial deformation modes. The resulting axial Cauchy stress components ($T_{1}$) for uniaxial tension and compression are given by:
\begin{subequations}
\begin{align}
T_{\text{prop}}^{(\text{tension})}(W({K_2,K_3}),K_3)\bigg|_{K_3=\tfrac{\pi}{6}}
&= \left(\frac{1}{\sqrt{6}} + \sqrt{\frac{2}{3}}\right)
   \frac{\partial W}{\partial K_2}
   \left.\frac{\partial \psi_{\text{prop}}}{\partial W}\right|_{K_3=\tfrac{\pi}{6}},
\\[4pt]
T_{\text{prop}}^{(\text{comp})}(W({K_2,K_3}),K_3)\bigg|_{K_3=-\tfrac{\pi}{6}}
&= -\left(\frac{1}{\sqrt{6}} + \sqrt{\frac{2}{3}}\right)
    \frac{\partial W}{\partial K_2}
    \left.\frac{\partial \psi_{\text{prop}}}{\partial W}\right|_{K_3=-\tfrac{\pi}{6}}.
\end{align}
\end{subequations}

The term $\partial W / \partial K_2$ in the above expressions is from the Prasad-Kannan formulation and is given in Eq.~\eqref{eq:gamma1}.

Since the experimental data are reported as engineering (i.e., first Piola–Kirchhoff) stress–stretch responses, we invoke the standard kinematic relation $\mathbf{P} = \,\mathbf{T}\mathbf{F}^{-\mathrm{T}}$ to obtain the axial components of the first Piola--Kirchhoff stress tensor under tension and compression:

\begin{subequations}
\begin{align}
\shortintertext{\textbf{Uniaxial tensile stress (proposed asymmetric failure model):}}
P_{\text{prop}}^{(\text{tension})}(W({K_2,K_3}),K_3)\bigg|_{K_3=\tfrac{\pi}{6}}
&= \frac{1}{\lambda}
\left(\frac{1}{\sqrt{6}}+\sqrt{\frac{2}{3}}\right)
\frac{\partial W}{\partial K_2}
\frac{\partial \psi_{\text{prop}}}{\partial W}\bigg|_{K_3=\tfrac{\pi}{6}},
\label{eq:uniaxial_tension_f}
\\[-2pt]
&\hspace{1.5em}\text{where }\;
\frac{\partial \psi_{\text{prop}}}{\partial W}\bigg|_{K_3=\tfrac{\pi}{6}}= \mathrm{Red_{\text{prop}}^{(\text{tension})}}(\lambda)= \frac{\partial \psi^+}{\partial W}
= \exp\!\left[-\!\left(\frac{W}{\Phi^{+}}\right)^{m^{+}}\right].
\label{eq:uniaxial_tension_final}
\\[6pt]
\shortintertext{\textbf{Uniaxial compressive stress (proposed asymmetric failure model):}}
P_{\text{prop}}^{(\text{comp})}(W({K_2,K_3}),K_3)\bigg|_{K_3=-\tfrac{\pi}{6}}
&= -\,\frac{1}{\lambda}
\left(\frac{1}{\sqrt{6}}+\sqrt{\frac{2}{3}}\right)
\frac{\partial W}{\partial K_2}
\frac{\partial \psi_{\text{prop}}}{\partial W}\bigg|_{K_3=-\tfrac{\pi}{6}},
\label{eq:uniaxial_compression_f}
\\[-2pt]
&\hspace{1.5em}\text{where }\;
\frac{\partial \psi_{\text{prop}}}{\partial W}\bigg|_{K_3=-\tfrac{\pi}{6}} =\mathrm{Red_{\text{prop}}^{(\text{comp})}}(\lambda)=\frac{\partial \psi^-}{\partial W} = \exp\!\left[-\!\left(\frac{W}{\Phi^{-}}\right)^{m^{-}}\right].
\label{eq:uniaxial_compression_final}
\end{align}
\end{subequations}

In Eqs.~\eqref{eq:uniaxial_tension_final}–\eqref{eq:uniaxial_compression_final}, $\mathrm{Red_{\text{prop}}^{(\text{ten})}(\lambda)}$ and $\mathrm{Red_{\text{prop}}^{(\text{comp})}(\lambda)}$ are mode-dependent \emph{stress-reduction factors} that regulate the transition from the intact elastic state to the damaged (softening) regime under uniaxial tension and compression, respectively (cf. Eq. (\ref{eq:Volokh-derivative-wrt-W})).

Having established the mode-specific stress–stretch responses under uniaxial deformation, we next generalize the formulation to capture a distinct distortional mode: pure shear.


\subsubsection{Pure shear deformation}
\label{subsubsec:pure_shear_deformation}

Pure shear deformation is defined kinematically by the principal stretches:
$\lambda_1 = \lambda, \quad \lambda_2 = 1, \quad \lambda_3 = \lambda^{-1}$ (Eq. ~\eqref{eq:F_ps}). The corresponding Lode invariants are $K_2 = \sqrt{2}|\ln\lambda|$ and $K_3 = 0$. Thus, the Lode-weighting function $\beta(S(K_3=0)) = \tfrac{1}{2}$. Under this condition, the proposed strain energy density in pure shear assumes the following form through Eq.~\eqref{eq:Psi_prop_compact}: 

\begin{equation}
\begin{aligned}
\psi_{\text{prop}}^{(\text{ps})}(W({K_2,K_3}),K_3)\bigg|_{K_3=0}
= \psi_{\mathrm{failure}}^{(\mathrm{ps})}(K_3=0)
- \psi_{\mathrm{elastic}}^{(\mathrm{ps})}(W(K_2, K_3=0), K_3=0)
\end{aligned}
\label{eq:psi_pureshear}
\end{equation}

The associated pure-shear-specific failure and elastic energy contributions are obtained from Eq.~\eqref{eq:psi_pm} as:
\begin{subequations}
\begin{align}
&\textrm{Failure energy (pure shear):} &
\psi_{f}^{(\mathrm{ps})}(K_3=0)
&= \frac{\psi_f^{+} + \psi_f^{-}}{2}, \\[3pt]
&\textrm{Elastic energy (pure shear):} &
\psi_{e}^{(\mathrm{ps})}(W({K_2,K_3=0}),K_3=0)
&= \frac{\psi_e^{+}(W(K_2,K_3=0)) + \psi_e^{-}(W(K_2,K_3=0))}{2}.
\end{align}
\label{eq:psi_ps_energies}
\end{subequations}

By substituting Eq.~\eqref{eq:psi_ps_energies} in Eq.~\eqref{eq:psi_pureshear} yields
\begin{equation}
\begin{aligned}
\psi_{\text{prop}}^{(\text{ps})}(W({K_2,K_3}),K_3)\bigg|_{K_3=0}
&= \frac{1}{2}\Bigg[
\frac{\Phi^{+}}{m^{+}}
\Bigg(
\Gamma\!\left(\frac{1}{m^{+}},0\right)
-
\Gamma\!\left(\frac{1}{m^{+}},\Bigl[\frac{W(K_2,K_3=0)}{\Phi^{+}}\Bigr]^{m^{+}}\right)
\Bigg)
\\
&\hspace{3.0em}
+
\frac{\Phi^{-}}{m^{-}}
\Bigg(
\Gamma\!\left(\frac{1}{m^{-}},0\right)
-
\Gamma\!\left(\frac{1}{m^{-}},\Bigl[\frac{W(K_2,K_3=0)}{\Phi^{-}}\Bigr]^{m^{-}}\right)
\Bigg)
\Bigg]
.
\end{aligned}
\label{eq:phi_pureshear}
\end{equation}

indicating that the proposed potential $\psi_{\text{prop}}^{(\mathrm{ps})}$ (Eq.~\eqref{eq:Psi_prop_compact}) represents the average of the tensile and compressive distortion modes under pure shear.

For $K_3 = 0$, the orthonormal kinematic tensors in the principal directions 
$\{\mathbf{q}_i\}$ are (see ~\ref{sec:appendix_A})
\begin{subequations}\label{eq:N1N2_ps}
\begin{align}
\mathbf{N}_1^{(\mathrm{ps})}
&=
\frac{1}{\sqrt{2}}
\left(
\mathbf{q}_1 \otimes \mathbf{q}_1
-
\mathbf{q}_3 \otimes \mathbf{q}_3
\right),
\label{eq:N1N2_ps_a}
\\
\mathbf{N}_2^{(\mathrm{ps})}
&=
\frac{1}{\sqrt{6}}
\left(
\mathbf{q}_1 \otimes \mathbf{q}_1
- 2\,\mathbf{q}_2 \otimes \mathbf{q}_2
+ \mathbf{q}_3 \otimes \mathbf{q}_3
\right).
\label{eq:N1N2_ps_b}
\end{align}
\end{subequations}

Substituting into the general stress representation 
(Eq.~\eqref{eq:general_stress_rule_proposed}) 
yields the principal Cauchy stresses
\begin{subequations}
\begin{align}
T_1^{(\mathrm{ps})} &= -p + \frac{1}{\sqrt{2}}\frac{\partial \psi_{\mathrm{prop}}}{\partial K_2} + \frac{1}{\sqrt{6}K_2}\frac{\partial \psi_{\mathrm{prop}}}{\partial K_3},
\label{eq:T1_AB}
\\
T_2^{(\mathrm{ps})} &= -p - \frac{2}{\sqrt{6}K_2}\frac{\partial \psi_{\mathrm{prop}}}{\partial K_3},
\label{eq:T2_AB}
\\
T_3^{(\mathrm{ps})} &= -p - \frac{1}{\sqrt{2}}\frac{\partial \psi_{\mathrm{prop}}}{\partial K_2}  + \frac{1}{\sqrt{6}K_2}\frac{\partial \psi_{\mathrm{prop}}}{\partial K_3}.
\label{eq:T3_AB}
\end{align}
\end{subequations}

Under pure shear, $\mathbf{T} = {T} _1^{(\mathrm{ps})}\,\mathbf{q}_1 \otimes \mathbf{q}_1 +{T} _2^{(\mathrm{ps})}\,\mathbf{q}_2 \otimes \mathbf{q}_2$, and imposing the traction‐free condition in the thickness direction ($T_{3}^{(\mathrm{ps})}=0$ in Eq.~\eqref{eq:T3_AB})  \cite{ogden1972large, criscione2000invariant}, the Lagrange multiplier ($p$) can be expressed as:

\begin{equation}
p^{(\mathrm{ps})}(W({K_2,K_3}),K_3)\bigg|_{K_3=0}
=
- \frac{1}{\sqrt{2}}\frac{\partial \psi_{\mathrm{prop}}}{\partial K_2}  + \frac{1}{\sqrt{6}K_2}\frac{\partial \psi_{\mathrm{prop}}}{\partial K_3}.
\label{eq:p_pure_shear}
\end{equation}

Substituting Eq.~\eqref{eq:p_pure_shear} in Eq.~\eqref{eq:T1_AB} yields the axial Cauchy stress along the direction of extension ($T_1^{(\mathrm{ps})}$) under pure shear deformation\footnote{
Although the second response function of the proposed model $\gamma_{2}^{(\mathrm{prop})}=\frac{1}{K_2}\frac{\partial \psi_{\text{prop}}}{\partial K_3}\neq 0$ at $K_3=0$,
the associated $\mathbf{N}_2$ doesn't contribute to the axial stress $T_1^{(\mathrm{ps})}$ under pure shear
once the traction-free condition $T_3^{(\mathrm{ps})}=0$ is enforced, as shown here. Consequently, the axial pure shear stress $T_1^{(\mathrm{ps})}$ depends solely on $\gamma_{1}^{(\mathrm{prop})} = \frac{\partial \psi_{\text{prop}}}{\partial K_2}$ (Eq.~\eqref{eq:psi_K2}). However, the second response function $\gamma_2^{(\mathrm{prop})}$, together with the  kinematic tensor $\mathbf{N}_2$, does influence the lateral stress component $T_2^{(\mathrm{ps})}$ through Eq.~\eqref{eq:T2_AB}.}:

\begin{equation}
T_{\text{prop}}^{(\mathrm{ps})}(W({K_2,K_3}),K_3)\bigg|_{K_3=0}
=\frac{2}{\sqrt{2}}\frac{\partial \psi_{\mathrm{prop}}}{\partial K_2}= \sqrt{2}\,
\frac{\partial W}{\partial K_2}
\frac{\partial \psi_{\text{prop}}}{\partial W}\bigg|_{K_3=0}.
\label{eq:Tps_intermediate}
\end{equation}

Using $\mathbf{P}=\mathbf{T}\mathbf{F}^{-\mathrm{T}}$, the corresponding axial component of the first Piola–Kirchhoff stress is derived as
\begin{subequations} \label{eq:Pps_final}
\begin{align}
\shortintertext{\textbf{{Pure shear axial stress (proposed asymmetric failure model):}}}
P_{\text{prop}}^{(\mathrm{ps})}(W({K_2,K_3}),K_3)\bigg|_{K_3=0}
&= \frac{T_{\text{prop}}^{(\mathrm{ps})}}{\lambda}
= \frac{\sqrt{2}}{\lambda}\,
\frac{\partial W}{\partial K_2}
\frac{\partial \psi_{\text{prop}}}{\partial W}\bigg|_{K_3=0},
\\[-2pt]
&\hspace{-12em}\text{where }\;
\frac{\partial \psi_{\text{prop}}}{\partial W}\bigg|_{K_3=0} =\mathrm{Red}_{\text{prop}}^{(\text{ps})}(\lambda) = \frac{1}{2}\left[\frac{\partial \psi^+}{\partial W} + \frac{\partial \psi^-}{\partial W}\right]
= \frac{1}{2}\!\left[
\exp\!\left[-\!\left(\frac{W}{\Phi^{+}}\right)^{m^{+}}\right]
+ \exp\!\left[-\!\left(\frac{W}{\Phi^{-}}\right)^{m^{-}}\right]
\right].
\end{align}
\end{subequations}


Overall, the specific asymmetric failure model proposed in Eq. \eqref{eq:Psi_prop_expanded}, when employing the Prasad--Kannan formulation \cite{prasad2020analysis} (Eqs. (\ref{eq:W_K2K3}-\ref{eq:gamma2})) as the intact hyperelastic baseline, comprises eight material parameters (see Table \ref{tab:params_model})---four elastic parameters $(\mu, a, b_{0}, b_{1})$ and four mode-specific failure parameters $(\Phi^{\pm}, m^{\pm})$. Nevertheless, the generalized bi-failure framework (Eq. (\ref{eq:Psi_prop_compact})) developed in this study remains agnostic to the specific choice of hyperelastic model; any formulation expressed in terms of the Lode invariants is fully compatible with the mode-dependent energy-limiting structure introduced here. Furthermore, once the uniaxial tension and compression responses are used to calibrate the failure parameters, the model provides a complete description of failure across the entire distortion space. Intermediate deformation modes---such as pure shear---are naturally captured through the Lode-weighting function $\beta(s(K_3))$.

\begin{table}[t] 
  \centering
  \caption{Model parameters for the proposed asymmetric failure model in Eq. \eqref{eq:Psi_prop_expanded}, which is based on a Lode-invariants-based bi-failure formulation. Note, our generalized formulation (Eq. (\ref{eq:Psi_prop_compact})) is agnostic to the choice of the hyperelastic model.}
  \label{tab:params_model}
  \renewcommand{\arraystretch}{1.2}
  \setlength{\tabcolsep}{6pt}
  \begin{tabular}{llc}
    \toprule
    \multicolumn{3}{l}{\textbf{Hyperelastic parameters (Prasad--Kannan model)}} \\
    \midrule
    \textbf{Symbol} & \textbf{Description}                                  & \textbf{Unit} \\ \midrule
    $\mu$           & Shear modulus                           & kPa           \\
    $a$             & Nonlinear strain-stiffening parameter                         & kPa           \\
    $b_0$           & \multirow{2}{*}{Mode-dependent modulus modulation parameters} & \multirow{2}{*}{--} \\
    $b_1$           &                                                       &               \\
    \midrule
    \multicolumn{3}{l}{\textbf{Mode-specific failure parameters (energy limiters)}} \\
    \midrule
    \multicolumn{3}{l}{\emph{Tension branch $(+)$}} \\
    \midrule
    $\Phi^{+}$      & Tensile pseudo--failure energy parameter                                & kPa           \\
    $m^{+}$         & Tensile softening sharpness parameter                  & --            \\
    \midrule
    \multicolumn{3}{l}{\emph{Compression branch $(-)$}} \\
    \midrule
    $\Phi^{-}$      & Compressive pseudo--failure energy parameter                           & kPa           \\
    $m^{-}$         & Compressive softening sharpness parameter              & --            \\
    \bottomrule
  \end{tabular}
\end{table}

\subsection{Behavior in the vicinity of $K_2=0$ and consistency with linear elasticity}
\label{subsec:K2_zero_behavior}

The limit $K_2 \to 0$ corresponds to vanishing distortion, i.e., 
$\|\operatorname{dev}(\mathbf E)\| \to 0$. In this regime, the deformation approaches a purely dilatational state and is governed solely by the volumetric invariant $K_1$, while the Lode invariant $K_3$ becomes indeterminate and loses independent physical significance. It is therefore essential to verify that the proposed constitutive model remains regular as $K_2 \to 0$. Accordingly, all deviatoric stress contributions must vanish continuously in the neighborhood of $K_2 = 0$. Mathematically, this requires that
\begin{equation}
\mathbf{T} \to -p\,\mathbf{I}
\quad \text{as } K_2 \to 0,
\end{equation}
which implies
\begin{equation}
\lim_{K_2 \to 0} \mathbf{T}:\mathbf{N}_1 = 0,
\qquad
\lim_{K_2 \to 0} \mathbf{T}:\mathbf{N}_2 = 0.
\label{eq:hydrostatic_condition}
\end{equation}

Using the stress representation of the proposed model in Section~\ref{subsec:stress_stretch} (Eq.~\eqref{eq:general_stress_rule_proposed}), the deviatoric stress is governed by the scalar response functions
\(\gamma_{1}^{(\mathrm{prop})}\) and \(\gamma_{2}^{(\mathrm{prop})}\). Since $\mathbf{N}_1$ and $\mathbf{N}_2$ are bounded orthonormal deviatoric tensors (see \ref{sec:appendix_A}), an equivalent condition to Eq. (\ref{eq:hydrostatic_condition}) is \cite{criscione2000invariant}
\begin{equation}
\lim_{K_2 \to 0} \gamma_{1}^{(\mathrm{prop})} = \lim_{K_2 \to 0} 
\frac{\partial \psi_{\mathrm{prop}}}{\partial K_2} = 0,
\qquad
\lim_{K_2 \to 0} \gamma_{2}^{(\mathrm{prop})} = \lim_{K_2 \to 0} 
\frac{1}{K_2}\frac{\partial \psi_{\mathrm{prop}}}{\partial K_3} = 0,
\label{eq:K2_conditions}
\end{equation}

In order to show that the proposed model follows this limiting condition on response functions $\gamma_{1}^{(\mathrm{prop})}$ and $\gamma_{2}^{(\mathrm{prop})}$, we examine the behavior of the following terms in Eqs. (\ref{eq:psi_K2}) and (\ref{eq:psi_K3}), respectively, at $K_2 \to 0$ i.e., (i) the response functions of the baseline intact hyperelastic model chosen in this work, the Prasad--Kannan model: $\gamma_{1,\mathrm{PK}}= \frac{\partial W_{\mathrm{PK}}}{\partial K_2}$ (Eq. (\ref{eq:gamma1})) and $\gamma_{2,\mathrm{PK}} = \frac{1}{K_2}\frac{\partial W_{\mathrm{PK}}}{\partial K_3}$ (Eq. (\ref{eq:gamma2})); and (ii) the additional contribution of $\beta'(s(K_3))$ in $\gamma_{2}^{(\mathrm{prop})}$: $\left[\frac{1}{K_2}\beta'(s(K_3))\left(\psi^{+}(W)-\psi^{-}(W)\right)\right]$. The proposed model will satisfy Eq. (\ref{eq:K2_conditions}) if both of these contributions vanish as $K_2 \to 0$ 

The Prasad--Kannan potential (Eq. (\ref{eq:W_K2K3})) admits the following small-$K_2$ expansion~\cite{prasad2020analysis}
\begin{equation}
W_{\mathrm{PK}}
=
\frac{\mu}{2}K_2^2
+
\frac{a}{6}K_2^3 \mathcal{G}(K_3)^2
+
O(K_2^4).
\label{eq:WPK_smallK2}
\end{equation}
Consequently, from Eqs.~\eqref{eq:gamma1}--\eqref{eq:gamma2}:
\begin{equation}
\gamma_{1,\mathrm{PK}} 
= \frac{\partial W_{\mathrm{PK}}}{\partial K_2}
= \mu K_2 
+ \frac{a}{2}\,\mathcal{G}(K_3)^2\,K_2^2 
+ O(K_2^3)
\quad \text{as } K_2 \to 0,
\end{equation}
and
\begin{equation}
\gamma_{2,\mathrm{PK}} =
\frac{1}{K_2}\frac{\partial W_{\mathrm{PK}}}{\partial K_3}
=
\frac{a}{3}K_2^2
\mathcal{G}(K_3)\mathcal{G}'(K_3)
+
O(K_2^3) 
\quad \text{as } K_2 \to 0,
\end{equation}
so that
\begin{equation}
\gamma_{1,\mathrm{PK}} = \frac{\partial W_{\mathrm{PK}}}{\partial K_2} = O(K_2),
\qquad
\gamma_{2,\mathrm{PK}} = \frac{1}{K_2}\frac{\partial W_{\mathrm{PK}}}{\partial K_3} = O(K_2^2) \quad \text{as } K_2 \to 0.
\label{eq:PK_response_fn}
\end{equation}
Hence, as demonstrated by Prasad and Kannan \cite{prasad2020analysis},
\begin{equation}
\lim_{K_2\to 0}\gamma_{1,\mathrm{PK}}=0,
\qquad
\lim_{K_2\to 0}\gamma_{2,\mathrm{PK}}=0.
\label{eq:WPK_limits}
\end{equation}

We now turn to the addition interpolation term of the form $\frac{1}{K_2}\beta'(s(K_3))(\psi^+ - \psi^-)$.
For each softening branch, neglecting higher-order terms in the small-$W$ asymptotic expansion given in Eqs.~\eqref{eq:psi_small_W} of~\ref{appendix:volokh_limit} yields
\begin{equation}
\psi^\pm(W)=W+O(W^{1+m^\pm}),
\quad \text{as } W \to 0.
\label{eq:Branch_limits}
\end{equation}
Clearly, the behavior of each branch-specific softening potential ($\psi^\pm(W)$) in the neighborhood of $K_2=0$ (and hence decay of $W$) explicitly depends on the branch-specific softening sharpness parameters $m^\pm$ of the chosen intact potential, as derived in~\ref{appendix:volokh_limit}.
Since Eq.~\eqref{eq:WPK_smallK2} implies that $W=W_{\mathrm{PK}}=O(K_2^2)$ as $K_2\to 0$, one obtains
\begin{equation}
\psi^+(W_{\mathrm{PK}})-\psi^-(W_{\mathrm{PK}})
=
O(K_2^{2(1+\bar m)}),
\qquad
\bar m=\min(m^+,m^-)>0.
\end{equation}
Hence, the interpolation contribution in $\gamma_{2}^{(\mathrm{prop})}$ (Eq. \eqref{eq:psi_K3}) satisfies
\begin{equation}
\frac{\beta'(s(K_3))}{K_2}
\big(\psi^+(W_{\mathrm{PK}})-\psi^-(W_{\mathrm{PK}})\big)
=
O(K_2^{1+2\bar m}) \to 0,
\qquad K_2 \to 0.
\label{eq:gamma2_limits}
\end{equation}
In other words,
\begin{equation}
\lim_{K_2\to 0}\frac{\beta'(s(K_3))}{K_2}
\big(\psi^+(W_{\mathrm{PK}})-\psi^-(W_{\mathrm{PK}})\big)
= 0.
\label{eq:gamma2_limits_lim}
\end{equation}

From Eq.~\eqref{eq:WPK_limits} and Eq.\eqref{eq:gamma2_limits_lim}, both conditions in Eq.~\eqref{eq:K2_conditions} are satisfied\footnote{From Eq.~\eqref{eq:dpsi_small_W} of~\ref{appendix:volokh_limit}, the weighted sum of the derivatives of branch-specific energies with respect to the intact strain energy density---$\left[(1-\beta(s(K_3)))\frac{\partial \psi^-}{\partial W} + \beta(s(K_3)) \frac{\partial \psi^+}{\partial W}\right]$---tends of unity as $K_2\to 0$. In the equations for the response functions $\gamma_{1}^{(\mathrm{prop})}$ and $\gamma_{2}^{(\mathrm{prop})}$ (Eqs. (\ref{eq:psi_K2}) and (\ref{eq:psi_K3})) this term appears as the coefficient of $\frac{\partial W_{\mathrm{PK}}}{\partial K_2}$ and $\frac{1}{K_2}\frac{\partial W_{\mathrm{PK}}}{\partial K_3}$, respectively.}. From this analysis, it is clear that the proposed model remains regular as $K_2 \to 0$, and the stress reduces to a purely hydrostatic response, consistent with isotropic symmetry. 

\paragraph{\textbf{Consistency with linear elasticity}}
The proposed constitutive relation should reduce to isotropic linear elasticity in the small-strain regime, where dilatation and distortions are small and rotations are negligible. In terms of the Lode-invariant representation, this condition implies that the leading order of the proposed strain energy density $\psi_{\mathrm{prop}}$ be quadratic in $K_2$ and independent of $K_3$ at quadratic order (see \cite{criscione2000invariant}).

To verify this, we make use of the previously established small-$K_2$ expansion of the intact Prasad--Kannan potential (Eq.~\eqref{eq:WPK_smallK2}), together with the asymptotic expansion of the softening branches in Eq.~\eqref{eq:Branch_limits}. This gives

\begin{equation}
\psi^\pm(W_{\mathrm{PK}})
=
W_{\mathrm{PK}}+O\!\left(K_2^{2(1+m^\pm)}\right).
\label{eq:psi_branch_m_scaling}
\end{equation}
Substituting Eq.~\eqref{eq:psi_branch_m_scaling} into the proposed bi-failure interpolation in Eq.~\eqref{eq:psi_pm} yields
\begin{equation}
\psi_{\mathrm{prop}}
=
\bigl(1-\beta(s(K_3))\bigr)
\left[W_{\mathrm{PK}}+O\!\left(K_2^{2(1+m^-)}\right)\right]
+
\beta(s(K_3))
\left[W_{\mathrm{PK}}+O\!\left(K_2^{2(1+m^+)}\right)\right].
\end{equation}
Expanding gives
\begin{equation}
\psi_{\mathrm{prop}}
=
\bigl[(1-\beta(s(K_3)))+\beta(s(K_3))\bigr]W_{\mathrm{PK}}
+
(1-\beta(s(K_3)))\,O\!\left(K_2^{2(1+m^-)}\right)
+
\beta(s(K_3))\,O\!\left(K_2^{2(1+m^+)}\right).
\end{equation}
Recall that $\beta(s(K_3))$ is independent of $K_2$ and bounded on its admissible domain. Therefore,
\begin{equation}
\psi_{\mathrm{prop}}
=
W_{\mathrm{PK}}
+
O\!\left(K_2^{2(1+\bar m)}\right),
\qquad
\bar m=\min(m^+,m^-)>0.
\label{eq:psi_prop_m_scaling}
\end{equation}
Finally, invoking Eq.~\eqref{eq:WPK_smallK2} and noting that $\bar m>0$, so that $2(1+\bar m)>2$, we obtain
\begin{equation}
\psi_{\mathrm{prop}}(K_2,K_3)
=
\frac{\mu}{2}K_2^2
+
O\!\left(K_2^{2(1+\bar m)}\right)
+
\frac{a}{6}K_2^3\mathcal{G}(K_3)^2
+
O(K_2^4),
\qquad K_2\to 0.
\label{eq:psi_prop_quadratic_limit_full}
\end{equation}

Therefore, the proposed strain energy density recovers the quadratic dependence in $K_2$ under infinitesimal strains in agreement with isotropic linear elasticity. In particular, the leading-order term is independent of $K_3$. Finally, using Eqs. (\ref{eq:PK_response_fn}) and (\ref{eq:gamma2_limits}) and the result in the footnote$^3$, it can be shown that the response functions follow $\gamma_{1}^{(\mathrm{prop})}=O\!\left(K_2\right)$ and $\gamma_{2}^{(\mathrm{prop})}=O\!\left(K_2^{2}\right) + O\!\left(K_2^{1+2\bar{m}}\right)$ as $K_2 \to 0$, again showing consistency with small-strain isotropic linear elasticity \cite{criscione2000invariant}.

Overall, the proposed Lode-angle-dependent bi-failure formulation yields bounded stresses throughout the $(K_2,K_3)$ invariant space satisfying symmetry requirements by reducing to a purely hydrostatic response in the limit of vanishing distortion, and also ensures consistency with linearized elasticity under infinitesimal strains.

With the constitutive framework and its associated behavioral characteristics now established, the following section details the experimental methodology employed to generate the data for model calibration and validation.

\section{Model Fitting and Validation Methodology}
\label{sec:methodology}

Calibration of the proposed constitutive model for asymmetric failure requires experimental data that sufficiently span the distortional response of the material—both in magnitude ($K_2$) and mode ($K_3$)—to enable independent identification of mode-dependent (tensile and compressive) hyperelastic and failure parameters. Accordingly, both uniaxial tension and compression tests were performed to capture the mode-dependent softening and failure behavior necessary for parameter calibration. The inclusion of both deformation modes is therefore intrinsic to the model formulation, ensuring that the full TCA is represented within the fitted response.

Agarose hydrogels were selected as the model material system due to their well-characterized polymer network structure, elastomeric behavior, and frequent use as soft-tissue phantoms. Three concentrations (1, 2, and 3 \% w/v) were investigated to provide a set of related “model materials” that span different network densities. These variations in polymer concentration introduce controlled differences in stiffness, failure strain, and microstructural mechanisms of deformation, thereby allowing evaluation of the framework’s predictive capability across materials with systematically varying architectures.

The overall methodology comprises four stages. First, uniaxial tension and compression experiments were conducted on the three agarose concentrations using standardized specimen preparation, fixturing, and testing protocols, with full-field strain measurement obtained via DIC. The model parameters were calibrated by solving a constrained nonlinear least-squares optimization problem using the combined uniaxial data, ensuring thermodynamic consistency and avoiding local minima through a multi-start strategy. Second, the calibrated model was validated under pure shear loading to assess predictive generalizability across deformation modes. Third, the energy landscape of the formulation was analyzed across the full $K_2$--$K_3$ invariant space to examine thermodynamic consistency and deformation-mode sensitivity beyond the specific loading paths. Finally, the relationship between the calibrated parameters and agarose concentration was characterized using power-law scaling, and the resulting trends were employed to predict the constitutive response of an intermediate, untested concentration (2.5 \% w/v) under both uniaxial and pure shear conditions.

This integrated methodology yields a unified, self-consistent framework for evaluating the proposed model’s performance across deformation modes, invariant space, and polymer network densities. The following subsections detail the experimental procedures for specimen preparation and mechanical testing, followed by the model calibration and fitting process. All corresponding analyses, validation, and discussions of results are presented subsequently in Section~\ref{sec:results_discussion}.

\subsection{Sample preparation}
\label{subsec:sample_prep}

Agarose specimens of all three concentrations (1, 2, and 3~\%~w/v) were prepared using a standardized protocol reported previously in Upadhyay et al. ~\cite{Upadhyay2020, yang2022mechanical}. Briefly, agarose powder (0.15\% sulfate; Cat. No. BP1356-500, Fisher Scientific, USA) was dispersed in deionized water at the required weight-to-volume ratios and heated in a microwave oven until fully dissolved, with intermittent agitation to ensure optical clarity and homogeneity. Heating duration varies depending on the polymer concentration and total solution volume. Immediately after heating, the solution was weighed and the evaporative water loss was replenished to restore the target concentration. The hot solution was then allowed to rest for 2--5 minutes to release entrapped air bubbles before being cast into custom-designed molds: rubber molds for cylindrical (compression) specimens and 3D-printed nylon carbon fiber molds for dog-bone (tension) and pure shear geometries. All specimens were stored in deionized water to maintain full hydration and were tested within 30~minutes of gelation to minimize evaporative losses and ensure consistent mechanical response.

\subsection{Mechanical testing}
\label{subsec:mech_testing}

\begin{figure}[ht]
    \centering
    \includegraphics[width=5.615in, height=2.501in]{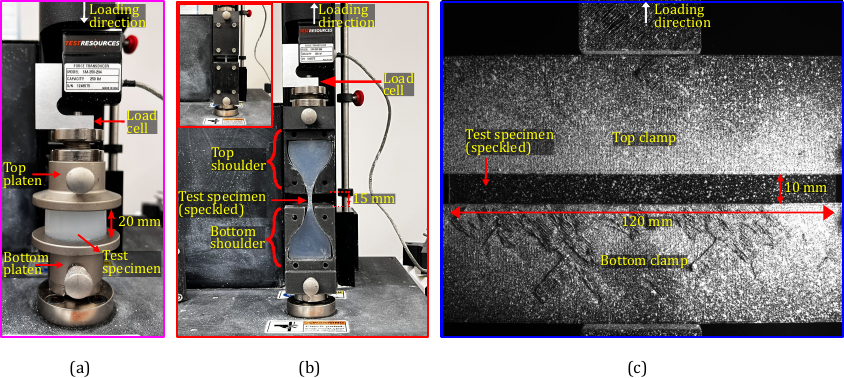}
    \caption{Experimental setup for mechanical testing of agarose gels: 
    (a) uniaxial compression between parallel platens; 
    (b) uniaxial tension of a dog-bone specimen (inset shows the assembled fixture with mounting plates); and 
    (c) pure shear of a rectangular strip clamped between rigid jaws. 
    Prior to loading, a high-contrast speckle pattern was applied on all specimens using white acrylic spray paint to facilitate DIC (excluding compression testing).}
    \label{Exp_setup}
\end{figure}

All mechanical tests, including compression, tension, and pure shear, were performed at ambient room temperature under displacement control using a single-column universal testing machine (Frame 240, TestResources, Inc., Shakopee, MN). A 1.1 kN load cell was used to record force data. All tests were conducted at a fixed engineering strain rate of \(0.01\,\text{s}^{-1}\), corresponding to a quasi-static loading regime \cite{yin2024rate, kwon2010compressive}, using appropriate fixtures for each test configuration, as illustrated in Fig.~\ref{Exp_setup}. For each gel concentration, four tests were conducted to ensure reproducibility. The specific configurations corresponding to each deformation mode are described below.

\subsubsection{Uniaxial compression}
\label{subsubsec:compression}

Uniaxial compression tests (Fig. \ref{Exp_setup}(a)) were performed on cylindrical agarose specimens with dimensions of 40~mm in diameter and 20~mm in height. Each specimen was placed between two polished steel platens; a thin layer of lubricant was applied to minimize frictional constraint and promote uniform axial deformation. Compression was imposed by translating the top platen downward at 12~mm/min, corresponding to an engineering strain rate of $0.01\,\text{s}^{-1}$. The instantaneous specimen height was computed using the real-time cross-head displacement values. Engineering stress and strain were computed from the load-cell force and platen displacement, respectively, and the test was continued until failure of the specimen.

\subsubsection{Uniaxial tension}
\label{subsubsec:tension}

The uniaxial tension setup, shown in Fig. \ref{Exp_setup}(b), employed a modified dog-bone specimen geometry (from Upadhyay et al. \cite{upadhyay2021validated}) designed to promote failure within the gauge section while minimizing stress concentrations near the grips. To prevent pre-damage during mounting, the curved shoulders of the specimen were supported using 3D-printed nylon carbon fiber clamps, following the procedure of Wale et~al.~\cite{wale2021applying}. The gauge section dimensions were 15~mm (length)~$\times$~4~mm (width)~$\times$~3~mm (thickness). Tensile loading was applied by translating the top shoulder vertically at a constant speed, corresponding to a nominal strain rate of $0.01\,\text{s}^{-1}$, while the bottom shoulder is fixed to the machine base plate. Engineering stress was obtained from load-cell measurements; however, accurate strain determination in such soft, slippery materials requires additional considerations~\cite{Upadhyay2020, upadhyay2021validated}, which are discussed in the following paragraph.

Accurate strain measurement in hydrogels during mechanical testing (excluding compression) is challenging due to their extremely compliant nature and low-friction surfaces. Conventional approaches such as crosshead displacement are unsuitable because slippage at the grips introduces substantial errors in estimating accurate material strain~\cite{kwon2010application}. Likewise, contact-based extensometers cannot be employed: even minimal clamping pressure from a clip-on device can induce localized deformation or trigger premature cracking in soft gels~\cite{subhash2011concentration, Upadhyay2020}. To overcome these limitations, the present study employs DIC—a non-contact optical technique that tracks a randomly distributed speckle pattern on the specimen surface to obtain full-field kinematic measurements. For uniaxial tension, a 3D stereo-DIC configuration was used to capture both in-plane and out-of-plane displacements. Synchronized 5~MP CCD cameras (BFS-U3-51S5M-C, FLIR), positioned at approximately $30^\circ$, acquired stereo images at 5~Hz. A set of 30 images of a standard calibration grid was used to perform the stereo-system calibration. The images were processed in MatchID (MatchID, Inc., Belgium) to compute axial engineering strain using a virtual extensometer placed on either side of the damage-initiation region within the gauge section. This approach ensured accurate strain quantification while avoiding artifacts associated with localized edge effects or out-of-plane motion.

This measurement protocol closely follows established procedures for similar soft materials, including PDMS elastomers and agarose hydrogels~\cite{upadhyay2021validated, Upadhyay2020}. These prior studies have extensively validated the assumption of nearly ideal uniaxial deformation within the gauge section, together with the use of material incompressibility ($K_1=0$) and isotropy—assumptions that remain well-supported and are not re-evaluated here.

\subsubsection{Pure shear}
\label{subsubsec:pureshear}

The pure shear configuration (Fig. \ref{Exp_setup}(c)) was implemented by clamping a wide and thin rectangular specimen between rigid jaws to approximate planar tension~\cite{brown2006physical, moreira2013comparison}. The specimen dimensions were 120~mm (width)~$\times$~70~mm (height)~$\times$~2~mm (thickness). Of the total height, 30 mm was clamped on each side, resulting in a gage section of 120~mm ~$\times$~10~mm ~$\times$~2~mm (see Fig. \ref{Exp_setup}(c)). This gage section geometry was chosen based on established guidelines~\cite{yeoh2001analysis} to minimize edge effects and ensure a uniform strain field. It promotes homogeneous thinning in the thickness direction while suppressing lateral contraction due to Poisson effect, thereby producing a near–pure-shear state in the central region of the specimen (Eq. (\ref{eq:F_ps})). Tests were performed under quasi-static loading at a nominal strain rate of $0.01\,\text{s}^{-1}$ by vertically translating the top clamp while keeping the bottom clamp fixed to the machine base.

Full-field strain measurements were obtained using 2D DIC with a single 5~MP camera operating at 30~Hz. The spatial resolution was $\sim$19 pixels/mm, such that around 190 pixels exist along the height of the specimen gage section. Acquired images were analyzed using the commercial MatchID software; subset size was selected for individual imaging data based on the speckle density and size (subset size $>$ (3 × speckle size)), and also the sigma value (one standard deviation confidence interval) of the subset tracking function. The use of 2D DIC was justified by the negligible out-of-plane deformation of the thin specimens and by maintaining a sufficiently large distance (at least 100~mm) between the camera lens and the specimen surface. Because in-plane strain errors in 2D DIC are approximately proportional to the ratio of out-of-plane displacement to the lens–specimen distance \cite{Sutton2008}, the combination of small specimen thickness (2~mm) and large imaging distance ensures that these errors remain negligible.

Accurate strain quantification in pure shear tests up to the onset of damage localization requires further methodological considerations. Although the full-specimen DIC area of interest (AOI) satisfies the ideal pure shear kinematics, failure in wide–thin agarose specimens can initiate closer to the grip-adjacent regions due to geometric constraints. Consequently, once strain localization develops, spatial averaging of full-field strain becomes increasingly unreliable. To address this issue, the present study extrapolates the average engineering strain measured during the homogeneous deformation regime (i.e., prior to localization) from full-specimen AOI (average) to the locally measured failure stretch at the rupture-zone. The corresponding engineering stress obtained from the load cell is then mapped to this extrapolated stretch, yielding a consistent and accurate engineering stress–stretch curve for pure shear loading up to failure.~\ref{appendix:pureshearextrapolation} details this data analysis procedure and demonstrates that the deformation state achieved in our tests accurately corresponds to the ideal pure shear condition.

\subsection{Fitting process}

The nominal stress--stretch relations (Eqs.~\eqref{eq:uniaxial_tension_f}--\eqref{eq:uniaxial_compression_final}) derived in Section~\ref{sec:prop_formulation} were implemented in \textsc{MATLAB} for parameter identification. 
For each agarose concentration (1, 2, and 3\% w/v), four independent specimens were tested in uniaxial tension and four in uniaxial compression up to failure. 
To construct representative combined datasets for model calibration, one tension dataset and one compression dataset were randomly paired to form a matched tension--compression set. 
This procedure was repeated to obtain four distinct combined datasets per concentration, ensuring that each calibration instance reflects independent experimental realizations of both loading modes.

The model parameters were then identified by simultaneously fitting the tension and compression branches of each combined dataset. 
This was achieved by minimizing an objective function based on the residual sum of squares (RSS), which quantifies the discrepancy between the model predictions and the corresponding experimental measurements. 
\begin{equation}
RSS = \sum_{i=1}^{k} \left( P_i^{\text{model}} - P_i^{\text{exp}} \right)^2,
\end{equation}
where \( k \) represents the number of experimental data points, \( P_i^{\text{exp}} \) is the experimentally measured engineering stress at the \( i \)th point, and \( P_i^{\text{model}} \) is the corresponding model-predicted engineering stress.

To ensure robust parameter identification, the fitting procedure begins by specifying initial guesses and bounds for both the hyperelastic and damage-related parameters. A Latin Hypercube Sampling scheme (\texttt{lhsdesign}) is employed to generate 500 randomly distributed initial parameter sets, thereby broadening the exploration of the search space and increasing the likelihood of identifying the global minimum. Each initialization is then used to launch a local optimization via \texttt{lsqnonlin} within a \texttt{MultiStart} framework. The \texttt{lsqnonlin} algorithm iteratively updates the parameters to minimize the RSS. Convergence is enforced when either the supremum norm of the objective-function gradient or the Newton step falls below $10^{-6}$. The \texttt{MultiStart} procedure terminates once all initial points have been processed and the optimal parameter set is selected based on the lowest residual error.

Along with the plots of the fitted responses, the relative errors were computed for each data point to assess the accuracy of the model, using the following expression:
\begin{equation}
\text{err}_{i} (\%) = \frac{|P_i^{\text{exp}} - P_i^{\text{model}}|}{\max\left\{0.1 \max(P^{\text{exp}}), |P_i^{\text{exp}}|\right\}} \times 100, \quad i = 1,2,\dots,m.
\end{equation}

The above equation quantifies the relative deviation between the experimental data \( P^{\text{exp}} \) and the model prediction \( P^{\text{model}} \) for the \( i \)th data point. The term $0.1 \max(P^{\text{exp}})$ is included in the denominator to prevent artificially large relative errors in the small strain regime (when stresses are very small). Expressed as a percentage, this metric provides a consistent measure of model accuracy \cite{upadhyay2020visco}. Accurate calibration therefore requires the stress--stretch relations derived in Section~\ref{sec:prop_formulation} to closely align with the corresponding experimental uniaxial data. The set of eight unknown model parameters identified through this calibration procedure—four governing the intact hyperelastic response and four describing the mode-specific failure behavior—is summarized in \autoref{tab:params_model}. The following constraints were enforced during the optimization:
\[
\mu > 0,\qquad 
a > 0,\qquad 
b_0 > 0,\qquad
b_1 > 100,\qquad
\Phi^{+} > 0,\qquad 
m^{+} > 0,\qquad 
\Phi^{-} > 0,\qquad 
m^{-} > 0.
\]
The lower bound on \(b_1\) follows the recommendation in \cite{prasad2020analysis}.


To assess the robustness and generalizability of the calibrated parameters, the model is subsequently evaluated under an independent distortional loading condition---pure shear. Specifically, the theoretical pure shear predictions given by Eq.~\eqref{eq:Pps_final} are compared directly against experimental test data, thereby confirming the model’s capability to capture multi-mode responses. 

With the parameter identification framework established, the next step is to assess the model’s performance against experimental data. The ensuing section presents a detailed comparison between experimental observations and model predictions, followed by a discussion of the calibrated parameters, their physical interpretation, and concentration-dependent trends.

\section{Results and Discussion}
\label{sec:results_discussion}

\subsection{Model calibration and performance under uniaxial deformation modes}
\label{sec:model_performance_uniaxial}

\begin{figure}[t]
    \centering
    \includegraphics[width=6.259in, height=5.631in]{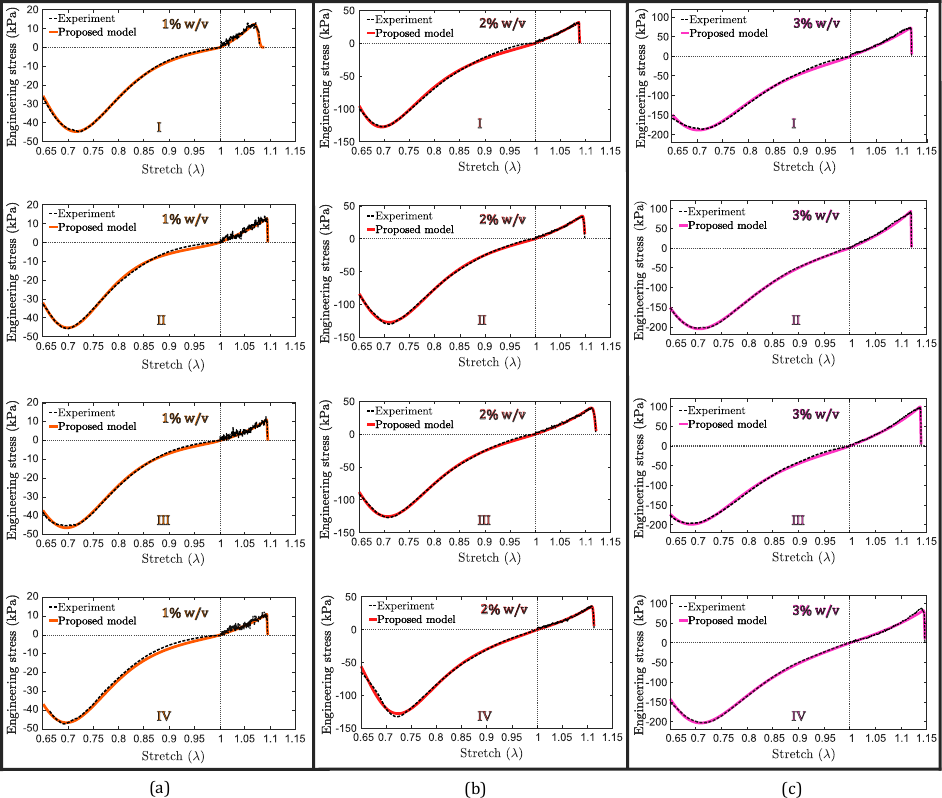}
    \caption{Combined experimental uniaxial engineering stress–stretch data (black) and the corresponding simultaneous fits of the proposed asymmetric failure model for agarose gels at (a) 1\% w/v (orange), (b) 2\% w/v (red), and (c) 3\% w/v (magenta), shown for each replicate (I--IV). Note: Y-axis scales differ between subplots for clarity.}
    \label{fig_fit}
\end{figure}

The proposed asymmetric failure constitutive model was calibrated using combined uniaxial tension and compression data obtained from four independent experimental replicates (\(n=4\)) for each agarose concentration (1, 2, and 3\% w/v). \autoref{fig_fit} presents the experimental stress–stretch responses (black) and the corresponding model predictions (orange, red, and magenta) for all three concentrations. Each fitting used one random combination of tensile and compressive datasets from the same concentration to ensure robustness of parameter identification. The consistency observed among the replicate measurements (i.e., I--IV) validates the reproducibility of the experimental procedure and supports the assumption of representative material behavior within each concentration group. From the figure, the proposed formulation successfully reproduces the characteristic asymmetric behavior between tension and compression, capturing both softening and failure transitions. The fitted parameters and average residual errors are summarized in \autoref{tab:params_all}, and the associated total failure energies in \autoref{tab:fracture_energy_only}.

Before analyzing the concentration-dependent trends, it is important to interpret the physical roles of the bulk hyperelastic parameters in the Prasad–Kannan formulation (Eq. (\ref{eq:W_K2K3})). The shear modulus $\mu$ governs the small-strain stiffness and represents the initial linear elastic response of the network. The coefficient $a$ controls the amplitude of the exponential growth of intact strain energy density at large strains (i.e., elastic strain-stiffening behavior) \cite{prasad2020analysis,sendova2005strong}. The constants $b_0$ and $b_1$ appear in the mode-dependent modulus function $\mathcal{G}(K_3)$, which governs the degree of TCA and enables the model to predict continuously varying effective moduli across distortion modes, consistent with experimental observations \cite{prasad2020analysis}.

\begin{table}[t]
\centering
\caption{Model parameters for the proposed asymmetric failure model obtained by fitting combined tension and compression data of agarose gels. The last column reports the overall average residual error of fit for each sample.}
\label{tab:params_all}
\renewcommand{\arraystretch}{1.15}
\setlength{\tabcolsep}{3.5pt}
\begin{tabular}{lccccccccc}
\toprule
\textbf{sample}
& $\mu$ (kPa) & $a$ (kPa) & $b_0$ & $b_1$
& $\Phi^{+}$ (kPa) & $m^{+}$ & $\Phi^{-}$ (kPa) & $m^{-}$ & $\overline{\text{err}}$ (\%) \\
\midrule
\multicolumn{10}{c}{\textbf{1\% w/v}} \\
\midrule
I   & 57.50 & 1.62 & 10.15 & 3867.21 & 0.52 & 168.26 & 0.13 & 0.22 & 9.74 \\
II  & 38.34 & 3.07 &  8.53 & 1116.54 & 0.42 & 185.26 & 0.30 & 0.24 & 8.91 \\
III & 40.80 & 2.31 &  9.08 & 1120.96 & 0.43 & 197.38 & 0.46 & 0.26 & 7.44 \\
IV  & 62.54 & 5.89 & 10.11 &  350.81 & 0.52 &   5.99 & 0.01 & 0.17 & 8.57 \\
\addlinespace
\textit{Avg.} & 49.79 & 3.22 & 9.47 & 1613.88 & 0.47 & 139.22 & 0.23 & 0.22 & 8.66 \\
\midrule
\multicolumn{10}{c}{\textbf{2\% w/v}} \\
\midrule
I   & 156.35 & 6.05 & 7.55 & 3767.95 & 1.85 &  43.36 & 5.45 & 0.36 & 7.46 \\
II  & 187.76 & 6.60 & 8.04 & 2863.50 & 1.63 &  33.06 & 3.31 & 0.31 & 8.35 \\
III & 174.31 & 6.65 & 7.52 & 4907.38 & 2.35 &  22.41 & 4.68 & 0.34 & 8.16 \\
IV  & 179.06 & 8.49 & 7.80 & 1443.68 & 1.29 & 287.34 & 2.69 & 0.30 & 6.99 \\
\addlinespace
\textit{Avg.} & 174.37 & 6.95 & 7.73 & 3245.63 & 1.78 & 96.54 & 4.03 & 0.33 & 7.74 \\
\midrule
\multicolumn{10}{c}{\textbf{3\% w/v}} \\
\midrule
I   & 305.11 & 15.29 & 6.35 & 1827.11 & 3.98 & 186.95 & 14.49 & 0.41 & 3.36 \\
II  & 326.93 & 20.03 & 6.86 & 4226.24 & 4.80 & 273.21 &  4.99 & 0.31 & 6.68 \\
III & 310.81 & 16.79 & 6.61 & 3291.09 & 5.95 & 234.84 &  7.15 & 0.33 & 4.81 \\
IV  & 261.55 & 16.04 & 5.85 & 4998.20 & 5.33 & 145.82 & 18.89 & 0.46 & 4.18 \\
\addlinespace
\textit{Avg.} & 301.10 & 17.04 & 6.42 & 3585.66 & 5.02 & 210.20 & 11.38 & 0.38 & 4.76 \\
\bottomrule
\end{tabular}
\end{table}

In Fig. \ref{fig_fit}(a), corresponding to the 1\%~w/v gel, both the compressive and tensile responses exhibit low stiffness, and the tensile response exhibits limited strain-stiffening. This behavior is consistent with the sparse polymer network and low cross-link density, which permit extensive chain mobility and low resistance to deformation \cite{Upadhyay2020, yang2022mechanical}. The fitted parameters (\autoref{tab:params_all}) reflect this: a low average $\mu$ of 49.79~kPa shows high compliance and low average $a$ of 3.22~kPa indicates limited strain-dependent reinforcement. 

Further, the small average tensile pseudo–failure energy ($\Phi^{+} = 0.47$~kPa) and large softening sharpness parameter ($m^{+} = 139.22$) imply an abrupt loss of load-carrying capacity near failure, characteristic of brittle-like rupture. Because $m^{+} \gg 1$, the pseudo–failure energy $\Phi^{+}$ closely approximates the true failure energy $\psi_f^{+}$ in tension (see Eq. (\ref{eq:Psi_f_pm})), as the stress drop occurs over a narrow energy range. In contrast, for compression, where $m^{-} \ll 1$, the compressive pseudo–failure energy $\Phi^{-}$ significantly underestimates the total dissipated energy since the actual failure energy $\psi_f^{-}$ depends jointly on $\Phi^{-}$ and $m^{-}$, reflecting a more gradual softening behavior. The associated failure energies (\autoref{tab:fracture_energy_only})—$\psi_f^{+}=0.46$~kPa and $\psi_f^{-}=8.76$~kPa—highlight a strong asymmetry, with failure energy nearly 19 times greater in compression than in tension.

At 2\%~w/v (Fig. \ref{fig_fit}(b)), both stiffness and nonlinearity increase markedly, consistent with higher cross-link density and reduced chain mobility. The higher average elastic parameters, $\mu = 174.37$~kPa and $a = 6.95$~kPa, suggest enhanced network reinforcement. The increase in average $\Phi^{+}$ and $\Phi^{-}$ (to 1.78~kPa and 4.03~kPa, respectively) demonstrates increased energy storage and delayed damage initiation. At the same time, the failure energies—$\psi_f^{+}=1.75$~kPa and $\psi_f^{-}=25.29$~kPa—show an order-of-magnitude increase from the 1\%~w/v gel, confirming stronger and more energy-dissipative network connectivity.

At 3\%~w/v (Fig. \ref{fig_fit}(c)), the gels exhibit pronounced nonlinearity, high stiffness, and more gradual softening before failure. The stiffer response ($\mu = 301.10$~kPa) and an elevated $a$ value of 17.04~kPa point to reduced molecular mobility and strain-hardening effects within a densely cross-linked network. The pseudo--failure energy parameters increase substantially ($\Phi^{+} = 5.02$~kPa, $\Phi^{-} = 11.38$~kPa), consistent with greater toughness and improved load-bearing capacity. The corresponding failure (i.e., saturation) energies (\autoref{tab:fracture_energy_only})—$\psi_f^{+}=5.00$~kPa and $\psi_f^{-}=43.74$~kPa—further highlight the large TCA.

Although the average $m^{+}$ at 3\%~w/v increased (to $\sim210$), no systematic trend in $m^{+}$ is observed across concentrations, indicating that the sharpness of tensile failure—i.e., the abruptness of post-peak decay—is largely insensitive to bulk gel concentration. In contrast, the average $m^{-}$ increases consistently from 1\%~w/v to 3\%~w/v. This trend does not imply a direct monotonic change in the steepness of compressive softening when comparing concentrations, since $m^{-}$ acts in concert with the intact stored energy $W(\lambda)$ and the pseudo–failure energy $\Phi^{-}$, both of which also increase with concentration. Rather, the fitted parameters indicate that stiffer, more densely cross-linked gels sustain higher compressive energies prior to the onset of relatively rapid softening, with $m^{-}$ modulating how sharply the stress-reduction factor $\mathrm{Red}(\lambda)$ (Eq. (\ref{eq:uniaxial_compression_final})) activates as $W(\lambda)$ approaches $\Phi^{-}$.

The role of softening sharpness parameters is further clarified in Supplementary Section~S.1, which presents a plot to illustrate the influence of $m^{\pm}$ on the evolution of stress-reduction factors with uniaxial stretch for fixed values of $\Phi^{\pm}$. From this figure, larger values of $m^{\pm}$ produce a sharper and more abrupt drop once the stored energy approaches the corresponding pseudo-failure energies ($\Phi^{\pm}$). This confirms that $m^{\pm}$ governs the activation and steepness of the softening profile. However, because the actual softening profile also depends on other model parameters, a monotonic reduction or increase in $m^{\pm}$ across different gel concentrations does not necessarily translate to more gradual or abrupt softening (across these groups). 

Across all concentrations, the compressive failure energies exceed their tensile counterparts by nearly an order of magnitude, confirming the strong TCA intrinsic to the agarose network microstructure \cite{drozdov2020tension}. The fitted residual errors remain below 10\% for all datasets, demonstrating stable and accurate calibration of the proposed asymmetric failure formulation.

\begin{table}[t]
\centering
\caption{Failure energies for each test sample and the averages within each concentration.}
\label{tab:fracture_energy_only}
\renewcommand{\arraystretch}{1.15}
\setlength{\tabcolsep}{6pt}
\begin{tabular}{lcc}
\toprule
\textbf{sample} & $\psi_f^{+}$ (kPa) & $\psi_f^{-}$ (kPa) \\
\midrule
\multicolumn{3}{c}{\textbf{1\% w/v}} \\
\midrule
I   & 0.52 & 8.18 \\
II  & 0.41 & 9.26 \\
III & 0.42 & 8.28 \\
IV  & 0.48 & 9.33 \\
\addlinespace
\textit{Avg.} & 0.46 & 8.76 \\
\midrule
\multicolumn{3}{c}{\textbf{2\% w/v}} \\
\midrule
I   & 1.83 & 25.09 \\
II  & 1.60 & 25.18 \\
III & 2.29 & 25.34 \\
IV  & 1.29 & 25.54 \\
\addlinespace
\textit{Avg.} & 1.75 & 25.29 \\
\midrule
\multicolumn{3}{c}{\textbf{3\% w/v}} \\
\midrule
I   & 3.97 & 43.79 \\
II  & 4.79 & 41.99 \\
III & 5.93 & 44.62 \\
IV  & 5.31 & 44.54 \\
\addlinespace
\textit{Avg.} & 5.00 & 43.74 \\
\bottomrule
\end{tabular}

\vspace{0.4ex}
{\footnotesize $\psi_f^\pm$ are derived from $\Phi^\pm$ and $m^\pm$ (Eq. \eqref{eq:Psi_f_pm}).}
\end{table}

Beyond the concentration-dependent trends in failure energy and softening sharpness, the experimental data further indicate that tension--compression asymmetry is evident in the small-strain regime. To clarify where agarose hydrogels lie within the broader spectrum of asymmetric soft-material responses, a Mooney-type asymmetry analysis is provided in Supplementary Section~S.2 and Fig.~S.2, following Reddipaga and Kannan~\cite{reddipaga2026construction}. This analysis shows that, within the small-deformation regime, agarose hydrogels do not exhibit the compression-stiffening behavior commonly reported for soft biological tissues such as brain~\cite{budday2020fifty} and liver~\cite{gao2010constitutive}. Instead, the 1\%~w/v gel shows pronounced tension-stiffening, whereas the 2\% and 3\%~w/v gels exhibit relatively mild tension-stiffening. In contrast, synthetic elastomers such as silicone rubber and soft polyurethane (S-PU)~\cite{falope2025experiments} exhibit nearly tension--compression symmetric behavior.

This observation provides useful context for the role of the Lode angle-dependent function \(\mathcal{G}(K_3)\) in the chosen intact potential. The function \(\mathcal{G}(K_3)\) was originally introduced to capture the elastic compression-stiffening behavior of human brain tissue \cite{prasad2020analysis}. Nevertheless, the intact response of agarose hydrogels is expected to transition toward compression-stiffening at larger finite deformations under compression. Consequently, the chosen function \(\mathcal{G}(K_3)\) provides the flexibility needed to represent this emergent behavior. Moreover, unlike the experimentally inferred tangents in the limit of infinitesimal strain, which differ between tension and compression, the Prasad--Kannan model approximates the small-strain response using a single modulus value.

Taken together, these results highlight a central advantage of the proposed asymmetric failure framework: each model parameter carries clear physical meaning and maps directly onto measurable aspects of macroscopic and microscopic behavior. The elastic parameters $\mu$ and $a$ govern the initial stiffness and nonlinear strain-stiffening response, reflecting polymer network connectivity and chain extensibility. $b_0$ and $b_1$ govern TCA in elastic response. The failure parameters $\Phi^{\pm}$ and $m^{\pm}$ link directly to the limiting strain-energy capacity and the abruptness of post-peak softening, providing quantitative access to experimentally derived failure energies ($\psi_f^{\pm}$). In this way, the model yields not only accurate macroscopic predictions but also mechanistic insight into microstructural features such as polymer density, cross-link distribution, and mode-dependent failure pathways.

For the agarose hydrogels examined here, these physical interpretations manifest clearly: increasing concentration systematically elevates $\mu$ and $a$, consistent with a stiffer, more tightly connected polymer network with reduced chain extensibility \cite{dusek2014constrained, jayawardena2023evaluation, ghebremedhin2021physics}; simultaneously, the growth of $\Phi^{\pm}$ and $\psi_f^{\pm}$ reflects enhanced toughness and greater energy dissipation prior to rupture \cite{bertula2019strain, normand2000new, barrangou2006textural}. A detailed quantitative assessment of these concentration-dependent trends and parameter interrelations is presented in Section \ref{subsec:powerlaw_validation}, using power-law scaling relationships.

\subsection{Comparison of softening mechanisms with the classical energy-limiting model}

\begin{figure}[h]
   \centering
   \includegraphics[width=6.019in, height=3.435in]{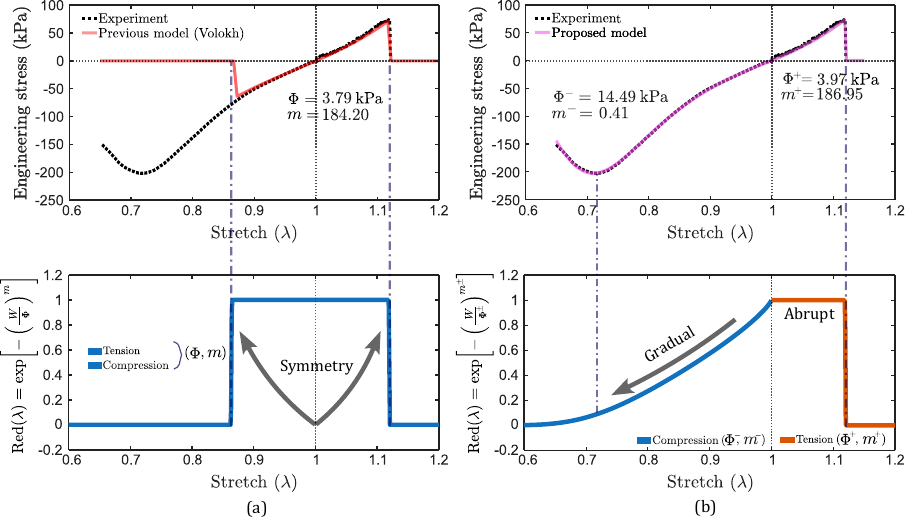}
   \caption{Comparison of experimental (black) and model-predicted (red and magenta) engineering stress–stretch responses for a 3\% agarose specimen.  
   (a) The classical Volokh model (red) employs a single set of failure parameters $(\Phi, m)$, yielding symmetric softening in tension and compression.
   (b) The proposed asymmetric failure formulation (magenta) introduces mode-specific pseudo--failure energies $(\Phi^{+},\Phi^{-})$ and corresponding softening sharpness parameters $(m^{+},m^{-})$.  
   The resulting reduction factors exhibit abrupt decay in tension and gradual decay in compression, consistent with the observed TCA.}
   \label{fig:asymm_fail}
\end{figure}

To further highlight the asymmetric softening mechanisms captured by the proposed model, we compare its predicted stress-reduction behavior with that of the classical energy-limiting hyperelastic formulation of Volokh (Section \ref{subsec:Classical_energy_limiters}). For this comparison, we calibrated the Volokh model parameters (paired with the Prasad–Kannan intact strain energy density) using only the tensile data from a representative 3\%~w/v agarose specimen, following standard practice for single-limiter energy-based models. The predicted tensile and compressive stress–stretch responses, together with the associated stress-reduction factor $\mathrm{Red}(\lambda)$ (Eq. (\ref{eq:Volokh-derivative-wrt-W})), are shown in \autoref{fig:asymm_fail}. The corresponding stress–stretch responses and stress-reduction factors (Eqs. (\ref{eq:uniaxial_tension_final}) (tension) and (\ref{eq:uniaxial_compression_final}) (compression)) from the proposed asymmetric formulation are also shown for comparison. In both formulations, the reduction factor governs the decay of the intact hyperelastic response during damage accumulation.

In the intact elastic regime, the stored energy $W$ remains small relative to the pseudo--failure energies (i.e., $\Phi$ in the Volokh model, and $\Phi^{\pm}$ in the proposed model). Consequently, both formulations yield stress-reduction factors close to unity (i.e., $\exp\!\left[-\left(\tfrac{W(\lambda)}{\Phi}\right)^{m}\right] \approx \exp\!\left[-\left(\tfrac{W(\lambda)}{\Phi^{\pm}}\right)^{m^{\pm}}\right] \approx 1$, indicating negligible softening. Beyond this small-strain regime, however, the two models diverge sharply in their predictions of tensile and compressive softening.

Because the Volokh model employs a single pseudo--failure energy $\Phi$ and a single softening sharpness parameter $m$ for all loading modes, it predicts identical decay of $\mathrm{Red}(\lambda)$ in tension and compression. As shown in Fig. \ref{fig:asymm_fail}(a), tensile and compressive responses exhibit a symmetric softening trajectory, and the predicted compressive response fails to reproduce the experimentally observed gradual loss of stiffness. Furthermore, the Volokh model underestimates the higher failure energy in compression, a hallmark of TCA in soft materials. These deficiencies arise from the inherent symmetry imposed by a single set of energy limiters.

In contrast, the proposed model introduces mode-specific pseudo--failure energies $\Phi^\pm$ and softening sharpness parameters $m^\pm$, enabling distinct decay patterns of $\mathrm{Red}(\lambda)$. In tension, a large $m^{+} \gg 1$ keeps $\mathrm{Red}(\lambda) \approx 1$ over most of the deformation path until $W \approx \Phi^+$, after which it drops abruptly, reproducing the sharp tensile rupture observed experimentally. In contrast, in compression, with $m^{-} < 1$, even modest increases in $W$ cause a smooth, gradual decay of $\mathrm{Red}(\lambda)$, matching the progressive softening and high failure energy seen in experiments. Figure \ref{fig:asymm_fail}(b) highlights this sharp contrast—an abrupt decay under tension and a gradual reduction under compression—demonstrating that the asymmetric formulation naturally encodes the experimentally observed TCA in softening and failure behavior.

\subsection{Generalizability of the proposed model: performance in pure shear}
\label{sec:validation}

\begin{figure}[t]
    \centering
    \includegraphics[width=3.4in, height=1.881in]{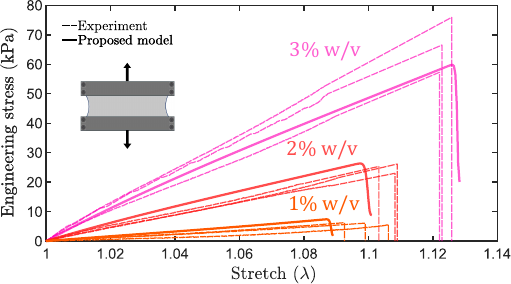}
    \caption{Engineering stress–stretch response for agarose gel specimens in pure shear. Dashed lines represent experimental data ($n=3$) for 1\%, 2\%, and 3\%~w/v specimens. Solid lines show the corresponding predictions of the proposed model using parameters calibrated solely from combined uniaxial compression–tension data.}
    \label{fig:pureshear}
\end{figure}

To evaluate the model’s generalizability beyond uniaxial loading, we next examined its predictive capability under pure shear—a deformation mode with fundamentally different distortional kinematics and stress state. For this analysis, the average calibrated parameters for each concentration (1, 2, and 3\%~w/v) obtained strictly from the combined tension–compression datasets (\autoref{tab:params_all}) were used directly to predict the pure shear response (using Eq. (\ref{eq:Pps_final})), without any additional fitting.

As shown in \autoref{fig:pureshear}, the predicted stress–stretch curves exhibit excellent agreement with experimental data across all concentrations (1\%, 2\%, and 3\%~w/v). The initial shear stiffness, the onset and progression of softening, and the failure stretch and stress values closely match the measured trends. The transition from intact elasticity to damage is governed by the same asymmetric pseudo--failure energies $\Phi^{\pm}$ and sharpness parameters $m^{\pm}$ identified from uniaxial tests, while the Lode-invariant weighting function $\beta(K_3)$ naturally adjusts their influence under pure shear. This produces a seamless extension of the uniaxially calibrated constitutive behavior into a distinct distortional mode.

Minor deviations in the post-peak response likely stem from localized instabilities or geometric effects not explicitly accounted for in the present homogeneous continuum formulation. Nonetheless, the model successfully captures the concentration-dependent increase in shear stiffness and failure stress, consistent with the higher cross-link density at elevated agarose concentrations.

Overall, this validation demonstrates that the proposed Lode-invariant-based asymmetric energy-limiting model, calibrated solely from uniaxial data, can accurately predict the mechanical response of agarose hydrogels under a fundamentally different distortional mode. The ability to reproduce both elastic and inelastic features across deformation paths without additional fitting highlights the robustness and generalizability of the proposed framework. Notably, this broader predictive capability is achieved by introducing only two additional energy-limiting parameters relative to the classical Volokh model, thereby extending a formulation originally restricted to a single failure branch into a unified constitutive framework capable of describing distortional-mode-dependent softening and failure throughout the \((K_2,K_3)\) invariant space.

\subsection{Free-energy landscape of the proposed model}
\label{subsec:energy_landscape}

Having established the predictive capability of the proposed model under pure shear, we now examine its free-energy landscape to reveal how TCA emerges within the Lode invariant space. Visualizing the evolution of the proposed potential \(\psi_{\text{prop}}\) (Eq. (\ref{eq:Psi_prop_expanded})) over the distortional invariants \(K_2\) (magnitude) and \(K_3\) (mode) provides direct mechanistic insight into how failure energy varies across distortion modes in a smooth and thermodynamically consistent framework—thereby linking the responses of uniaxial tension, pure shear, and uniaxial compression within a unified representation. 

The proposed strain energy density \(\psi_{\text{prop}}\) was evaluated by first computing the intact stored energy density \(W(K_2, K_3)\) from the Prasad–Kannan hyperelastic framework \cite{prasad2020analysis}, followed by mapping this energy through the proposed energy-limiting function (Eq. (\ref{eq:Psi_prop_expanded})). This mapping, governed by the mode-dependent energy limiters \(\Phi^{\pm}\) and $m^\pm$ and the Lode-angle-based weighting function \(\beta(s(K_3))\in[0,1]\), ensures a continuous interpolation between the tensile and compressive failure energies. As a result, \(\psi_{\text{prop}}\) depends intrinsically on the deformation mode, reflecting the physical variation of attainable failure energy with the underlying distortional state rather than imposing a single, mode-independent saturation limit.

\begin{figure}[b!]
    \centering
    \includegraphics[width=6.206in, height=7.423in]{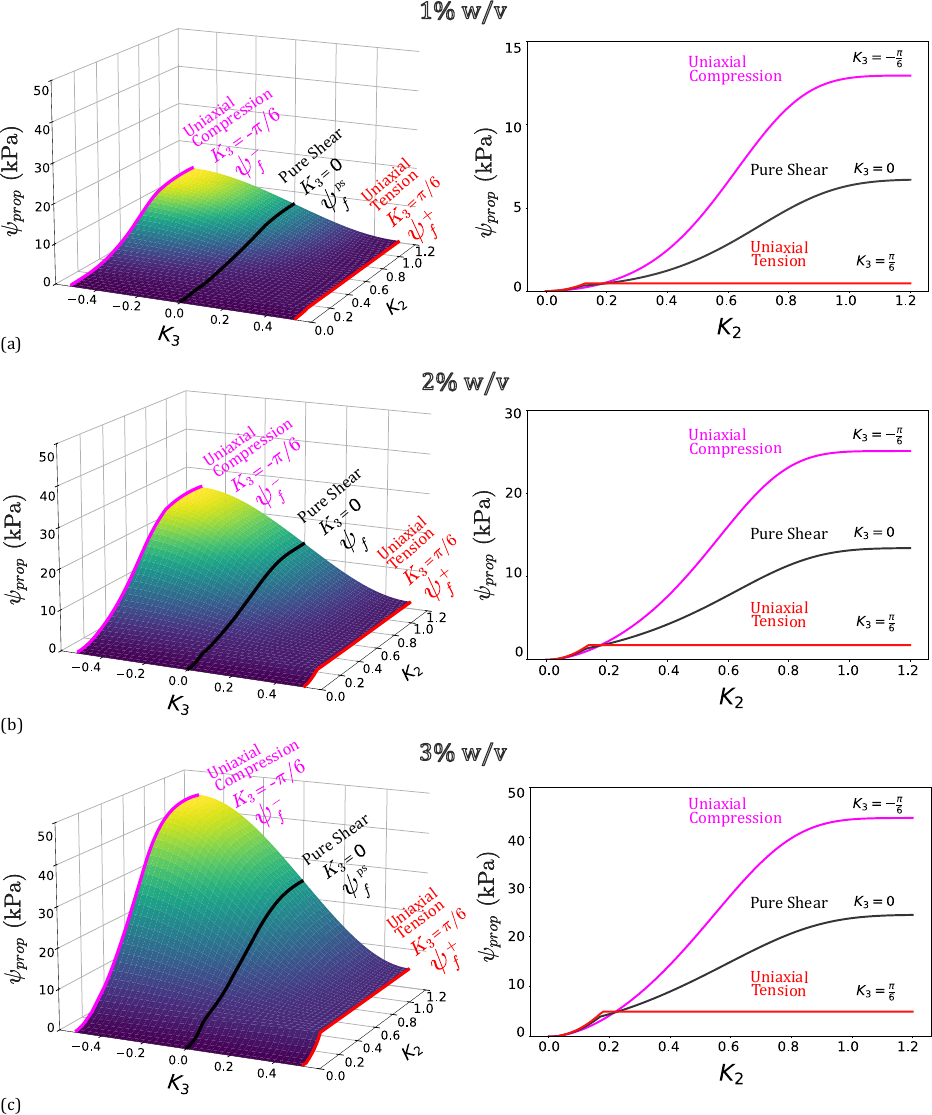}
    \caption{Free-energy landscapes of \(\psi_{\mathrm{prop}}\) for agarose gels at (a) 1\%~w/v, (b) 2\%~w/v, and (c) 3\%~w/v. Left column: \(\psi_{\mathrm{prop}}(K_2,K_3)\) surfaces over the \(K_2\!-\!K_3\) invariant space. Right column: corresponding \(\psi_{\mathrm{prop}}(K_2)\) slices at \(K_3=\{-\pi/6,\,0,\,\pi/6\}\), representing uniaxial compression (magenta), pure shear (black), and uniaxial tension (red).}
    \label{fig:energy_surface}
\end{figure}

\autoref{fig:energy_surface} presents the resulting energy landscapes for agarose gels at 1\%, 2\%, and 3\%~w/v concentrations. The left panels illustrate the three-dimensional variation of \(\psi_{\text{prop}}\) over the \((K_2, K_3)\) invariant domain, while the right panels show characteristic two-dimensional \(\psi_{\text{prop}}-K_2\) slices at \(K_3=\{-\pi/6,\,0,\,\pi/6\}\), corresponding respectively to uniaxial compression, pure shear, and uniaxial tension. The computational domain spans the physically admissible range of Lode invariants: \(K_2\) varies from zero (undeformed) to a large distortional magnitude (i.e., \(K_2=1.2\)), and \(K_3\) spans from compression- to tension-dominant states.

Across all concentrations, the surfaces reveal a clear and systematic TCA: the saturation energy is consistently highest in compression (\(K_3=-\pi/6\)) and lowest in tension (\(K_3=+\pi/6\)), mirroring the experimental observation that agarose sustains and dissipates substantially greater energy before failure in compression. Intermediate deformation modes transition smoothly between these bounds, confirming that the Lode-invariant-based interpolation introduces no discontinuities or artificial artifacts.


Importantly, singularity or discontinuous curvature are observed across the energy surfaces, even near the failure regions. This indicates that the constitutive formulation maintains bounded and smoothly varying energy derivatives 
\((\partial\psi/\partial K_{2},\, \partial\psi/\partial K_{3})\), preventing nonphysical responses such as abrupt stiffness changes. The continuous and smooth topology of the energy landscape thus validates the internal consistency of the model and confirms that the Lode-invariant interpolation remains thermodynamically consistent across all distortion paths \cite{prasad2020analysis, hill1970constitutive, neff2016geometry}. 


While the energy landscape remains smooth and bounded, this regularity does not imply global constitutive stability. Instead, the onset of failure is associated with a progressive loss of convexity of the strain energy potential as the deformation approaches the mode-dependent energy limit.

In the pre-softening regime, the present formulation asymptotically recovers the intact Prasad--Kannan hyperelastic potential \(W_{\mathrm{PK}}(K_2,K_3)\), which satisfies the Baker--Ericksen and Hill inequalities by construction 
\cite{prasad2020analysis}. Accordingly, the model preserves the physically admissible ordering of principal stresses and stretches \cite{baker1954inequalities}. In the present Hencky strain setting, this ordering may be written as \cite{prasad2020analysis}

\begin{equation}
(T_i - T_j)\bigl(\ln(\lambda_i) - \ln(\lambda_j)\bigr) \geq 0, \qquad i,j=1,2,3.
\end{equation}

Moreover, under plane-stress conditions, the Cauchy stress remains a monotone increasing function of the Hencky strain, consistent with Hill's inequalities \cite{hill1970constitutive, wollner2025search}. These inequalities are known to yield a physically reasonable material response and are, in turn, implied by stronger notions of incremental stability, such as strong ellipticity through the Legendre--Hadamard condition \cite{marsden1994mathematical, ghiba2015ellipticity}.

However, the introduction of an energy limiter fundamentally alters this behavior. As the stored energy approaches the mode-dependent saturation level, the stress-reduction factor \(\partial \psi_{\mathrm{prop}}/\partial W\) decreases, and the 
higher-order derivatives \(\partial^2 \psi_{\mathrm{prop}}/\partial W^2\) may change sign, leading to a degradation of the tangent moduli. This loss of stiffness is the mechanism by which softening and failure are produced.

Consistent with the general theory of energy-limiting elasticity, such formulations necessarily violate strong ellipticity as failure is approached. As emphasized by Volokh \cite{volokh2017loss}, the imposition of an energy bound limits the attainable stress and energy, and the resulting response must exhibit a loss of ellipticity, reflecting the onset of failure localization and the inability of the material to sustain stable wave propagation \cite{sawyers1978speed, zubov2011criterion, marsden1994mathematical}. In this sense, the loss of ellipticity is not a deficiency of the model but a constitutive signature of failure \cite{knowles1978failure, abeyaratne1980discontinuous}.

Within the present framework, failure is identified through the violation of strong ellipticity. In a planar setting, these conditions can be evaluated using the material tangent moduli via the necessary and sufficient criteria established by Dacorogna \cite{dacorogna2001necessary}. The deformation states at which ellipticity is lost define a constitutive instability surface in principal stretch space, whose locus forms the theoretical failure envelope, as shown in~\ref{appendix:failure_envelope}.

Overall, the proposed model preserves classical constitutive admissibility within the intact regime, satisfying both the Baker--Ericksen and Hill inequalities. At the same time, it admits a regulated and physically motivated loss of ellipticity as the energy capacity is approached, thereby enabling the description of both the onset of failure and the subsequent transition into the softening regime.

Referring again to Fig. \ref{fig:energy_surface}, for any fixed deformation mode \(K_{3}\), the potential increases monotonically with \(K_{2}\), demonstrating that \(K_{2}\) governs the magnitude of stored energy accumulation, while \(K_{3}\) modulates the attainable energy limit and the manner of failure (e.g., gradual versus abrupt) across distortion modes. The pure shear response (\(K_{3}=0\)) naturally interpolates between the tensile and compressive bounds, yielding an intermediate softening trajectory consistent with the experimental observations presented earlier.

As the concentration increases from 1\% to 3\%~w/v, both the magnitude and curvature of \(\psi_{\text{prop}}\) increase, reflecting the stiffness and enhanced failure thresholds of denser agarose networks \cite{ghebremedhin2021physics}. The enhanced energy storage capacity and greater asymmetry at higher concentrations are consistent with the densification and strengthening of the polymer network \cite{jayawardena2023evaluation, normand2000new, drozdov2020tension}, as previously observed in the stress–stretch data (Figs. \ref{fig_fit} and \ref{fig:pureshear}).

Overall, the free-energy landscapes show that the proposed Lode-invariant-based formulation produces a physically interpretable, mode-sensitive, and thermodynamically consistent potential. By unifying the tensile, compressive, and intermediate mixed-mode energy responses within a single framework, the model provides a mechanistic foundation for the asymmetric softening and failure behaviors quantified in the preceding sections.

\subsection{Power-law parameter scaling and predictive cross-concentration validation at 2.5\% w/v}
\label{subsec:powerlaw_validation}

\begin{figure}[b!]
    \centering
    \includegraphics[width=5.981in, height=5.487in]{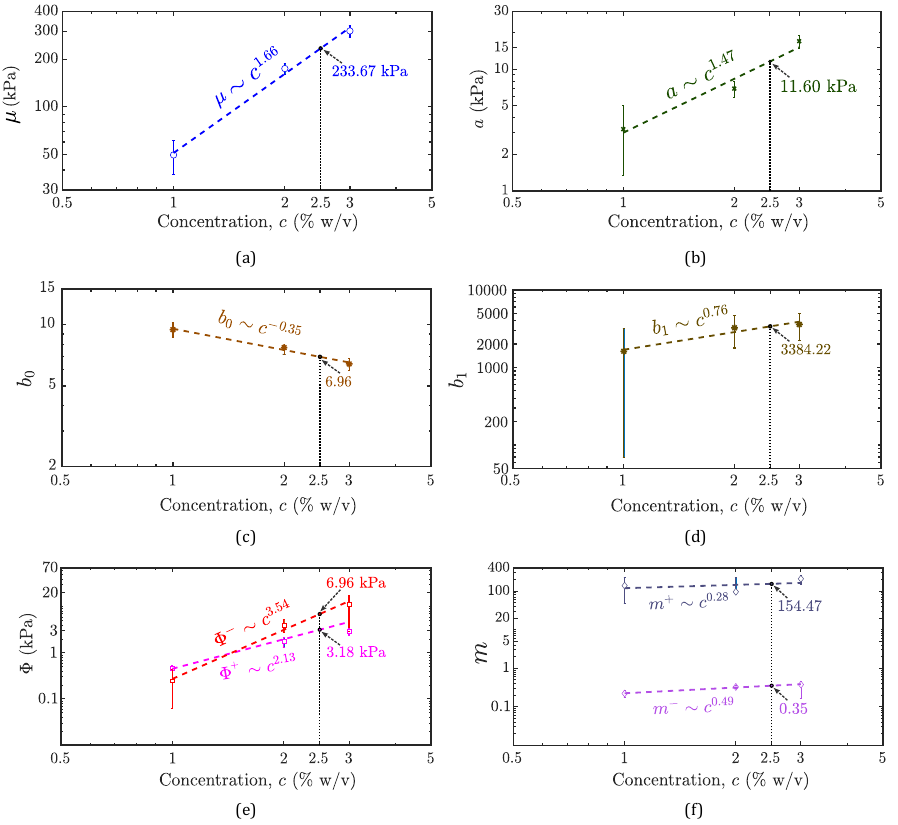}
    \caption{%
    Evolution of the proposed model's parameters with agarose gel concentration (from Table \ref{tab:params_all}).
    (a) Shear modulus \( \mu \);
    (b) nonlinear strain-stiffening parameter \( a \);
    (c) \( b_{0} \) and (d) \( b_{1} \), parameters defining the mode-dependent modulus function $\mathcal{G}(K_3)$;
    (e) tensile and compressive pseudo--failure energy parameters \( \Phi^{+} \) and \( \Phi^{-} \);
    and (f) tensile and compressive softening sharpness parameters \( m^{+} \) and \( m^{-} \).
    Remember, all parameters were obtained from simultaneous tension–compression fitting. Error bars represent one standard deviation.
    Dotted lines show power-law fits; power law exponents and interpolated values at 2.5\%~w/v are highlighted.}
    \label{fig:powerlaw}
\end{figure}

Having established the model’s accuracy and thermodynamic consistency across deformation modes, we now assess its scalability across agarose gel concentrations. This subsection aims to (i) quantify how each calibrated model parameter evolves with gel concentration using power-law scaling, and (ii) use these scaling relationships to predict the constitutive response of an intermediate concentration (2.5\%~w/v) that was not used during calibration. Successful prediction of uniaxial and pure shear responses at this interpolated composition provides a stringent test of cross-concentration generalizability.

The concentration-dependent evolution of each model parameter (\(Y\)) was fitted using a power-law relationship of the form:
\begin{equation}
Y = K\, c^{n},
\end{equation}
where \(c\) denotes gel concentration, \(K\) is a scaling constant, and \(n\) represents the power-law exponent. The data were linearized using logarithmic transformation,
\begin{equation}
\log(Y) = \log(K) + n \log(c),
\end{equation}
and fitted in MATLAB using a linear model (\texttt{poly1}). Here, \(n\) is the slope of the log--log plot, and $\log(K)$ is the intercept, which yields the scaling factor via \(K = \mathrm{exp}\left( \log(K)\right)\).

\autoref{fig:powerlaw} shows the resulting power-law fits for the eight model parameters—\(\mu, a, b_0, b_1, \Phi^{\pm}\), and \(m^{\pm}\)—for 1\%, 2\%, and 3\%~w/v agarose gels (parameters are listed in Table \ref{tab:params_all}). Error bars represent one standard deviation. The shear modulus \(\mu\) exhibits a strong positive scaling (\(\mu \sim c^{1.66}\)), consistent with the expected stiffening behavior of polymer networks with concentration \cite{Upadhyay2020, jones1990rigid, normand2000new, dormoy1991transient, fujii2000scaling}. Similarly, the strain-stiffening parameter \(a \sim c^{1.47}\) indicates increasing nonlinear reinforcement with concentration, reflecting a progressive amplification of strain-dependent shear response \cite{drozdov2020tension}. The modulus-function constants \(b_0\) and \(b_1\) exhibit weaker trends (\(b_0 \sim c^{-0.35}\), \(b_1 \sim c^{0.76}\)), implying that the shape of the mode-dependent stiffness modulation varies only modestly across concentrations. 

\begin{table}[b]
\centering
\caption{Estimated model parameters of 2.5\% w/v agarose from power-law relationship.}
\label{tab:estimated_params}
\begin{tabular}{cccccccc} 
\toprule
$\mu$ (kPa) & $a$ (kPa) & $b_{0}$ & $b_{1}$ & $\Phi^+$ (kPa) & $m^+$ & $\Phi^-$ (kPa) & $m^-$ \\
\midrule
233.67 & 11.60 & 6.96 & 3384.22 & 3.18 & 154.47 & 6.96 & 0.35 \\
\bottomrule
\end{tabular}
\end{table}

The tensile and compressive pseudo--failure energy parameters \(\Phi^{+} \sim c^{2.13}\) and \(\Phi^{-} \sim c^{3.54}\) grow substantially with concentration, reflecting increased failure energies and enhanced load-transfer capacity in denser networks. Further, they exhibit distinct scaling trends, with \(\Phi^{-}\) showing a steeper scaling with concentration, echoing the stronger energy-dissipation capacity in compression, with microstructural densification and enhanced load transfer through the polymer network at higher concentrations (see Section \ref{sec:classical-energy-limiters}). The tensile softening sharpness parameter $m^+$ shows no systematic trend while the compressive softening sharpness parameter $m^-$ shows a weak positive scaling, consistent with earlier observations.

\begin{figure}[t]
    \centering
    \includegraphics[width=6.066in, height=3.467in]{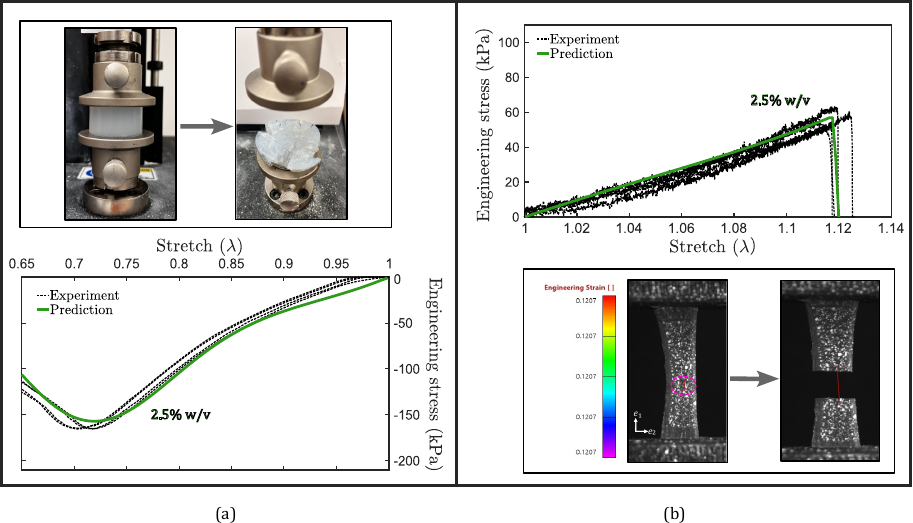}
    \caption{Experimental (n=4) and predicted engineering stress versus stretch plots of 2.5\% w/v agarose hydrogel in (a) compression and (b) tension. Predictions use parameters obtained solely from power-law interpolation of 1–3\% w/v data. Tensile strain was measured using DIC–based virtual extensometer.}
    \label{fig:conc}
\end{figure}

The interpolated parameter values at 2.5\%~w/v, listed in \autoref{tab:estimated_params} and highlighted in the figure, provide an intermediate reference dataset for predictive validation. Using these interpolated values, we predicted the mechanical response of 2.5\%~w/v gels under both uniaxial and pure shear loading without any additional fitting. As shown in Fig. \ref{fig:conc}, the predicted tensile and compressive responses agree closely with experiments, accurately capturing the elastic response, softening behavior, and failure onset. This agreement confirms that the power-law scaling effectively bridges concentration-dependent material behavior within the proposed asymmetric energy limiters framework.




\begin{figure}[h]
    \centering
    \includegraphics[width=3.4in, height=1.90in]{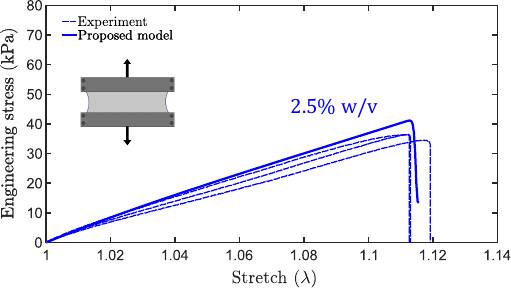}
    \caption{Engineering stress–stretch response of 2.5\%~w/v agarose hydrogel in pure shear. Dashed lines represent experimental data (n=3), while the solid line represents prediction from the proposed model using power-law–interpolated parameters.}
    \label{fig:pureshear_2.5p}
\end{figure}

\autoref{fig:pureshear_2.5p} extends this validation to the pure shear, a deformation path that was not used in the original parameter calibration. The ability of the interpolated parameters to accurately reproduce the elastic, softening, and failure responses under this unseen distortional mode demonstrates the robustness of the concentration-scaling framework. Together with the uniaxial results, the close agreement in pure shear confirms the framework’s \emph{cross-concentration} and \emph{cross-deformation-mode} generalizability.

Overall, the power-law scaling analysis provides a physically grounded and compact description of how elastic and failure parameters evolve with concentration. Combined with the validated cross-mode predictions, these results underscore the model’s capability to unify stiffness evolution, softening behavior, and failure energetics across both compositional and kinematic spaces, setting the stage for future application to other soft polymeric materials.

\section{Summary and Conclusion}
\label{sec:summary_conclusion}

Soft materials such as elastomers, hydrogels, and biological tissues exhibit pronounced TCA in their damage and failure responses—an aspect inadequately captured by most existing hyperelastic softening formulations. Motivated by this limitation, the present work introduces a generalized constitutive framework that extends Volokh’s energy-limiting theory to explicitly incorporate TCA through a physically motivated, distortion–mode–dependent softening formulation. The proposed model exploits the physical interpretability of the invariants of the logarithmic (Hencky) strain tensor—particularly the Lode invariants \((K_2, K_3)\)~\cite{prasad2020analysis, kulwant2023semi, chen2012general, criscione2000invariant}—to decompose the failure energy into distinct tensile and compressive branches. This bifurcation establishes an analytically driven switching criterion that allows independent control of distortion-mode–sensitive softening, thereby improving model generality and predictive capability across different loading modes while maintaining thermodynamic consistency. 
While a mode-dependent energy-limiting formulation may also be possible within classical principal-invariant-based \((I_1,I_2)\) or principal-stretch-based \((\lambda_i)\) descriptions, the Lode invariants of the logarithmic strain, \((K_2,K_3)\), offer a particularly convenient representation for the present bi-failure framework. Specifically, \(K_2\) quantifies the magnitude of distortional deformation, whereas \(K_3\) identifies the associated distortion mode. This separation provides a straightforward interpretation of mode dependence and enables a bounded, physically interpretable interpolation between tensile and compressive failure branches governed by the mode variable \(K_3\). In contrast, classical \((I_1,I_2)\)-based descriptions generally entangle deformation magnitude and mode \cite{chen2012general, criscione2000invariant}.

To represent the intact (undamaged) response, the proposed asymmetric failure formulation incorporates the Prasad–Kannan hyperelastic potential~\cite{prasad2020analysis}, originally developed from brain tissue experimental data and derived through an \textit{a priori} enforcement of the B–E inequalities~\cite{baker1954inequalities}. Within this framework, the complete constitutive description comprises eight parameters—four governing the elastic response and four regulating failure energetics. The model was calibrated using combined uniaxial tension and compression data for agarose gels at concentrations from 1–3\% w/v. Unlike conventional single-mode calibrations, the simultaneous fitting across both deformation modes is enabled by the Lode-invariant energy limiters, enabling a unified and physically motivated, distortion-mode-aware representation of both the intact hyperelastic response and its associated softening behavior. 

The framework demonstrated two levels of predictive capability. First, a multi-deformation-mode generalizability was established, wherein the model accurately predicted the pure shear response at each concentration (1, 2, and 3 \% w/v) using parameters calibrated solely from combined uniaxial tension–compression data. Second, an independent predictive validation was performed at 2.5 \% w/v by computing model parameters through power-law interpolation of fitted parameters from the 1–3 \% w/v range. The predicted stress–stretch responses for all three primary distortion modes showed excellent agreement with experiments, confirming both cross-concentration and cross-mode predictive accuracy of the proposed theory.

A key feature of the framework is the introduction of mode-dependent energy limiters (i.e., pseudo--failure energies \((\Phi^{\pm})\) and softening sharpness parameters $(m^\pm)$ ) governing both elastic storage and dissipation. This establishes a direct and physically interpretable link between microstructural failure mechanisms and macroscopic continuum-level behavior—without invoking any internal state variables. The calibrated parameters reliably represent essential microstructural mechanisms associated with rupture and energy dissipation, yielding a continuous and mechanistically grounded representation of softening. Such interpretability, long recognized as central to constitutive modeling~\cite{dal2021performance, guan2025hyperelastic, xiang2018general}, enhances both predictive fidelity and transparency while offering insight into how network architecture influences progressive damage. The agarose gels studied here serve as bio-mimetic surrogates, establishing a foundation for probing tissue-like failure mechanisms in a controlled environment \cite{pervin2011mechanically, hong2016localized, aghayan2022experimental}.

Looking ahead, the proposed framework can be extended to construct a comprehensive three-dimensional failure map over the \((K_2, K_3)\) invariant space. This advancement would require systematic multi-axial testing--such as combined compression–shear loading or energetically exhaustive protocols like unequal biaxial deformation~\cite{falope2024energetic}—to capture a complete representation of the energy surface and unveil the entire failure envelope of the material. Such experiments would enable direct evaluation of the Lode-weighting function $\beta(s(K_3))$, which in the present study was assumed to vary as a cubic smoothstep function based on validation across the three primary distortional modes. Incorporating these experimental insights would refine this assumption and facilitate a more accurate, data-driven characterization of mode-dependent failure energetics, thereby advancing toward a unified and comprehensive description of soft-material damage across the entire distortional domain. 
A related limitation concerns the representation of the tension-compression response in the limit of infinitesimal strain. The experimentally observed branch-wise infinitesimal moduli are distinct, whereas the intact hyperelastic backbone adopted in the present framework represents this behavior through a continuously varying stiffness with a single modulus at infinitesimal strain. Future extensions aimed at capturing this feature more accurately could therefore incorporate a bi-modulus formulation~\cite{du2020tension,du2014variational} to represent the piecewise, non-smooth constitutive response explicitly.

In addition, the present formulation is restricted to a time-independent constitutive description under quasi-static monotonic loading. While this is consistent with the loading conditions and hydrogel systems considered here, many biological tissues exhibit pronounced viscoelastic and poroelastic effects that can influence both the pre-failure response and damage evolution. Extending the proposed Lode-invariant-based asymmetric failure framework to incorporate rate dependence, for example through viscoelastic or poro-viscoelastic generalizations of the strain energy potential, constitutes an important direction for future work. Such developments would enable the model to capture time-dependent dissipation while retaining the mode-dependent failure structure, thereby broadening its applicability to rate-sensitive soft biological materials.

\section*{Declaration of competing Interest} \label{Declaration}
The authors declare that they have no known competing financial interests or personal relationships that could have appeared to influence the work reported in this paper. 

\section*{Acknowledgements} \label{Acknowledgements}
This material is based upon work supported by the National Science Foundation under Grant No. 2623467. The authors gratefully acknowledge the anonymous reviewers for their insightful and constructive feedback on an earlier version of this paper, which significantly improved the clarity and overall quality of this work.

\section*{Data availability} \label{Data-availability}
Data will be made available on request.

\appendix

\section{Tensor basis components of the Cauchy stress in Lode-invariants-based hyperelasticity}
\label{sec:appendix_A}
Mutually orthogonal unit kinematic tensors derived from Lode invariants and the left stretch tensor $\mathbf{V}$ for an incompressible material are defined as follows \cite{prasad2020analysis, reddipaga2026construction, kulwant2023semi, chen2012general}:
      
\begin{align}
  \mathbf{N}_1 &= \frac{\partial K_2}{\partial\ln\mathbf{V}}
                \;=\; \frac{\operatorname{dev}(\ln\mathbf{V})}{K_2},
                \label{eq:N1_definition} \\[6pt]
  \mathbf{N}_2 &= K_2\,\frac{\partial K_3}{\partial\ln\mathbf{V}}
                \;=\; \frac{\sqrt{6}}{\cos(3K_3)}
                  \Bigl(
                  \mathbf{N}_1^{\,2}
                  -\frac{1}{3}\mathbf{I}
                  -\operatorname{tr}\!\bigl(\mathbf{N}_1^{\,3}\bigr)\,\mathbf{N}_1
                  \Bigr)
                \label{eq:N2_definition}
\end{align}

which satisfy
$\mathbf{N}_1:\mathbf{N}_2=0$ and $\|\mathbf{N}_1\|=\|\mathbf{N}_2\|=1$.

\section{On irreversibility in hyperelasticity with energy limiters}
\label{appendix:irreversibility}

The energy-limiting formulation proposed by Volokh~\cite{volokh2007hyperelasticity, volokh2010modeling, volokh2013review} in Eq. (\ref{eq:volokh}) provides a framework for incorporating material failure into hyperelastic models. However, in its standard reversible form, the softening response remains path-independent: upon unloading, the material can elastically recover to its original state. This reversibility limits its physical realism, as actual material failure is inherently irreversible once the energy threshold is exceeded.

To impose irreversibility and path dependency, Volokh~\cite{volokh2014irreversibility} introduced a switch parameter $\zeta$ through a Heaviside step function that permanently restricts the recoverable energy once the failure energy density $\psi_f$ is reached. The modified stored-energy function is expressed as:
\begin{equation}
\psi(W(\mathbf{F}),\zeta) = \psi_{f} - H(\zeta)\,\psi_{e}\left(W(\mathbf{F})\right),
\label{eq:Psi_prop_irreversible}
\end{equation}
where $\psi_{e}$ denotes the elastic strain energy and $H(\zeta)$ is the Heaviside function governing the transition between reversible and irreversible states.

The evolution of the switch parameter $\zeta \in (-\infty, 0]$ is defined by
\begin{equation}
\dot{\zeta} = -H\!\left(\epsilon - \frac{\psi_{e}}{\psi_{f}} \right), 
\qquad 
\zeta(0) = 0,
\label{eq:zeta_evolution}
\end{equation}
where $\epsilon$ is a small dimensionless tolerance parameter ($0 < \epsilon \ll 1$) introduced to avoid numerical singularities near the energy limit.

In this formulation, the Heaviside function $H(\zeta)$ acts as a state-locking mechanism: 
\[
H(\zeta) = 
\begin{cases}
1, & \text{for } \zeta \geq 0 \quad \text{(reversible region)},\\[3pt]
0, & \text{for } \zeta < 0 \quad \text{(irreversible region)}.
\end{cases}
\]

When $\psi_{e} < \psi_{f}$, the system remains in the reversible regime ($\zeta = 0$, $H(\zeta)=1$), and deformation is purely hyperelastic. Once $\psi_{e}$ attains the failure threshold $\psi_{f}$, the variable $\zeta$ becomes negative, enforcing $H(\zeta)=0$. The strain energy is then locked at $\psi_{f}$, thereby preventing further recovery upon unloading and ensuring irreversible energy dissipation \cite{volokh2014irreversibility, mythravaruni2018failure, lev2019thermoelastic}.

Thus, $\zeta$ does not evolve continuously like an internal damage variable; rather, it acts as a discrete state switch that governs the transition from reversible to irreversible deformation. This treatment introduces path dependence into the otherwise elastic framework, ensuring that once failure occurs, the material cannot regain its original stored-energy capacity.

\section{Pure Shear stress--stretch reconstruction up to failure}
\label{appendix:pureshearextrapolation}
\renewcommand{\thefigure}{C.\arabic{figure}}
\setcounter{figure}{0}

\begin{figure}[h]
    \centering
    \includegraphics[width=5.898in, height=4.234in]{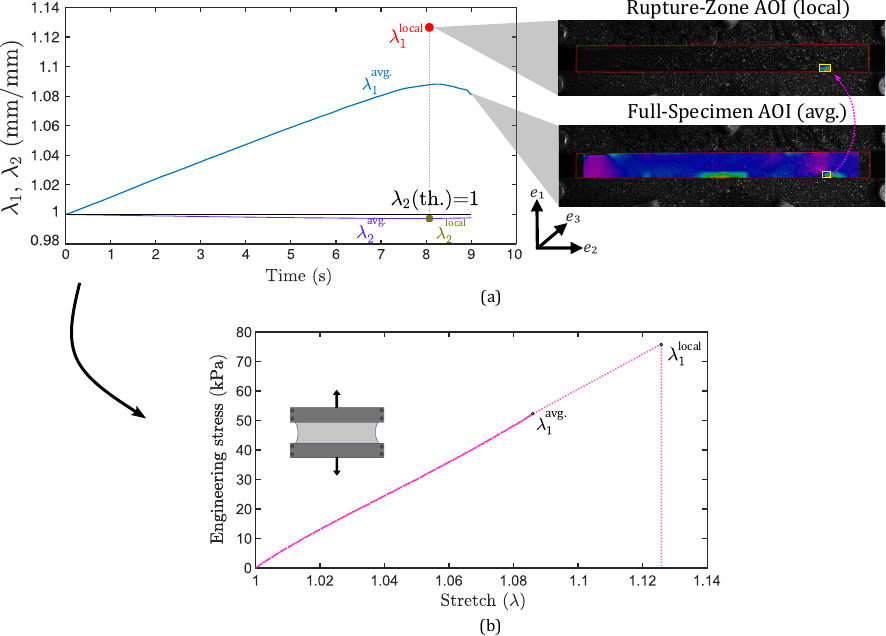}
    \caption{(a) In-plane principal stretches \(\lambda_1\) (axial) and \(\lambda_2\) (transverse) extracted from the average full-specimen AOI (avg.) and from a small local AOI positioned at the eventual rupture site (local).  
    Both AOIs satisfy the pure shear condition \(\lambda_2 \approx 1\), consistent with the mapping introduced in Section~\ref{sec:kinematics}.  
    The local AOI provides the avg.\ stretch immediately prior to rupture.  
    (b) Engineering stress--stretch response reconstructed by pairing the avg.\ engineering stress with the avg.\ stretch during the homogeneous regime and extrapolated to the local rupture stretch \(\lambda_1^{\mathrm{local}}\) at the failure instant.}
    \label{fig:pureshear_extra}
\end{figure}

\autoref{fig:pureshear_extra} summarizes the procedure to obtain a consistent pure shear stress--stretch response up to rupture.  
The method relies on two key observations:

\begin{enumerate}
    \item \textbf{Pure shear kinematics in both AOIs.}  
     The evolution of the in-plane strain components 
     (\(\varepsilon_{11} = \lambda_1 - 1\), \(\varepsilon_{22} = \lambda_2 - 1\)) extracted from the 
     average strain contour of the full-specimen AOI shows that the transverse stretch remains close 
     to unity (\(\lambda_2^{\mathrm{avg}} \approx 1\)) throughout the loading history, confirming that the specimen maintains an ideal pure shear state during the homogeneous deformation phase.  
     Immediately before rupture, the locally measured stretches at the failure site 
     (\(\lambda_1^{\mathrm{local}}, \lambda_2^{\mathrm{local}}\)) also satisfy 
    \(\lambda_2^{\mathrm{local}} \approx 1\), indicating that the deformation field at the rupture 
     location remains consistent with the incompressible pure shear mapping 
     \(\{\lambda,1,\lambda^{-1}\}\) introduced in Section~\ref{sec:kinematics} (Eqs.~\eqref{eq:pure_shear_map}--\eqref{eq:F_ps}).  Accordingly, both AOIs provide kinematically admissible descriptions of the deformation state: the avg.\ AOI captures the global homogeneous response, whereas the local AOI furnishes the appropriate stretch measure at the point of failure.

    \item \textbf{Localization and local stretch at failure.}  
    As rupture approaches, strain localizes in a small region of the specimen, causing the local axial stretch to deviate from the global, spatially averaged response.  
    The avg.\ AOI continues to characterize the homogeneous deformation portion of the test, while the local AOI isolates the stretch at the failure site.  
    The rupture stretch \(\lambda_1^{\mathrm{local}}\) is therefore taken directly from the local AOI at the final pre-rupture frame and appended to the avg.\ stretch history to complete the deformation trajectory.
\end{enumerate}

\paragraph{\textbf{Importance of Using Average Engineering Stress--Stretch Before Localization}}
The engineering stress during pure shear is obtained from the measured load-cell force divided by the cross-sectional area (specimen width times thickness; see Fig. \ref{fig:pure_shear_setup}).  
Because this load corresponds to the \emph{global} planar tensile force acting on the entire section, the associated strain measure must likewise represent a global deformation.  
Pairing a global stress with a local strain value would be inconsistent and would artificially distort the stress--stretch response, especially in materials such as hydrogels where deformation fields may be spatially non-uniform due to material heterogeneity or slight loading inhomogeneities \cite{millar2021improved, cantrell2017experimental}.

2D-DIC provides a full-field measure of the deformation and enables the computation of spatially averaged principal stretches across the entire gauge region.  
As reported in prior analyses of shear tests by Millar et al. \cite{millar2021improved}, averaging the strain field over the full-specimen AOI yields the most reliable representation of the global shear response, particularly when shear and normal strain distributions vary across the width.

Once the avg.\ engineering stretch \(\lambda_1^{\mathrm{avg}}\) has been paired consistently with the corresponding global engineering stress during the homogeneous regime, the final local rupture stretch \(\lambda_1^{\mathrm{local}}\) is appended to the avg.\ curve to extend the stress--stretch response to the rupture point.  
This procedure ensures that:
\begin{itemize}
    \item the stress is always paired with a physically meaningful strain measure,  
    \item global deformation is represented accurately up to the onset of localization, and  
    \item the final point on the curve reflects the true local strain at the site of material failure.
\end{itemize}

The resulting complete stress--stretch response, shown in Fig. \ref{fig:pureshear_extra}(b), is therefore consistent with pure shear kinematics throughout loading and remains suitable for validation of the proposed constitutive model up to failure.



\section{Small-strain limit of the Volokh-type branch specific energies}
\label{appendix:volokh_limit}

We now derive the asymptotic results used in Section~\ref{subsec:K2_zero_behavior}, with particular emphasis on the small-strain behavior of the Volokh-type bounded softening potentials.

Consider the branch-specific softening potential (Eq. (\ref{eq:psi_pm_exp})), defined using Eqs.~\eqref{eq:Psi_f_pm}--\eqref{eq:Psi_e_pm_1} in terms of the upper incomplete gamma function $\Gamma(s,x)$,
\begin{equation}
\psi^\pm(W)
=
\frac{\Phi^\pm}{m^\pm}\,
\Gamma\!\left(\frac{1}{m^\pm},\,0\right)
-
\frac{\Phi^\pm}{m^\pm}
\Gamma\!\left(
\frac{1}{m^\pm},
\left[\frac{W}{\Phi^\pm}\right]^{m^\pm}
\right).
\end{equation}
Using the identity $\Gamma(s)=\Gamma(s,x)+\gamma(s,x)$, one obtains
\begin{equation}
\Gamma\!\left(\frac{1}{m^\pm},\,0\right)
=
\Gamma\!\left(
\frac{1}{m^\pm},
\left[\frac{W}{\Phi^\pm}\right]^{m^\pm}
\right)
+
\gamma\!\left(
\frac{1}{m^\pm},
\left[\frac{W}{\Phi^\pm}\right]^{m^\pm}
\right).
\end{equation}
Accordingly, the bounded branch energy may equivalently be expressed in terms of the lower incomplete gamma function $\gamma(s,x)$ as
\begin{equation}
\psi^\pm(W)
=
\frac{\Phi^\pm}{m^\pm}
\gamma\!\left(
\frac{1}{m^\pm},
\left[\frac{W}{\Phi^\pm}\right]^{m^\pm}
\right).
\label{eq:branch_lower_gamma}
\end{equation}
Here, the lower incomplete gamma function is defined by
\begin{equation}
\gamma(s,x)
=
\int_0^x t^{s-1} e^{-t}\,dt.
\end{equation}

For convenience, let
\begin{equation}
s = \frac{1}{m^\pm},
\qquad
x = \left(\frac{W}{\Phi^\pm}\right)^{m^\pm}.
\label{eq:s_x_def}
\end{equation}
As $W \to 0$, it follows that $x \to 0$. Hence, for small $x$, the lower incomplete gamma function admits the asymptotic expansion
\begin{equation}
\gamma(s,x)
=
\frac{x^s}{s}
-
\frac{x^{s+1}}{s+1}
+
O(x^{s+2}).
\label{eq:gamma_series}
\end{equation}
Using Eq.~\eqref{eq:s_x_def}, it follows that
\begin{equation}
x^s
=
\frac{W}{\Phi^\pm},
\label{eq:x_power_s}
\end{equation}
and the substitution of \eqref{eq:gamma_series}--\eqref{eq:x_power_s} into \eqref{eq:branch_lower_gamma} gives
\begin{equation}
\psi^\pm(W)
=
W
-
\frac{1}{m^\pm+1}
\frac{W^{m^\pm+1}}{(\Phi^\pm)^{m^\pm}}
+
O(W^{2m^\pm+1}),
\qquad W \to 0.
\label{eq:psi_small_W}
\end{equation}
Further, differentiation yields
\begin{equation}
\frac{\partial \psi^\pm}{\partial W}
=
1
-
\frac{W^{m^\pm}}{(\Phi^\pm)^{m^\pm}}
+
O(W^{2m^\pm}),
\qquad W \to 0.
\label{eq:dpsi_small_W}
\end{equation}

Eqs.~\eqref{eq:psi_small_W}--\eqref{eq:dpsi_small_W} establish the small-$W$ asymptotic structure of the branch energies and justify the regularity and linearization results presented in Section~\ref{subsec:K2_zero_behavior}.


\section{Construction of the theoretical failure envelope in principal stretch space}
\label{appendix:failure_envelope}
\renewcommand{\thefigure}{D.\arabic{figure}}
\setcounter{figure}{0}

The proposed asymmetric energy-limiting formulation, once calibrated, provides a constitutive 
description of softening and failure under arbitrary distortional states, extending beyond the 
specific distortion modes used for calibration and validation. A useful way to visualize this broader predictive capability is through a theoretical failure envelope constructed in the space of principal stretches.

For a homogeneous incompressible planar deformation, the state of deformation is fully 
described by the two in-plane principal stretches \(\lambda_1\) and \(\lambda_2\), with the 
out-of-plane stretch constrained kinematically by $\lambda_3 = (\lambda_1\lambda_2)^{-1}$. 
Within the present Lode-invariant-based formulation, the strain energy density is expressed in 
terms of the distortional invariants \(K_2\) and \(K_3\), which can be written explicitly in terms 
of the principal stretches through the Hencky strain tensor \(\ln \mathbf{V}\). These invariants can be expressed as \cite{prasad2020analysis}
\begin{align}
K_2 &= \sqrt{2\left(\ln^2(\lambda_1) + \ln^2(\lambda_2) + \ln(\lambda_1)\ln(\lambda_2)\right)}, \\
K_3 &= -\frac{1}{3}\sin^{-1}\!\left(
\frac{3\sqrt{6}\,\ln(\lambda_1)\ln(\lambda_2)\bigl(\ln(\lambda_1)+\ln(\lambda_2)\bigr)}
{K_2^3}
\right).
\end{align}
These relations establish a direct mapping from the principal-stretch space \((\lambda_1,\lambda_2)\) to the Lode invariant space \((K_2,K_3)\). Consequently, the proposed strain energy density \(\psi_{\mathrm{prop}}(K_2,K_3)\) may be equivalently interpreted as a reduced potential \(\hat{\psi}_{\mathrm{prop}}(\lambda_1,\lambda_2)\).

Within the framework of hyperelasticity with energy limiters, failure is identified through the violation of the strong ellipticity condition \cite{mythravaruni2018failure,volokh2017loss, mythravaruni2019incompressibility, volokh2024modeling}. In the present planar setting, this loss of strong ellipticity corresponds to a loss of ellipticity of the quasi-static governing equations, signaling the onset of deformation localization and constitutive instability \cite{volokh2017loss, yan2025virtual, landauer2019experimental, triantafyllidis1986gradient, bigoni2012nonlinear}. To assess the onset of such instabilities for a given set of material parameters, the present study computes the material moduli $\frac{\partial^2\psi_{\mathrm{prop}}}{\partial F_{ij}\partial F_{kl}}$ as functions of the Lode invariants $(K_2,K_3)$ and then evaluates the necessary and sufficient criteria for strong ellipticity established by Dacorogna \cite{dacorogna2001necessary}. This provides a systematic constitutive criterion for identifying the onset of failure in the proposed model, consistent with prior studies based on energy-limiting formulations \cite{mythravaruni2018failure,mythravaruni2019incompressibility}.

It is worth noting that, in addition to strong-ellipticity-based failure criteria used in energy-limiter formulations, related studies have also used alternative criteria, including the vanishing determinant of the Hessian matrix of the strain energy density, to construct theoretical failure envelopes in principal-stretch space~\cite{volokh2013review,volokh2019constitutive,volokh2020new,balakhovsky2012inflation}. Such a construction is useful for visualizing the loss of local convexity of the energy potential along prescribed homogeneous deformation paths~\cite{xiao2020granular}. In contrast, the Legendre--Hadamard condition enforces strong ellipticity by requiring the acoustic tensor to remain positive definite for all rank-one perturbations \cite{hill1962acceleration,hadamard1903leccons}. Violation of this condition corresponds to the vanishing of a real wave speed for a superimposed plane wave along at least one propagation direction and polarization mode, indicating constitutive instability associated with the onset of failure~\cite{volokh2017loss,ball1976convexity,marsden1994mathematical,ogden1997non,mythravaruni2018failure,chen1991strong, d2016localization}.

Accordingly, in this appendix, the failure envelope is constructed using the strong-ellipticity-based Legendre--Hadamard condition. A complementary version based on the vanishing determinant of the Hessian of the reduced strain energy density is provided in Supplementary Section~S.3, where failure envelopes for both the proposed formulation and Volokh's classical framework are plotted in Fig.~S.3. For the calibrated case considered here, the Hessian-based envelope is qualitatively similar to the strong-ellipticity-based envelope, but the two criteria are not generally equivalent: the former probes loss of local convexity in a prescribed homogeneous principal-stretch subspace, whereas the latter tests stability with respect to all rank-one perturbations.

Governed by the loss of strong ellipticity, the failure condition used in this work defines a continuous locus in the \((\lambda_1,\lambda_2)\)-plane, which represents a theoretical failure envelope marking the onset of instability. In the present study, this envelope was generated numerically using the calibrated material parameters of 3\%~w/v agarose obtained from the combined uniaxial tension–compression fit. This serves as a representative case to illustrate the mode-dependent failure predictions of the proposed framework.

\begin{figure}[t]
    \centering
    \includegraphics[width=3.767in, height=3.656in]{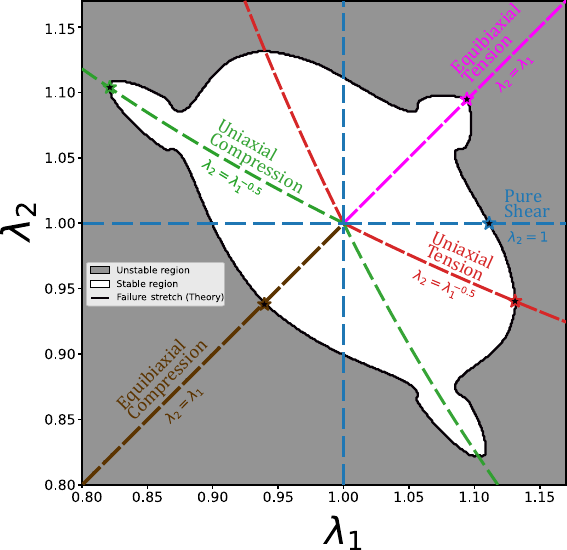}
    \caption{Theoretical failure envelope in principal stretch space, computed from the calibrated material parameters for the representative 3\%~w/v agarose gel. The black curve denotes the locus of constitutive instability defined by the loss of strong ellipticity conditions derived by Dacorogna \cite{dacorogna2001necessary}. Dashed lines indicate characteristic deformation paths in the 
    \((\lambda_1,\lambda_2)\)-plane: uniaxial tension 
    \((\lambda_2=\lambda_1^{-0.5}<1)\), uniaxial compression 
    \((\lambda_2=\lambda_1^{-0.5}>1)\), pure shear \((\lambda_2=1)\), equibiaxial tension 
    \((\lambda_2=\lambda_1>1)\), and equibiaxial compression \((\lambda_2=\lambda_1 <1)\).  
    The intersections of these loading paths with the envelope identify the corresponding 
    theoretically predicted failure stretches.}
    \label{fig:failure_envelope}
\end{figure}

Figure~\ref{fig:failure_envelope} presents the resulting failure envelope together with several canonical loading paths, namely uniaxial tension, uniaxial compression, pure shear, equibiaxial tension, and equibiaxial compression. The intersections of these loading paths with the instability envelope determine the corresponding theoretical failure stretches for each deformation mode.

Several important features emerge from this construction. First, the envelope is distinctly non-circular and strongly asymmetric, reflecting the intrinsic TCA embedded in the calibrated energy-limiting potential. Second, the predicted failure stretches vary substantially across deformation modes. In particular, the critical stretches associated with pure shear and equibiaxial tension differ markedly from those obtained under uniaxial loading, even though all of these deformation states are described within the same constitutive framework. This indicates that, in the proposed model, failure is not governed by a single universal critical stretch but instead by a mode-dependent instability condition that arises naturally from the underlying strain energy density.

These predictions are consistent with the experimentally observed failure stretches for the uniaxial tension and pure shear modes considered in the present study (see Section~\ref{sec:results_discussion}). More broadly, these results align with the findings of Rosendahl et al.~\cite{rosendahl2019equivalent}, who demonstrated that traditional failure criteria often cannot capture the complex shapes required to accurately represent the failure surfaces of many hyperelastic materials. Furthermore, Rosendahl et al.\ established that the incompressibility constraint defines a curved manifold in principal stretch space. Consequently, when failure criteria are projected onto the $(\lambda_1,\lambda_2)$-plane under the condition of incompressibility, the resulting failure envelopes can exhibit nonconvexity. The proposed failure envelope is therefore qualitatively consistent with experimental trends reported in multiaxial tests on incompressible hyperelastic materials, such as natural rubber. Collectively, these observations underscore an important conceptual point: the commonly used notion of ``material strength'' inferred solely from uniaxial data is not sufficient to characterize failure under general multiaxial loading. The limitations of such a viewpoint have attracted increasing attention in the recent literature; see, for example, \cite{de2022nucleation, lopez2025classical}. Consequently, constitutive predictions based exclusively on uniaxial strength measures can be misleading when extended to general three-dimensional deformation states, as also noted in related energy-limiter studies \cite{volokh2019constitutive}.

Within this broader context, it is important to emphasize that the present envelope represents a theoretical constitutive prediction obtained under an idealized homogeneous deformation setting. In related studies on energy-limiting formulations, experimentally measured critical stretches were found to be slightly lower than the corresponding theoretical predictions, likely due to material and geometric imperfections that are not captured in the idealized continuum description \cite{volokh2013review, lev2019thermoelastic}. Accordingly, the envelope obtained here should be interpreted as an ideal constitutive prediction, with similar deviations expected in the presence of experimental imperfections. Furthermore, the present analysis is restricted to instability associated with distortional deformation and does not account for failure mechanisms associated with hydrostatic stress. Extending the framework to incorporate interactions between distortional and dilatational failure surfaces therefore remains an important direction for future study.

Although derived for a specific representative case, the present envelope demonstrates how the proposed Lode-invariant-based asymmetric failure model can map failure continuously across principal stretch space. It thereby provides a foundation for constructing more comprehensive three-dimensional, mode-dependent failure surfaces informed by broader multiaxial experimental datasets.

 \bibliographystyle{elsarticle-num} 
 \bibliography{cas-refs}

@article{Upadhyay2020,
	author = {Kshitiz Upadhyay and Ghatu Subhash and Douglas Spearot},
	doi = {10.1016/j.ijengsci.2020.103314},
	issn = {00207225},
	journal = {International Journal of Engineering Science},
	month = {9},
	publisher = {Elsevier Ltd},
	title = {Hyperelastic constitutive modeling of hydrogels based on primary deformation modes and validation under 3D stress states},
	volume = {154},
	year = {2020},
}

@article{Sutton2008,
author = {Sutton, M.A. and Yan, J.H. and Tiwari, V. and Schreier, H.W. and Orteu, J.J.},
doi = {10.1016/j.optlaseng.2008.05.005},
issn = {01438166},
journal = {Optics and Lasers in Engineering},
month = {oct},
number = {10},
pages = {746--757},
publisher = {Elsevier Ltd},
title = {{The effect of out-of-plane motion on 2D and 3D digital image correlation measurements}},
volume = {46},
year = {2008}
}

@article{Moerman2016,
	author = {Kevin M. Moerman and Ciaran K. Simms and Thomas Nagel},
	doi = {10.1016/j.jmbbm.2015.11.027},
	issn = {18780180},
	journal = {Journal of the Mechanical Behavior of Biomedical Materials},
	month = {3},
	pages = {218-228},
	pmid = {26719933},
	publisher = {Elsevier Ltd},
	title = {Control of tension-compression asymmetry in Ogden hyperelasticity with application to soft tissue modelling},
	volume = {56},
	year = {2016},
}

@incollection{holzapfel2017modeling,
  title={Modeling of damage in soft biological tissues},
  author={Holzapfel, Gerhard A and Fereidoonnezhad, Behrooz},
  booktitle={Biomechanics of living organs},
  pages={101--123},
  year={2017},
  publisher={Elsevier}
}

@article{budday2017mechanical,
	title={Mechanical characterization of human brain tissue},
	author={Budday, Silvia and Sommer, Gerhard and Birkl, Christoph and Langkammer, Christian and Haybaeck, Johannes and Kohnert, Julius and Bauer, Melanie and Paulsen, Friedrich and Steinmann, Paul and Kuhl, Ellen and others},
	journal={Acta biomaterialia},
	volume={48},
	pages={319--340},
	year={2017},
	publisher={Elsevier}
}

@article{budday2017rheological,
	title={Rheological characterization of human brain tissue},
	author={Budday, Silvia and Sommer, Gerhard and Haybaeck, Johannes and Steinmann, Paul and Holzapfel, Gerhard A and Kuhl, Ellen},
	journal={Acta biomaterialia},
	volume={60},
	pages={315--329},
	year={2017},
	publisher={Elsevier}
}

@article{Lohr2022,
	author = {Matthew J. Lohr and Gabriella P. Sugerman and Sotirios Kakaletsis and Emma Lejeune and Manuel K. Rausch},
	doi = {10.1098/rsta.2021.0365},
	issn = {1364503X},
	issue = {2234},
	journal = {Philosophical Transactions of the Royal Society A: Mathematical, Physical and Engineering Sciences},
	month = {10},
	pmid = {36031838},
	publisher = {Royal Society Publishing},
	title = {An introduction to the Ogden model in biomechanics: Benefits, implementation tools and limitations},
	volume = {380},
	year = {2022},
}

@article{ogden1972large,
	title={Large deformation isotropic elasticity--on the correlation of theory and experiment for incompressible rubberlike solids},
	author={Ogden, Raymond William},
	journal={Proceedings of the Royal Society of London. A. Mathematical and Physical Sciences},
	volume={326},
	number={1567},
	pages={565--584},
	year={1972},
	publisher={The Royal Society London}
}

@article{Trapper2010,
	author = {P. Trapper and K. Y. Volokh},
	doi = {10.1016/j.ijsolstr.2010.08.016},
	issn = {00207683},
	issue = {25-26},
	journal = {International Journal of Solids and Structures},
	month = {12},
	pages = {3389-3396},
	title = {Elasticity with energy limiters for modeling dynamic failure propagation},
	volume = {47},
	year = {2010},
}

@article{volokh2021new,
  title={New approaches to modeling failure and fracture of rubberlike materials},
  author={Volokh, KY},
  journal={Fatigue Crack Growth in Rubber Materials: Experiments and Modelling},
  pages={131--151},
  year={2021},
  publisher={Springer}
}

@incollection{volokh2022modeling,
  title={Modeling failure and fracture in soft biological tissues},
  author={Volokh, Konstantin Y},
  booktitle={Solid (Bio) mechanics: Challenges of the Next Decade: A Book Dedicated to Professor Gerhard A. Holzapfel},
  pages={391--406},
  year={2022},
  publisher={Springer}
}

@article{volokh2007hyperelasticity,
	title={Hyperelasticity with softening for modeling materials failure},
	author={Volokh, KY},
	journal={Journal of the Mechanics and Physics of Solids},
	volume={55},
	number={10},
	pages={2237--2264},
	year={2007},
	publisher={Elsevier}
}

@article{volokh2013review,
	title={Review of the energy limiters approach to modeling failure of rubber},
	author={Volokh, KY},
	journal={Rubber Chemistry and Technology},
	volume={86},
	number={3},
	pages={470--487},
	year={2013},
	publisher={The Rubber Division, American Chemical Society}
}

@article{volokh2004nonlinear,
  title={Nonlinear elasticity for modeling fracture of isotropic brittle solids},
  author={Volokh, Konstantin Yu},
  journal={J. Appl. Mech.},
  volume={71},
  number={1},
  pages={141--143},
  year={2004}
}

@article{volokh2008multiscale,
  title={Multiscale modeling of material failure: From atomic bonds to elasticity with energy limiters},
  author={Volokh, Konstantin Y},
  journal={International Journal for Multiscale Computational Engineering},
  volume={6},
  number={5},
  year={2008},
  publisher={Begel House Inc.}
}

@article{wale2021applying,
  title={Applying ASTM standards to tensile tests of musculoskeletal soft tissue: Methods to reduce grip failures and promote reproducibility},
  author={Wale, Madison E and Nesbitt, Derek Q and Henderson, Bradley S and Fitzpatrick, Clare K and Creechley, Jaremy J and Lujan, Trevor J},
  journal={Journal of Biomechanical Engineering},
  volume={143},
  number={1},
  pages={011011},
  year={2021},
  publisher={American Society of Mechanical Engineers}
}

@article{kwon2010application,
  title={Application of digital image correlation method to biogel},
  author={Kwon, Hyock-Ju and Rogalsky, Allan D and Kovalchick, Christopher and Ravichandran, Guruswami},
  journal={Polymer Engineering \& Science},
  volume={50},
  number={8},
  pages={1585--1593},
  year={2010},
  publisher={Wiley Online Library}
}

@article{valanis1967strain,
  title={The strain-energy function of a hyperelastic material in terms of the extension ratios},
  author={Valanis, KC and Landel, Robert F},
  journal={Journal of Applied Physics},
  volume={38},
  number={7},
  pages={2997--3002},
  year={1967},
  publisher={AIP Publishing}
}

@article{rivlin1948large,
  title={Large elastic deformations of isotropic materials IV. Further developments of the general theory},
  author={Rivlin, Ronald S},
  journal={Philosophical transactions of the royal society of London. Series A, Mathematical and physical sciences},
  volume={241},
  number={835},
  pages={379--397},
  year={1948},
  publisher={The Royal Society London}
}

@article{volokh2014irreversibility,
  title={On irreversibility and dissipation in hyperelasticity with softening},
  author={Volokh, KY},
  journal={Journal of Applied Mechanics},
  volume={81},
  number={7},
  pages={074501},
  year={2014},
  publisher={American Society of Mechanical Engineers}
}

@article{moreira2013comparison,
  title={Comparison of simple and pure shear for an incompressible isotropic hyperelastic material under large deformation},
  author={Moreira, DC and Nunes, LCS},
  journal={Polymer Testing},
  volume={32},
  number={2},
  pages={240--248},
  year={2013},
  publisher={Elsevier}
}

@book{brown2006physical,
  title={Physical testing of rubber},
  author={Brown, Roger},
  year={2006},
  publisher={Springer Science \& Business Media}
}

@article{ambartsumyan1986elasticity,
  title={Elasticity theory of different moduli},
  author={Ambartsumyan, SA},
  journal={Trans. RF Wu and YZ Zhang. China Railway, Beijing},
  year={1986}
}

@article{du2020tension,
  title={Tension-compression asymmetry at finite strains: A theoretical model and exact solutions},
  author={Du, Zongliang and Zhang, Gang and Guo, Tianfu and Tang, Shan and Guo, Xu},
  journal={Journal of the Mechanics and Physics of Solids},
  volume={143},
  pages={104084},
  year={2020},
  publisher={Elsevier}
}

@article{volokh2019constitutive,
  title={Constitutive model of human artery adventitia enhanced with a failure description},
  author={Volokh, KY},
  journal={Mechanics of Soft Materials},
  volume={1},
  pages={1--9},
  year={2019},
  publisher={Springer}
}

@article{latorre2017understanding,
  title={Understanding the need of the compression branch to characterize hyperelastic materials},
  author={Latorre, Marcos and De Rosa, Erica and Mont{\'a}ns, Francisco J},
  journal={International Journal of Non-Linear Mechanics},
  volume={89},
  pages={14--24},
  year={2017},
  publisher={Elsevier}
}

@article{weisbecker2012layer,
  title={Layer-specific damage experiments and modeling of human thoracic and abdominal aortas with non-atherosclerotic intimal thickening},
  author={Weisbecker, Hannah and Pierce, David M and Regitnig, Peter and Holzapfel, Gerhard A},
  journal={Journal of the mechanical behavior of biomedical materials},
  volume={12},
  pages={93--106},
  year={2012},
  publisher={Elsevier}
}

@article{anssari2024continuous,
  title={Continuous Softening as a State of Hyperelasticity: Examples of Application to the Softening Behavior of the Brain Tissue},
  author={Anssari-Benam, Afshin and Saccomandi, Giuseppe},
  journal={Journal of Biomechanical Engineering},
  volume={146},
  number={9},
  year={2024},
  publisher={American Society of Mechanical Engineers Digital Collection}
}

@article{anssari2023continuous,
  title={Continuous softening up to the onset of failure: A hyperelastic modelling approach with intrinsic softening for isotropic incompressible soft solids},
  author={Anssari-Benam, Afshin},
  journal={Mechanics Research Communications},
  volume={132},
  pages={104183},
  year={2023},
  publisher={Elsevier}
}

@article{gao1998numerical,
  title={Numerical simulation of crack growth in an isotropic solid with randomized internal cohesive bonds},
  author={Gao, Huajian and Klein, Patrick},
  journal={Journal of the Mechanics and Physics of Solids},
  volume={46},
  number={2},
  pages={187--218},
  year={1998},
  publisher={Elsevier}
}

@article{xiao2021micromechanical,
  title={Micromechanical modeling of the multi-axial deformation behavior in double network hydrogels},
  author={Xiao, Rui and Mai, Thanh-Tam and Urayama, Kenji and Gong, Jian Ping and Qu, Shaoxing},
  journal={International Journal of Plasticity},
  volume={137},
  pages={102901},
  year={2021},
  publisher={Elsevier}
}

@article{mai2019damage,
  title={Damage cross-effect and anisotropy in tough double network hydrogels revealed by biaxial stretching},
  author={Mai, Thanh-Tam and Matsuda, Takahiro and Nakajima, Tasuku and Gong, Jian Ping and Urayama, Kenji},
  journal={Soft matter},
  volume={15},
  number={18},
  pages={3719--3732},
  year={2019},
  publisher={Royal Society of Chemistry}
}

@article{anssari2024unified,
  title={A unified pseudo-elastic model of continuous and discontinuous softening in the finite deformation of isotropic soft solids},
  author={Anssari-Benam, Afshin and Hossain, Mokarram},
  journal={International Journal of Solids and Structures},
  volume={290},
  pages={112670},
  year={2024},
  publisher={Elsevier}
}

@article{dargazany2009network,
  title={A network evolution model for the anisotropic Mullins effect in carbon black filled rubbers},
  author={Dargazany, Roozbeh and Itskov, Mikhail},
  journal={International Journal of Solids and Structures},
  volume={46},
  number={16},
  pages={2967--2977},
  year={2009},
  publisher={Elsevier}
}

@article{zhang2019fracture,
  title={Fracture in tension--compression-asymmetry solids via phase field modeling},
  author={Zhang, Gang and Guo, Tian Fu and Guo, Xu and Tang, Shan and Fleming, Mark and Liu, Wing Kam},
  journal={Computer Methods in Applied Mechanics and Engineering},
  volume={357},
  pages={112573},
  year={2019},
  publisher={Elsevier}
}

@article{drozdov2020tension,
  title={Tension--compression asymmetry in the mechanical response of hydrogels},
  author={Drozdov, Aleksey D and Christiansen, J deC},
  journal={Journal of the Mechanical Behavior of Biomedical Materials},
  volume={110},
  pages={103851},
  year={2020},
  publisher={Elsevier}
}

@article{voyiadjis2018hyperelastic,
  title={Hyperelastic modeling of the human brain tissue: effects of no-slip boundary condition and compressibility on the uniaxial deformation},
  author={Voyiadjis, George Z and Samadi-Dooki, Aref},
  journal={Journal of the mechanical behavior of biomedical materials},
  volume={83},
  pages={63--78},
  year={2018},
  publisher={Elsevier}
}

@article{upadhyay2019thermodynamics,
  title={Thermodynamics-based stability criteria for constitutive equations of isotropic hyperelastic solids},
  author={Upadhyay, Kshitiz and Subhash, Ghatu and Spearot, Douglas},
  journal={Journal of the Mechanics and Physics of Solids},
  volume={124},
  pages={115--142},
  year={2019},
  publisher={Elsevier}
}

@article{ogden1999pseudo,
  title={A pseudo--elastic model for the Mullins effect in filled rubber},
  author={Ogden, Ray W and Roxburgh, David George},
  journal={Proceedings of the Royal Society of London. Series A: Mathematical, Physical and Engineering Sciences},
  volume={455},
  number={1988},
  pages={2861--2877},
  year={1999},
  publisher={The Royal Society}
}

@article{mythravaruni2019incompressibility,
  title={On incompressibility constraint and crack direction in soft solids},
  author={Mythravaruni, P and Volokh, KY},
  journal={Journal of Applied Mechanics},
  volume={86},
  number={10},
  pages={101004},
  year={2019},
  publisher={American Society of Mechanical Engineers}
}

@article{long2012crack,
  title={Crack buckling in soft gels under compression},
  author={Long, Rong and Hui, Chung-Yuen},
  journal={Acta Mechanica Sinica},
  volume={28},
  pages={1098--1105},
  year={2012},
  publisher={Springer}
}

@article{sun2022rheology,
  title={Rheology of fibrous gels under compression},
  author={Sun, Chuanpeng and Purohit, Prashant K},
  journal={Extreme Mechanics Letters},
  volume={54},
  pages={101757},
  year={2022},
  publisher={Elsevier}
}

@article{miller2021microstructurally,
  title={A microstructurally motivated constitutive description of collagenous soft biological tissue towards the description of their non-linear and time-dependent properties},
  author={Miller, Christopher and Gasser, T Christian},
  journal={Journal of the Mechanics and Physics of Solids},
  volume={154},
  pages={104500},
  year={2021},
  publisher={Elsevier}
}

@article{miller2022bottom,
  title={A bottom-up approach to model collagen fiber damage and failure in soft biological tissues},
  author={Miller, Christopher and Gasser, T Christian},
  journal={Journal of the Mechanics and Physics of Solids},
  volume={169},
  pages={105086},
  year={2022},
  publisher={Elsevier}
}

@article{upadhyay2021validated,
  title={Validated tensile characterization of the strain rate dependence in soft materials},
  author={Upadhyay, Kshitiz and Spearot, Douglas and Subhash, Ghatu},
  journal={International Journal of Impact Engineering},
  volume={156},
  pages={103949},
  year={2021},
  publisher={Elsevier}
}

@article{upadhyay2020visco,
  title={Visco-hyperelastic constitutive modeling of strain rate sensitive soft materials},
  author={Upadhyay, Kshitiz and Subhash, Ghatu and Spearot, Douglas},
  journal={Journal of the Mechanics and Physics of Solids},
  volume={135},
  pages={103777},
  year={2020},
  publisher={Elsevier}
}

@article{Upadhyay2022_headinjurymodel,
author = {Upadhyay, Kshitiz and Alshareef, Ahmed and Knutsen, Andrew K. and Johnson, Curtis L. and Carass, Aaron and Bayly, Philip V. and Pham, Dzung L. and Prince, Jerry L. and Ramesh, K T},
doi = {10.1098/rsif.2022.0561},
isbn = {0000000213},
issn = {1742-5662},
journal = {Journal of The Royal Society Interface},
month = {oct},
number = {195},
pages = {1--27},
title = {{Development and validation of subject-specific 3D human head models based on a nonlinear visco-hyperelastic constitutive framework}},
volume = {19},
year = {2022}
}

@article{Bayly2021,
author = {Bayly, Philip V. and Alshareef, Ahmed and Knutsen, Andrew K. and Upadhyay, Kshitiz and Okamoto, Ruth J. and Carass, Aaron and Butman, John A. and Pham, Dzung L. and Prince, Jerry L. and Ramesh, K. T. and Johnson, Curtis L.},
doi = {10.1007/s10439-021-02820-0},
issn = {15739686},
journal = {Annals of Biomedical Engineering},
number = {10},
pages = {2677--2692},
pmid = {34212235},
title = {{MR Imaging of Human Brain Mechanics In Vivo: New Measurements to Facilitate the Development of Computational Models of Brain Injury}},
volume = {49},
year = {2021}
}

@article{volokh2010modeling,
  title={On modeling failure of rubber-like materials},
  author={Volokh, KY},
  journal={Mechanics Research Communications},
  volume={37},
  number={8},
  pages={684--689},
  year={2010},
  publisher={Elsevier}
}

@article{volokh2017loss,
  title={Loss of ellipticity in elasticity with energy limiters},
  author={Volokh, KY},
  journal={European Journal of Mechanics-A/Solids},
  volume={63},
  pages={36--42},
  year={2017},
  publisher={Elsevier}
}

@article{liang2017phase,
  title={Phase transitions during compression and decompression of clots from platelet-poor plasma, platelet-rich plasma and whole blood},
  author={Liang, Xiaojun and Chernysh, Irina and Purohit, Prashant K and Weisel, John W},
  journal={Acta biomaterialia},
  volume={60},
  pages={275--290},
  year={2017},
  publisher={Elsevier}
}

@article{filla2023multiscale,
  title={A multiscale framework for modeling fibrin fiber networks: Theory development and validation},
  author={Filla, Nicholas and Hou, Jixin and Li, He and Wang, Xianqiao},
  journal={Journal of the Mechanics and Physics of Solids},
  volume={179},
  pages={105392},
  year={2023},
  publisher={Elsevier}
}

@article{islam2018effect,
  title={Effect of network architecture on the mechanical behavior of random fiber networks},
  author={Islam, MR and Picu, RC},
  journal={Journal of Applied Mechanics},
  volume={85},
  number={8},
  pages={081011},
  year={2018},
  publisher={American Society of Mechanical Engineers}
}

@article{rivlin1953rupture,
  title={Rupture of rubber. I. Characteristic energy for tearing},
  author={Rivlin, Ronald S and Thomas, A Gr},
  journal={Journal of polymer science},
  volume={10},
  number={3},
  pages={291--318},
  year={1953},
  publisher={Wiley Online Library}
}

@article{tutwiler2020rupture,
  title={Rupture of blood clots: Mechanics and pathophysiology},
  author={Tutwiler, Valerie and Singh, Jaspreet and Litvinov, Rustem I and Bassani, John L and Purohit, Prashant K and Weisel, John W},
  journal={Science advances},
  volume={6},
  number={35},
  pages={eabc0496},
  year={2020},
  publisher={American Association for the Advancement of Science}
}

@article{lu2020pseudo,
  title={A pseudo-elasticity theory to model the strain-softening behavior of tough hydrogels},
  author={Lu, Tongqing and Wang, Zhongtong and Tang, Jingda and Zhang, Wenlei and Wang, Tiejun},
  journal={Journal of the Mechanics and Physics of Solids},
  volume={137},
  pages={103832},
  year={2020},
  publisher={Elsevier}
}

@article{pena2011prediction,
  title={Prediction of the softening and damage effects with permanent set in fibrous biological materials},
  author={Pe{\~n}a, Estefan{\'\i}a},
  journal={Journal of the Mechanics and Physics of Solids},
  volume={59},
  number={9},
  pages={1808--1822},
  year={2011},
  publisher={Elsevier}
}

@book{bergstrom2015mechanics,
  title={Mechanics of solid polymers: theory and computational modeling},
  author={Bergstrom, Jorgen S},
  year={2015},
  publisher={William Andrew}
}

@article{liu2018predicting,
  title={Predicting the mixed-mode I/II spatial damage propagation along 3D-printed soft interfacial layer via a hyperelastic softening model},
  author={Liu, Lei and Li, Yaning},
  journal={Journal of the Mechanics and Physics of Solids},
  volume={116},
  pages={17--32},
  year={2018},
  publisher={Elsevier}
}

@article{bremer2024ballistic,
  title={Ballistic and blast-relevant, high-rate material properties of physically and chemically crosslinked hydrogels},
  author={Bremer-Sai, EC and Yang, J and McGhee, A and Franck, C},
  journal={Experimental Mechanics},
  volume={64},
  number={4},
  pages={587--592},
  year={2024},
  publisher={Springer}
}

@article{jones1990rigid,
  title={Rigid polymer network models},
  author={Jones, JL and Marques, CM},
  journal={Journal de Physique},
  volume={51},
  number={11},
  pages={1113--1127},
  year={1990},
  publisher={Soci{\'e}t{\'e} Fran{\c{c}}aise de Physique}
}

@article{prasad2020analysis,
  title={An analysis driven construction of distortional-mode-dependent and Hill-Stable elastic potential with application to human brain tissue},
  author={Prasad, Durga and Kannan, K.},
  journal={Journal of the Mechanics and Physics of Solids},
  volume={134},
  pages={103752},
  year={2020},
  publisher={Elsevier}
}

@article{falope2024energetic,
  title={Energetic exhaustiveness for the direct characterization of energy forms of hyperelastic isotropic materials},
  author={Falope, Federico Oyedeji and Lanzoni, Luca and Tarantino, Angelo Marcello},
  journal={Journal of the Mechanics and Physics of Solids},
  volume={193},
  pages={105885},
  year={2024},
  publisher={Elsevier}
}

@article{criscione2000invariant,
  title={An invariant basis for natural strain which yields orthogonal stress response terms in isotropic hyperelasticity},
  author={Criscione, John C and Humphrey, Jay D and Douglas, Andrew S and Hunter, William C},
  journal={Journal of the Mechanics and Physics of Solids},
  volume={48},
  number={12},
  pages={2445--2465},
  year={2000},
  publisher={Elsevier}
}

@article{chen2012general,
  title={General invariant representations of the constitutive equations for isotropic nonlinearly elastic materials},
  author={Chen, MX and Tan, YW and Wang, BF},
  journal={International journal of solids and structures},
  volume={49},
  number={2},
  pages={318--327},
  year={2012},
  publisher={Elsevier}
}

@article{yeoh2001analysis,
  title={Analysis of deformation and fracture of ‘pure shear’rubber testpiece},
  author={Yeoh, OH},
  journal={Plastics, rubber and composites},
  volume={30},
  number={8},
  pages={389--397},
  year={2001},
  publisher={SAGE Publications Sage UK: London, England}
}

@article{kulwant2023semi,
  title={A semi-analytical inverse method to obtain the hyperelastic potential using experimental data},
  author={Kulwant, Vijay and Arvind, K and Prasad, Durga and Sreejith, P and Mohankumar, KV and Kannan, K},
  journal={Journal of the Mechanics and Physics of Solids},
  volume={181},
  pages={105431},
  year={2023},
  publisher={Elsevier}
}

@article{lev2019thermoelastic,
  title={Thermoelastic deformation and failure of rubberlike materials},
  author={Lev, Y and Faye, A and Volokh, KY},
  journal={Journal of the Mechanics and Physics of Solids},
  volume={122},
  pages={538--554},
  year={2019},
  publisher={Elsevier}
}

@article{ogden2004fitting,
  title={Fitting hyperelastic models to experimental data},
  author={Ogden, Raymond W and Saccomandi, Giuseppe and Sgura, Ivonne},
  journal={Computational Mechanics},
  volume={34},
  number={6},
  pages={484--502},
  year={2004},
  publisher={Springer}
}

@article{cortes2012extra,
  title={Extra-fibrillar matrix mechanics of annulus fibrosus in tension and compression},
  author={Cortes, Daniel H and Elliott, Dawn M},
  journal={Biomechanics and modeling in mechanobiology},
  volume={11},
  number={6},
  pages={781--790},
  year={2012},
  publisher={Springer}
}

@article{myers2015continuous,
  title={A continuous fiber distribution material model for human cervical tissue},
  author={Myers, Kristin M and Hendon, Christine P and Gan, Yu and Yao, Wang and Yoshida, Kyoko and Fernandez, Michael and Vink, Joy and Wapner, Ronald J},
  journal={Journal of biomechanics},
  volume={48},
  number={9},
  pages={1533--1540},
  year={2015},
  publisher={Elsevier}
}

@article{carrillo2013nonlinear,
  title={Nonlinear elasticity: from single chain to networks and gels},
  author={Carrillo, Jan-Michael Y and MacKintosh, Fred C and Dobrynin, Andrey V},
  journal={Macromolecules},
  volume={46},
  number={9},
  pages={3679--3692},
  year={2013},
  publisher={ACS Publications}
}

@article{holzapfel2025biomechanics,
  title={Biomechanics of soft biological tissues and organs, mechanobiology, homeostasis and modelling},
  author={Holzapfel, Gerhard A and Humphrey, Jay D and Ogden, Ray W},
  journal={Journal of the Royal Society Interface},
  volume={22},
  number={222},
  pages={20240361},
  year={2025},
  publisher={The Royal Society}
}

@article{franceschini2006brain,
  title={Brain tissue deforms similarly to filled elastomers and follows consolidation theory},
  author={Franceschini, Giulia and Bigoni, Davide and Regitnig, Peter and Holzapfel, Gerhard A},
  journal={Journal of the Mechanics and Physics of Solids},
  volume={54},
  number={12},
  pages={2592--2620},
  year={2006},
  publisher={Elsevier}
}

@article{mendis1995constitutive,
  title={A constitutive relationship for large deformation finite element modeling of brain tissue},
  author={Mendis, KK and Stalnaker, RL and Advani, SH},
  journal={Journal of Biomechanical Engineering},
  year={1995}
}

@article{simo1987fully,
  title={On a fully three-dimensional finite-strain viscoelastic damage model: formulation and computational aspects},
  author={Simo, Juan Carlos},
  journal={Computer methods in applied mechanics and engineering},
  volume={60},
  number={2},
  pages={153--173},
  year={1987},
  publisher={Elsevier}
}

@article{martonova2025generalized,
  title={Generalized invariants meet constitutive neural networks: A novel framework for hyperelastic materials},
  author={Martonov{\'a}, Denisa and Goriely, Alain and Kuhl, Ellen},
  journal={Journal of the Mechanics and Physics of Solids},
  pages={106352},
  year={2025},
  publisher={Elsevier}
}

@article{tang1997stress,
  title={Stress-strain relationships for gellan gels in tension, compression and torsion},
  author={Tang, Juming and Tung, Marvin A and Lelievre, John and Zeng, Yanyin},
  journal={Journal of Food Engineering},
  volume={31},
  number={4},
  pages={511--529},
  year={1997},
  publisher={Elsevier}
}

@article{normand2000new,
  title={New insight into agarose gel mechanical properties},
  author={Normand, Val{\'e}ry and Lootens, Didier L and Amici, Eleonora and Plucknett, Kevin P and Aymard, Pierre},
  journal={Biomacromolecules},
  volume={1},
  number={4},
  pages={730--738},
  year={2000},
  publisher={ACS Publications}
}

@article{pasumarthy2016mechanical,
  title={Mechanical and optical characterization of a tissue surrogate polymer gel},
  author={Pasumarthy, Raj Krishna Abhiram and Tippur, HV},
  journal={Polymer Testing},
  volume={55},
  pages={219--229},
  year={2016},
  publisher={Elsevier}
}

@article{goswami2025mechanics,
  title={Mechanics of Physically Cross-Linked Hydrogels: Experiments and Theoretical Modeling},
  author={Goswami, Mohit and Dutta, Agniva and Kulshreshtha, Rishi and Vasilyev, Gleb and Zussman, Eyal and Volokh, Konstantin},
  journal={Macromolecules},
  volume={58},
  number={9},
  pages={4478--4487},
  year={2025},
  publisher={ACS Publications}
}

@article{volokh2015cavitation,
  title={Cavitation instability as a trigger of aneurysm rupture},
  author={Volokh, KY},
  journal={Biomechanics and modeling in mechanobiology},
  volume={14},
  number={5},
  pages={1071--1079},
  year={2015},
  publisher={Springer}
}

@article{volokh2011modeling,
  title={Modeling failure of soft anisotropic materials with application to arteries},
  author={Volokh, KY},
  journal={Journal of the mechanical behavior of biomedical materials},
  volume={4},
  number={8},
  pages={1582--1594},
  year={2011},
  publisher={Elsevier}
}

@article{volokh2008prediction,
  title={Prediction of arterial failure based on a microstructural bi-layer fiber--matrix model with softening},
  author={Volokh, KY},
  journal={Journal of Biomechanics},
  volume={41},
  number={2},
  pages={447--453},
  year={2008},
  publisher={Elsevier}
}

@article{volokh2011cavitation,
  title={Cavitation instability in rubber},
  author={Volokh, KY},
  journal={International Journal of Applied Mechanics},
  volume={3},
  number={02},
  pages={299--311},
  year={2011},
  publisher={World Scientific}
}

@article{volokh2005modified,
  title={On the modified virtual internal bond method},
  author={Volokh, KY and Gao, H},
  journal={Journal of Applied mechanics},
  volume={72},
  number={6},
  pages={969--971},
  year={2005},
  publisher={American Society of Mechanical Engineers Digital Collection}
}

@article{goriely2015mechanics,
  title={Mechanics of the brain: perspectives, challenges, and opportunities},
  author={Goriely, Alain and Geers, Marc GD and Holzapfel, Gerhard A and Jayamohan, Jayaratnam and J{\'e}rusalem, Antoine and Sivaloganathan, Sivabal and Squier, Waney and van Dommelen, Johannes AW and Waters, Sarah and Kuhl, Ellen},
  journal={Biomechanics and modeling in mechanobiology},
  volume={14},
  number={5},
  pages={931--965},
  year={2015},
  publisher={Springer}
}

@article{anssari2022modelling,
  title={Modelling brain tissue elasticity with the Ogden model and an alternative family of constitutive models},
  author={Anssari-Benam, Afshin and Destrade, Michel and Saccomandi, Giuseppe},
  journal={Philosophical Transactions of the Royal Society A},
  volume={380},
  number={2234},
  pages={20210325},
  year={2022},
  publisher={The Royal Society}
}

@article{tang2019phase,
  title={Phase field modeling of fracture in nonlinearly elastic solids via energy decomposition},
  author={Tang, Shan and Zhang, Gang and Guo, Tian Fu and Guo, Xu and Liu, Wing Kam},
  journal={Computer Methods in Applied Mechanics and Engineering},
  volume={347},
  pages={477--494},
  year={2019},
  publisher={Elsevier}
}

@article{du2014variational,
  title={Variational principles and the related bounding theorems for bi-modulus materials},
  author={Du, Zongliang and Guo, Xu},
  journal={Journal of the Mechanics and Physics of Solids},
  volume={73},
  pages={183--211},
  year={2014},
  publisher={Elsevier}
}

@article{liu2025tube,
  title={A tube-based constitutive model of brain tissue with inner pressure},
  author={Liu, Wei and Yu, Zefeng and Elkhodary, Khalil I and Xiao, Hanlin and Tang, Shan and Guo, Tianfu and Guo, Xu},
  journal={Journal of the Mechanics and Physics of Solids},
  volume={196},
  pages={105993},
  year={2025},
  publisher={Elsevier}
}

@article{konale2025modeling,
  title={On modeling fracture of soft polymers},
  author={Konale, Aditya and Srivastava, Vikas},
  journal={Mechanics of Materials},
  volume={206},
  pages={105346},
  year={2025},
  publisher={Elsevier}
}

@article{destrade2017methodical,
  title={Methodical fitting for mathematical models of rubber-like materials},
  author={Destrade, Michel and Saccomandi, Giuseppe and Sgura, Ivonne},
  journal={Proceedings of the Royal Society A: Mathematical, Physical and Engineering Sciences},
  volume={473},
  number={2198},
  pages={20160811},
  year={2017},
  publisher={The Royal Society Publishing}
}

@article{volokh2017fracture,
  title={Fracture as a material sink},
  author={Volokh, KY},
  journal={Materials Theory},
  volume={1},
  number={1},
  pages={3},
  year={2017},
  publisher={Springer}
}

@article{huang2016effects,
  title={Effects of tension--compression asymmetry on the surface wrinkling of film--substrate systems},
  author={Huang, Xiao and Li, Bo and Hong, Wei and Cao, Yan-Ping and Feng, Xi-Qiao},
  journal={Journal of the Mechanics and Physics of Solids},
  volume={94},
  pages={88--104},
  year={2016},
  publisher={Elsevier}
}

@article{mooney1940theory,
  title={A theory of large elastic deformation},
  author={Mooney, Melvin},
  journal={Journal of applied physics},
  volume={11},
  number={9},
  pages={582--592},
  year={1940}
}

@article{rivlin1951large,
  title={Large elastic deformations of isotropic materials VII. Experiments on the deformation of rubber},
  author={Rivlin, Ronald S and Saunders, DW0042},
  journal={Philosophical Transactions of the Royal Society of London. Series A, Mathematical and Physical Sciences},
  volume={243},
  number={865},
  pages={251--288},
  year={1951},
  publisher={The Royal Society London}
}

@article{notbohm2015microbuckling,
  title={Microbuckling of fibrin provides a mechanism for cell mechanosensing},
  author={Notbohm, Jacob and Lesman, Ayelet and Rosakis, Phoebus and Tirrell, David A and Ravichandran, Guruswami},
  journal={Journal of The Royal Society Interface},
  volume={12},
  number={108},
  pages={20150320},
  year={2015},
  publisher={The Royal Society}
}

@article{rivlin1949largeb,
  title={Large elastic deformations of isotropic materials VI. Further results in the theory of torsion, shear and flexure},
  author={Rivlin, Ronald Samuel},
  journal={Philosophical Transactions of the Royal Society of London. Series A, Mathematical and Physical Sciences},
  volume={242},
  number={845},
  pages={173--195},
  year={1949},
  publisher={The Royal Society London}
}

@article{lu2025waviness,
  title={A waviness-centered damage model for collagenous soft tissues},
  author={Lu, Jia and He, Xuehuan and Auricchio, Ferdinando},
  journal={Acta Biomaterialia},
  volume={195},
  pages={134--143},
  year={2025},
  publisher={Elsevier}
}

@article{li2018constitutive,
  title={Constitutive laws with damage effect for the human great saphenous vein},
  author={Li, Wenguang},
  journal={Journal of the Mechanical Behavior of Biomedical Materials},
  volume={81},
  pages={202--213},
  year={2018},
  publisher={Elsevier}
}

@article{li2016invariant,
  title={An invariant-based damage model for human and animal skins},
  author={Li, Wenguang and Luo, Xiaoyu Y},
  journal={Annals of biomedical engineering},
  volume={44},
  number={10},
  pages={3109--3122},
  year={2016},
  publisher={Springer}
}

@article{lemaitre1984use,
  title={How to use damage mechanics},
  author={Lemaitre, Jean},
  journal={Nuclear engineering and design},
  volume={80},
  number={2},
  pages={233--245},
  year={1984},
  publisher={Elsevier}
}

@article{voyiadjis2015investigation,
  title={Investigation of the damage variable basic issues in continuum damage and healing mechanics},
  author={Voyiadjis, George Z and Kattan, Peter I},
  journal={Mechanics Research Communications},
  volume={68},
  pages={89--94},
  year={2015},
  publisher={Elsevier}
}

@article{kachanov1958time,
  title={Time of rupture process under deep conditions},
  author={Kachanov, LM},
  journal={Izv. Akad. Nauk SSSR},
  volume={8},
  pages={26},
  year={1958}
}

@article{rabotnov1980elements,
  title={Elements of hereditary solid mechanics},
  author={Rabotnov, I︠U︡},
  journal={(No Title)},
  year={1980}
}

@article{keller2011thermodynamic,
  title={On the thermodynamic consistency of the equivalence principle in continuum damage mechanics},
  author={Keller, A and Hutter, K},
  journal={Journal of the Mechanics and Physics of Solids},
  volume={59},
  number={5},
  pages={1115--1120},
  year={2011},
  publisher={Elsevier}
}

@article{gurtin1981simple,
  title={Simple rate-independent model for damage},
  author={Gurtin, Morton E and Francis, Eugene C},
  journal={Journal of Spacecraft and Rockets},
  volume={18},
  number={3},
  pages={285--286},
  year={1981}
}

@article{chagnon2004relevance,
  title={On the relevance of continuum damage mechanics as applied to the Mullins effect in elastomers},
  author={Chagnon, Gr{\'e}gory and Verron, Erwan and Gornet, Laurent and Marckmann, Gilles and Charrier, Pierre},
  journal={Journal of the Mechanics and Physics of Solids},
  volume={52},
  number={7},
  pages={1627--1650},
  year={2004},
  publisher={Elsevier}
}

@article{mullins1969softening,
  title={Softening of rubber by deformation},
  author={Mullins, Leonard},
  journal={Rubber chemistry and technology},
  volume={42},
  number={1},
  pages={339--362},
  year={1969}
}

@article{schmidt2014statistical,
  title={Statistical approach for a continuum description of damage evolution in soft collagenous tissues},
  author={Schmidt, Thomas and Balzani, Daniel and Holzapfel, Gerhard A},
  journal={Computer Methods in Applied Mechanics and Engineering},
  volume={278},
  pages={41--61},
  year={2014},
  publisher={Elsevier}
}

@article{gasser2011irreversible,
  title={An irreversible constitutive model for fibrous soft biological tissue: a 3-D microfiber approach with demonstrative application to abdominal aortic aneurysms},
  author={Gasser, T Christian},
  journal={Acta Biomaterialia},
  volume={7},
  number={6},
  pages={2457--2466},
  year={2011},
  publisher={Elsevier}
}

@article{rodriguez2006stochastic,
  title={A stochastic-structurally based three dimensional finite-strain damage model for fibrous soft tissue},
  author={Rodr{\'\i}guez, Jos{\'e} F and Cacho, Fernando and Bea, Jos{\'e} A and Doblar{\'e}, Manuel},
  journal={Journal of the Mechanics and Physics of Solids},
  volume={54},
  number={4},
  pages={864--886},
  year={2006},
  publisher={Elsevier}
}

@article{rodriguez2008finite,
  title={Finite element implementation of a stochastic three dimensional finite-strain damage model for fibrous soft tissue},
  author={Rodriguez, Jose F and Alastrue, Victor and Doblare, Manuel},
  journal={Computer methods in applied mechanics and engineering},
  volume={197},
  number={9-12},
  pages={946--958},
  year={2008},
  publisher={Elsevier}
}

@article{holzapfel2020damage,
  title={A damage model for collagen fibres with an application to collagenous soft tissues},
  author={Holzapfel, Gerhard A and Ogden, Ray W},
  journal={Proceedings of the Royal Society A},
  volume={476},
  number={2236},
  pages={20190821},
  year={2020},
  publisher={The Royal Society Publishing}
}

@article{hamedzadeh2018constitutive,
  title={On the constitutive modelling of recruitment and damage of collagen fibres in soft biological tissues},
  author={Hamedzadeh, Amir and Gasser, T Christian and Federico, Salvatore},
  journal={European Journal of Mechanics-A/Solids},
  volume={72},
  pages={483--496},
  year={2018},
  publisher={Elsevier}
}

@article{sun2025damage,
  title={Damage-induced energy dissipation in artificial soft tissues},
  author={Sun, WK and Yin, BB and Liew, KM},
  journal={Journal of the Mechanics and Physics of Solids},
  volume={194},
  pages={105933},
  year={2025},
  publisher={Elsevier}
}

@article{ji2004mechanical,
  title={Mechanical properties of nanostructure of biological materials},
  author={Ji, Baohua and Gao, Huajian},
  journal={Journal of the Mechanics and Physics of Solids},
  volume={52},
  number={9},
  pages={1963--1990},
  year={2004},
  publisher={Elsevier}
}

@article{martins1998numerical,
  title={A numerical model of passive and active behavior of skeletal muscles},
  author={Martins, JAC and Pires, EB and Salvado, R and Dinis, PB},
  journal={Computer methods in applied mechanics and engineering},
  volume={151},
  number={3-4},
  pages={419--433},
  year={1998},
  publisher={Elsevier}
}

@article{pena2009mullins,
  title={On the Mullins effect and hysteresis of fibered biological materials: A comparison between continuous and discontinuous damage models},
  author={Pe{\~n}a, Estefan{\'\i}a and Pe{\~n}a, Juan A and Doblar{\'e}, Manuel},
  journal={International Journal of Solids and Structures},
  volume={46},
  number={7-8},
  pages={1727--1735},
  year={2009},
  publisher={Elsevier}
}

@article{rausch2017modeling,
  title={Modeling soft tissue damage and failure using a combined particle/continuum approach},
  author={Rausch, MK and Karniadakis, GE and Humphrey, JD},
  journal={Biomechanics and modeling in mechanobiology},
  volume={16},
  number={1},
  pages={249--261},
  year={2017},
  publisher={Springer}
}

@article{bui2021localized,
  title={A localized mass-field damage model with energy decomposition: Formulation and FE implementation},
  author={Bui, Tinh Quoc and Tran, Hung Thanh},
  journal={Computer Methods in Applied Mechanics and Engineering},
  volume={387},
  pages={114134},
  year={2021},
  publisher={Elsevier}
}

@article{pena2014computational,
  title={Computational aspects of the numerical modelling of softening, damage and permanent set in soft biological tissues},
  author={Pe{\~n}a, Estefan{\'\i}a},
  journal={Computers \& Structures},
  volume={130},
  pages={57--72},
  year={2014},
  publisher={Elsevier}
}

@article{harwood1965stress,
  title={Stress softening in natural rubber vulcanizates. Part II. Stress softening effects in pure gum and filler loaded rubbers},
  author={Harwood, JAC and Mullins, Leonard and Payne, Arthur R},
  journal={Journal of Applied Polymer Science},
  volume={9},
  number={9},
  pages={3011--3021},
  year={1965},
  publisher={Wiley Online Library}
}

@article{valanis2022valanis,
  title={The Valanis--Landel strain energy function Elasticity of incompressible and compressible rubber-like materials},
  author={Valanis, Kirk C},
  journal={International Journal of Solids and Structures},
  volume={238},
  pages={111271},
  year={2022},
  publisher={Elsevier}
}

@article{martin2018non,
  title={A non-ellipticity result, or the impossible taming of the logarithmic strain measure},
  author={Martin, Robert J and Ghiba, Ionel-Dumitrel and Neff, Patrizio},
  journal={International Journal of Non-Linear Mechanics},
  volume={102},
  pages={147--158},
  year={2018},
  publisher={Elsevier}
}

@article{criscione2002direct,
  title={Direct tensor expression for natural strain and fast, accurate approximation},
  author={Criscione, John C},
  journal={Computers \& structures},
  volume={80},
  number={25},
  pages={1895--1905},
  year={2002},
  publisher={Elsevier}
}

@article{faye2019effect,
  title={The effect of local inertia around the crack-tip in dynamic fracture of soft materials},
  author={Faye, Anshul and Lev, Yoav and Volokh, KY},
  journal={Mechanics of Soft Materials},
  volume={1},
  number={1},
  pages={4},
  year={2019},
  publisher={Springer}
}

@article{pena2010constitutive,
  title={A constitutive formulation of vascular tissue mechanics including viscoelasticity and softening behaviour},
  author={Pe{\~n}a, E and Alastru{\'e}, V and Laborda, A and Mart{\'\i}nez, MA and Doblar{\'e}, M},
  journal={Journal of biomechanics},
  volume={43},
  number={5},
  pages={984--989},
  year={2010},
  publisher={Elsevier}
}

@article{calvo2009modelling,
  title={On modelling damage process in vaginal tissue},
  author={Calvo, B and Pena, E and Martins, P and Mascarenhas, T and Doblare, M and Jorge, RM Natal and Ferreira, A},
  journal={Journal of Biomechanics},
  volume={42},
  number={5},
  pages={642--651},
  year={2009},
  publisher={Elsevier}
}

@article{pena2011mechanical,
  title={Mechanical characterization of the softening behavior of human vaginal tissue},
  author={Pe{\~n}a, E and Martins, P and Mascarenhas, T and Jorge, RM Natal and Ferreira, A and Doblar{\'e}, M and Calvo, B},
  journal={Journal of the mechanical behavior of biomedical materials},
  volume={4},
  number={3},
  pages={275--283},
  year={2011},
  publisher={Elsevier}
}

@article{munoz2008experimental,
  title={An experimental study of the mouse skin behaviour: damage and inelastic aspects},
  author={Munoz, MJ and Bea, JA and Rodr{\'\i}guez, JF and Ochoa, I and Grasa, J and del Palomar, A P{\'e}rez and Zaragoza, P and Osta, R and Doblar{\'e}, M},
  journal={Journal of biomechanics},
  volume={41},
  number={1},
  pages={93--99},
  year={2008},
  publisher={Elsevier}
}

@article{alastrue2008experimental,
  title={Experimental study and constitutive modelling of the passive mechanical properties of the ovine infrarenal vena cava tissue},
  author={Alastru{\'e}, V and Pe{\~n}a, E and Mart{\'\i}nez, MA and Doblar{\'e}, M},
  journal={Journal of biomechanics},
  volume={41},
  number={14},
  pages={3038--3045},
  year={2008},
  publisher={Elsevier}
}

@article{pena2019failure,
  title={Failure damage mechanical properties of thoracic and abdominal porcine aorta layers and related constitutive modeling: phenomenological and microstructural approach},
  author={Pe{\~n}a, Juan A and Mart{\'\i}nez, Miguel A and Pe{\~n}a, Estefan{\'\i}a},
  journal={Biomechanics and Modeling in Mechanobiology},
  volume={18},
  number={6},
  pages={1709--1730},
  year={2019},
  publisher={Springer}
}

@article{landauer2019experimental,
  title={Experimental characterization and hyperelastic constitutive modeling of open-cell elastomeric foams},
  author={Landauer, Alexander K and Li, Xiuqi and Franck, Christian and Henann, David L},
  journal={Journal of the Mechanics and Physics of Solids},
  volume={133},
  pages={103701},
  year={2019},
  publisher={Elsevier}
}

@article{neff2016geometry,
  title={Geometry of logarithmic strain measures in solid mechanics},
  author={Neff, Patrizio and Eidel, Bernhard and Martin, Robert J},
  journal={Archive for Rational Mechanics and Analysis},
  volume={222},
  number={2},
  pages={507--572},
  year={2016},
  publisher={Springer}
}

@misc{holzapfel2002nonlinear,
  title={Nonlinear solid mechanics: a continuum approach for engineering science},
  author={Holzapfel, Gerhard A},
  year={2002},
  publisher={Kluwer Academic Publishers Dordrecht}
}

@article{li2019multiscale,
  title={Multiscale modeling of fiber recruitment and damage with a discrete fiber dispersion method},
  author={Li, Kewei and Holzapfel, Gerhard A},
  journal={Journal of the Mechanics and Physics of Solids},
  volume={126},
  pages={226--244},
  year={2019},
  publisher={Elsevier}
}

@article{treloar1975physics,
  title={The physics of rubber elasticity},
  author={Treloar, LR G},
  year={1975},
  publisher={OUP Oxford}
}

@article{steck2023multiscale,
  title={Multiscale stress deconcentration amplifies fatigue resistance of rubber},
  author={Steck, Jason and Kim, Junsoo and Kutsovsky, Yakov and Suo, Zhigang},
  journal={Nature},
  volume={624},
  number={7991},
  pages={303--308},
  year={2023},
  publisher={Nature Publishing Group UK London}
}

@article{sun2012highly,
  title={Highly stretchable and tough hydrogels},
  author={Sun, Jeong-Yun and Zhao, Xuanhe and Illeperuma, Widusha RK and Chaudhuri, Ovijit and Oh, Kyu Hwan and Mooney, David J and Vlassak, Joost J and Suo, Zhigang},
  journal={Nature},
  volume={489},
  number={7414},
  pages={133--136},
  year={2012},
  publisher={Nature Publishing Group UK London}
}

@article{treloar1943elasticity,
  title={The elasticity of a network of long-chain molecules. I},
  author={Treloar, Leslie Ronald George},
  journal={Transactions of the Faraday Society},
  volume={39},
  pages={36--41},
  year={1943},
  publisher={Royal Society of Chemistry}
}

@article{treloar1943elasticity2,
  title={The elasticity of a network of long-chain molecules—II},
  author={Treloar, Leslie RG},
  journal={Transactions of the Faraday Society},
  volume={39},
  pages={241--246},
  year={1943},
  publisher={Royal Society of Chemistry}
}

@article{flory1943statistical,
  title={Statistical mechanics of cross-linked polymer networks II. Swelling},
  author={Flory, Paul J and Rehner Jr, John},
  journal={The journal of chemical physics},
  volume={11},
  number={11},
  pages={521--526},
  year={1943},
  publisher={American Institute of Physics}
}

@article{flory1944network,
  title={Network Structure and the Elastic Properties of Vulcanized Rubber.},
  author={Flory, Paul J},
  journal={Chemical reviews},
  volume={35},
  number={1},
  pages={51--75},
  year={1944},
  publisher={ACS Publications}
}

@article{wang20243d,
  title={3D observations provide striking findings in rubber elasticity},
  author={Wang, Zifan and Das, Shuvrangsu and Joshi, Akshay and Shaikeea, Angkur JD and Deshpande, Vikram S},
  journal={Proceedings of the National Academy of Sciences},
  volume={121},
  number={24},
  pages={e2404205121},
  year={2024},
  publisher={National Academy of Sciences}
}

@article{stewart2016wrinkling,
  title={Wrinkling, creasing, and folding in fiber-reinforced soft tissues},
  author={Stewart, Peter S and Waters, Sarah L and El Sayed, Tamer and Vella, Dominic and Goriely, Alain},
  journal={Extreme Mechanics Letters},
  volume={8},
  pages={22--29},
  year={2016},
  publisher={Elsevier}
}

@article{dal2021performance,
  title={On the performance of isotropic hyperelastic constitutive models for rubber-like materials: a state of the art review},
  author={Dal, H{\"u}sn{\"u} and A{\c{c}}{\i}kg{\"o}z, Kemal and Badienia, Yashar},
  journal={Applied Mechanics Reviews},
  volume={73},
  number={2},
  pages={020802},
  year={2021},
  publisher={American Society of Mechanical Engineers}
}

@article{guan2025hyperelastic,
  title={Hyperelastic modeling based on generalized Landau invariants and multi-stage calibration},
  author={Guan, Jiashen and Li, Xin and Yuan, Hongyan and Liu, Ju},
  journal={Journal of the Mechanics and Physics of Solids},
  pages={106338},
  year={2025},
  publisher={Elsevier}
}

@article{xiang2018general,
  title={A general constitutive model of soft elastomers},
  author={Xiang, Yuhai and Zhong, Danming and Wang, Peng and Mao, Guoyong and Yu, Honghui and Qu, Shaoxing},
  journal={Journal of the Mechanics and Physics of Solids},
  volume={117},
  pages={110--122},
  year={2018},
  publisher={Elsevier}
}

@article{yang2022mechanical,
  title={Mechanical characterization of agarose hydrogels and their inherent dynamic instabilities at ballistic to ultra-high strain-rates via inertial microcavitation},
  author={Yang, Jin and Cramer III, Harry C and Bremer, Elizabeth C and Buyukozturk, Selda and Yin, Yue and Franck, Christian},
  journal={Extreme Mechanics Letters},
  volume={51},
  pages={101572},
  year={2022},
  publisher={Elsevier}
}

@article{mythravaruni2018failure,
  title={Failure of rubber bearings under combined shear and compression},
  author={Mythravaruni, P and Volokh, KY},
  journal={Journal of Applied Mechanics},
  volume={85},
  number={7},
  year={2018},
  publisher={American Society of Mechanical Engineers Digital Collection}
}

@article{zhang2002numerical,
  title={Numerical simulation of cohesive fracture by the virtual-internal-bond model},
  author={Zhang, P and Klein, P and Huang, Y and Gao, H and Wu, PD},
  journal={Computer Modeling in Engineering and Sciences},
  volume={3},
  number={2},
  pages={263--277},
  year={2002},
  publisher={Tech Science Press}
}

@article{thiagarajan2004finite,
  title={Finite element implementation of virtual internal bond model for simulating crack behavior},
  author={Thiagarajan, Ganesh and Hsia, K Jimmy and Huang, Yonggang},
  journal={Engineering Fracture Mechanics},
  volume={71},
  number={3},
  pages={401--423},
  year={2004},
  publisher={Elsevier}
}

@article{klein1998crack,
  title={Crack nucleation and growth as strain localization in a virtual-bond continuum},
  author={Klein, P and Gao, H},
  journal={Engineering fracture mechanics},
  volume={61},
  number={1},
  pages={21--48},
  year={1998},
  publisher={Elsevier}
}

@book{park2009potential,
  title={Potential-based fracture mechanics using cohesive zone and virtual internal bond modeling},
  author={Park, Kyoungsoo},
  year={2009},
  publisher={University of Illinois at Urbana-Champaign}
}

@inproceedings{born1940stability,
  title={On the stability of crystal lattices. I},
  author={Born, Max},
  booktitle={Mathematical Proceedings of the Cambridge Philosophical Society},
  volume={36},
  pages={160--172},
  year={1940},
  organization={Cambridge University Press}
}

@article{klein2001physics,
  title={Physics-based modeling of brittle fracture: cohesive formulations and the application of meshfree methods},
  author={Klein, PA and Foulk, JW and Chen, EP and Wimmer, SA and Gao, HJ},
  journal={Theoretical and Applied Fracture Mechanics},
  volume={37},
  number={1-3},
  pages={99--166},
  year={2001},
  publisher={Elsevier}
}

@article{park2009unified,
  title={A unified potential-based cohesive model of mixed-mode fracture},
  author={Park, Kyoungsoo and Paulino, Glaucio H and Roesler, Jeffery R},
  journal={Journal of the Mechanics and Physics of Solids},
  volume={57},
  number={6},
  pages={891--908},
  year={2009},
  publisher={Elsevier}
}

@article{gao2003modeling,
  title={Modeling fracture in nanomaterials via a virtual internal bond method},
  author={Gao, Huajian and Ji, Baohua},
  journal={Engineering Fracture Mechanics},
  volume={70},
  number={14},
  pages={1777--1791},
  year={2003},
  publisher={Elsevier}
}

@article{gao1997elastic,
  title={Elastic waves in a hyperelastic solid near its plane-strain equibiaxial cohesive limit},
  author={Gao, Huajian},
  journal={Philosophical magazine letters},
  volume={76},
  number={5},
  pages={307--314},
  year={1997},
  publisher={Taylor \& Francis}
}

@article{volokh2006approach,
  title={An approach to multi-body interactions in a continuum-atomistic context: Application to analysis of tension instability in carbon nanotubes},
  author={Volokh, KY and Ramesh, KT},
  journal={International Journal of Solids and Structures},
  volume={43},
  number={25-26},
  pages={7609--7627},
  year={2006},
  publisher={Elsevier}
}

@article{ramzi1998structure,
  title={Structure- properties relation for agarose thermoreversible gels in binary solvents},
  author={Ramzi, Mohamed and Rochas, Cyrille and Guenet, Jean-Michel},
  journal={Macromolecules},
  volume={31},
  number={18},
  pages={6106--6111},
  year={1998},
  publisher={ACS Publications}
}

@inproceedings{guenet2006agarose,
  title={Agarose sols and gels revisited},
  author={GUENET, Jean-Michel and ROCHAS, Cyrille},
  booktitle={Macromolecular symposia},
  pages={65--70},
  year={2006}
}

@book{ogden1997non,
  title={Non-linear elastic deformations},
  author={Ogden, Raymond W},
  year={1997},
  publisher={Courier Corporation}
}

@article{ball1976convexity,
  title={Convexity conditions and existence theorems in nonlinear elasticity},
  author={Ball, John M},
  journal={Archive for rational mechanics and Analysis},
  volume={63},
  number={4},
  pages={337--403},
  year={1976},
  publisher={Springer}
}

@article{jayawardena2023evaluation,
  title={Evaluation of techniques used for visualisation of hydrogel morphology and determination of pore size distributions},
  author={Jayawardena, Imanda and Turunen, Petri and Garms, Bruna Cambraia and Rowan, Alan and Corrie, Simon and Gr{\o}ndahl, Lisbeth},
  journal={Materials Advances},
  volume={4},
  number={2},
  pages={669--682},
  year={2023},
  publisher={Royal Society of Chemistry}
}

@article{linka2018fatigue,
  title={Fatigue of soft fibrous tissues: multi-scale mechanics and constitutive modeling},
  author={Linka, Kevin and Hillg{\"a}rtner, Markus and Itskov, Mikhail},
  journal={Acta biomaterialia},
  volume={71},
  pages={398--410},
  year={2018},
  publisher={Elsevier}
}

@article{sun2010review,
  title={A review on the research of mechanical problems with different moduli in tension and compression},
  author={Sun, Jun-yi and Zhu, Hai-qiao and Qin, Shi-hong and Yang, Da-lin and He, Xiao-ting},
  journal={Journal of mechanical science and technology},
  volume={24},
  number={9},
  pages={1845--1854},
  year={2010},
  publisher={Springer}
}

@article{ghebremedhin2021physics,
  title={Physics of agarose fluid gels: Rheological properties and microstructure},
  author={Ghebremedhin, Marta and Seiffert, Sebastian and Vilgis, Thomas A},
  journal={Current Research in Food Science},
  volume={4},
  pages={436--448},
  year={2021},
  publisher={Elsevier}
}

@article{dormoy1991transient,
  title={Transient electric birefringence study of highly dilute agarose solutions},
  author={Dormoy, Y and Candau, S},
  journal={Biopolymers: Original Research on Biomolecules},
  volume={31},
  number={1},
  pages={109--117},
  year={1991},
  publisher={Wiley Online Library}
}

@article{fujii2000scaling,
  title={Scaling analysis on elasticity of agarose gel near the sol--gel transition temperature},
  author={Fujii, T and Yano, T and Kumagai, H and Miyawaki, O},
  journal={Food hydrocolloids},
  volume={14},
  number={4},
  pages={359--363},
  year={2000},
  publisher={Elsevier}
}

@article{dusek2014constrained,
  title={Constrained swelling of polymer networks: characterization of vapor-deposited cross-linked polymer thin films},
  author={Du{\v{s}}ek, Karel and Choukourov, Andrei and Du{\v{s}}kov{\'a}-Smr{\v{c}}kov{\'a}, Miroslava and Biederman, Hynek},
  journal={Macromolecules},
  volume={47},
  number={13},
  pages={4417--4427},
  year={2014},
  publisher={ACS Publications}
}

@article{bertula2019strain,
  title={Strain-stiffening of agarose gels},
  author={Bertula, Kia and Martikainen, Lahja and Munne, Pauliina and Hietala, Sami and Klefstr{\"o}m, Juha and Ikkala, Olli and Nonappa},
  journal={ACS Macro Letters},
  volume={8},
  number={6},
  pages={670--675},
  year={2019},
  publisher={ACS Publications}
}

@article{barrangou2006textural,
  title={Textural properties of agarose gels. I. Rheological and fracture properties},
  author={Barrangou, Lisa M and Daubert, Christopher R and Foegeding, E Allen},
  journal={Food hydrocolloids},
  volume={20},
  number={2-3},
  pages={184--195},
  year={2006},
  publisher={Elsevier}
}

@article{zhu2019visco,
  title={A visco-hyperelastic model of brain tissue incorporating both tension/compression asymmetry and volume compressibility},
  author={Zhu, Zhongmeng and Jiang, Chengkai and Jiang, Han},
  journal={Acta Mechanica},
  volume={230},
  number={6},
  pages={2125--2135},
  year={2019},
  publisher={Springer}
}

@article{baker1954inequalities,
  title={Inequalities restricting the form of the stress-deformation relations for isotropic elastic solids and Reiner-Rivlin fluids},
  author={Baker, M and Ericksen, JL63235},
  journal={Journal of the Washington Academy of Sciences},
  volume={44},
  number={2},
  pages={33--35},
  year={1954},
  publisher={JSTOR}
}

@article{neff2015exponentiated,
  title={The exponentiated Hencky-logarithmic strain energy. Part I: Constitutive issues and rank-one convexity},
  author={Neff, Patrizio and Ghiba, Ionel-Dumitrel and Lankeit, Johannes},
  journal={Journal of Elasticity},
  volume={121},
  number={2},
  pages={143--234},
  year={2015},
  publisher={Springer}
}

@article{flory1961thermodynamic,
  title={Thermodynamic relations for high elastic materials},
  author={Flory, PJ128117},
  journal={Transactions of the Faraday Society},
  volume={57},
  pages={829--838},
  year={1961},
  publisher={Royal Society of Chemistry}
}

@article{anssari2024generalisation,
  title={A generalisation of the Pucci--Saccomandi model of rubber elasticity},
  author={Anssari-Benam, Afshin},
  journal={International Journal of Non-Linear Mechanics},
  volume={158},
  pages={104578},
  year={2024},
  publisher={Elsevier}
}

@article{diani2005combining,
  title={Combining the logarithmic strain and the full-network model for a better understanding of the hyperelastic behavior of rubber-like materials},
  author={Diani, Julie and Gilormini, Pierre},
  journal={Journal of the Mechanics and Physics of Solids},
  volume={53},
  number={11},
  pages={2579--2596},
  year={2005},
  publisher={Elsevier}
}

@article{freed1995natural,
  title={Natural strain},
  author={Freed, Alan D},
  journal={Journal of Engineering Materials and Technology},
  volume={117},
  number={4},
  pages={379--385},
  year={1995},
  publisher={American Society of Mechanical Engineers Digital Collection}
}

@article{hencky1931law,
  title={The law of elasticity for isotropic and quasi-isotropic substances by finite deformations},
  author={Hencky, H},
  journal={Journal of Rheology},
  volume={2},
  number={2},
  pages={169--176},
  year={1931},
  publisher={The Society of Rheology}
}

@article{subhash2011concentration,
  title={Concentration dependence of tensile behavior in agarose gel using digital image correlation},
  author={Subhash, G and Liu, Q and Moore, DF and Ifju, PG and Haile, MA},
  journal={Experimental Mechanics},
  volume={51},
  number={2},
  pages={255--262},
  year={2011},
  publisher={Springer}
}

@article{millar2021improved,
  title={An improved direct shear characterisation technique for soft gelatinous and elastomeric materials},
  author={Millar, David and Mennu, Matlock and Upadhyay, Kshitiz and Morley, Cameron and Ifju, Peter},
  journal={Strain},
  volume={57},
  number={3},
  pages={e12383},
  year={2021},
  publisher={Wiley Online Library}
}

@article{cantrell2017experimental,
  title={Experimental characterization of the mechanical properties of 3D-printed ABS and polycarbonate parts},
  author={Cantrell, Jason T and Rohde, Sean and Damiani, David and Gurnani, Rishi and DiSandro, Luke and Anton, Josh and Young, Andie and Jerez, Alex and Steinbach, Douglas and Kroese, Calvin and others},
  journal={Rapid Prototyping Journal},
  volume={23},
  number={4},
  pages={811--824},
  year={2017},
  publisher={Emerald Publishing Limited}
}

@article{prasad2023new,
  title={A new viscoelastic model for human brain tissue using Lode invariants based rate-type thermodynamic framework},
  author={Prasad, Durga and Sreejith, P and Kannan, K},
  journal={Applications in Engineering Science},
  volume={15},
  pages={100130},
  year={2023},
  publisher={Elsevier}
}

@article{reddipaga2026construction,
  title={On the construction of a viscoelastic constitutive model for brain tissue maximizing tension--compression asymmetry},
  author={Reddipaga, Mani and Kannan, K},
  journal={International Journal of Engineering Science},
  volume={218},
  pages={104415},
  year={2026},
  publisher={Elsevier}
}

@article{sendova2005strong,
  title={On strong ellipticity for isotropic hyperelastic materials based upon logarithmic strain},
  author={Sendova, Tsvetanka and Walton, Jay R},
  journal={International Journal of Non-Linear Mechanics},
  volume={40},
  number={2-3},
  pages={195--212},
  year={2005},
  publisher={Elsevier}
}

@article{hill1970constitutive,
  title={Constitutive inequalities for isotropic elastic solids under finite strain},
  author={Hill, Robert},
  journal={Proceedings of the Royal Society of London. A. Mathematical and Physical Sciences},
  volume={314},
  number={1519},
  pages={457--472},
  year={1970},
  publisher={The Royal Society London}
}

@article{bohringer2025compression,
  title={Compression-Tension-Asymmetry and Stiffness Nonlinearity of Collagen-Matrigel Composite Hydrogels},
  author={B{\"o}hringer, David and Hinrichsen, Jan and Gataulin, Radik and Wiedenmann, Sandra and Sp{\"o}rrer, Marina and Sherifova, Selda and Steinmann, Paul and Holzapfel, Gerhard A and Fabry, Ben and Budday, Silvia},
  journal={Advanced Healthcare Materials},
  pages={e03052},
  year={2025},
  publisher={Wiley Online Library}
}

@article{ed2021poroviscoelasticity,
  title={Poroviscoelasticity and compression-softening of agarose hydrogels},
  author={Ed-Daoui, Abderrahim and Snabre, Patrick},
  journal={Rheologica Acta},
  volume={60},
  number={6},
  pages={327--351},
  year={2021},
  publisher={Springer}
}

@article{doi2009gel,
  title={Gel dynamics},
  author={Doi, Masao},
  journal={Journal of the Physical Society of Japan},
  volume={78},
  number={5},
  pages={052001},
  year={2009},
  publisher={The Physical Society of Japan}
}

@article{hong2016localized,
  title={Localized tissue surrogate deformation due to controlled single bubble cavitation},
  author={Hong, Yu and Sarntinoranont, Malisa and Subhash, G and Canchi, S and King, MA},
  journal={Experimental Mechanics},
  volume={56},
  number={1},
  pages={97--109},
  year={2016},
  publisher={Springer}
}

@inproceedings{pervin2011mechanically,
  title={Mechanically similar gel simulants for brain tissues},
  author={Pervin, Farhana and Chen, Weinong W},
  booktitle={Dynamic Behavior of Materials, Volume 1: Proceedings of the 2010 Annual Conference on Experimental and Applied Mechanics},
  pages={9--13},
  year={2011},
  organization={Springer}
}

@article{aghayan2022experimental,
  title={Experimental and numerical investigation of dynamic cavitation in agarose gel as a soft tissue simulant},
  author={Aghayan, Sam and Weinberg, Kerstin},
  journal={Mechanics of Materials},
  volume={175},
  pages={104486},
  year={2022},
  publisher={Elsevier}
}

@article{yang2020inelasticity,
  title={Inelasticity increases the critical strain for the onset of creases on hydrogels},
  author={Yang, Jiawei and Illeperuma, Widusha and Suo, Zhigang},
  journal={Extreme Mechanics Letters},
  volume={40},
  pages={100966},
  year={2020},
  publisher={Elsevier}
}

@book{marsden1994mathematical,
  title={Mathematical foundations of elasticity},
  author={Marsden, Jerrold E and Hughes, Thomas JR},
  year={1994},
  publisher={Courier Corporation}
}

@article{zubov2011criterion,
  title={A criterion for the strong ellipticity of the equilibrium equations of an isotropic non-linearly elastic material},
  author={Zubov, LM and Rudev, AN},
  journal={Journal of applied mathematics and mechanics},
  volume={75},
  number={4},
  pages={432--446},
  year={2011},
  publisher={Elsevier}
}

@article{sawyers1978speed,
  title={On the speed of propagation of waves in a deformed compressible elastic material},
  author={Sawyers, KN and Rivlin, RS},
  journal={Zeitschrift f{\"u}r angewandte Mathematik und Physik ZAMP},
  volume={29},
  number={2},
  pages={245--251},
  year={1978},
  publisher={Springer}
}

@article{knowles1978failure,
  title={On the failure of ellipticity and the emergence of discontinuous deformation gradients in plane finite elastostatics},
  author={Knowles, James K and Sternberg, Eli},
  journal={Journal of Elasticity},
  volume={8},
  number={4},
  pages={329--379},
  year={1978},
  publisher={Springer}
}

@article{ghiba2015ellipticity,
  title={An ellipticity domain for the distortional Hencky logarithmic strain energy},
  author={Ghiba, Ionel-Dumitrel and Neff, Patrizio and Martin, Robert J},
  journal={Proceedings of the Royal Society A: Mathematical, Physical and Engineering Sciences},
  volume={471},
  number={2184},
  year={2015},
  publisher={The Royal Society}
}

@article{chen1991strong,
  title={On strong ellipticity and the Legendre-Hadamard condition},
  author={Chen, Yi-Chao},
  journal={Archive for rational mechanics and analysis},
  volume={113},
  number={2},
  pages={165--175},
  year={1991},
  publisher={Springer}
}

@article{wollner2025search,
  title={In search of constitutive conditions in isotropic hyperelasticity: polyconvexity versus true-stress-true-strain monotonicity},
  author={Wollner, Maximilian P and Holzapfel, Gerhard A and Neff, Patrizio},
  journal={Journal of the Mechanics and Physics of Solids},
  pages={106465},
  year={2025},
  publisher={Elsevier}
}

@article{triantafyllidis1986gradient,
  title={A gradient approach to localization of deformation. I. Hyperelastic materials},
  author={Triantafyllidis, Nicolas and Aifantis, Elias C},
  journal={Journal of Elasticity},
  volume={16},
  number={3},
  pages={225--237},
  year={1986},
  publisher={Springer}
}

@book{bigoni2012nonlinear,
  title={Nonlinear solid mechanics: bifurcation theory and material instability},
  author={Bigoni, Davide},
  year={2012},
  publisher={Cambridge University Press}
}

@article{yan2025virtual,
  title={A Virtual Fields Method-Genetic Algorithm (VFM-GA) calibration framework for isotropic hyperelastic constitutive models with application to an elastomeric foam material},
  author={Yan, Zicheng and Tao, Jialiang and Franck, Christian and Henann, David L},
  journal={arXiv preprint arXiv:2510.07683},
  year={2025}
}

@article{dacorogna2001necessary,
  title={Necessary and sufficient conditions for strong ellipticity of isotropic functions in any dimension},
  author={Dacorogna, Bernard},
  journal={Discrete and Continuous Dynamical Systems-B},
  volume={1},
  number={2},
  pages={257--263},
  year={2001},
  publisher={Discrete and Continuous Dynamical Systems-B}
}

@article{rosendahl2019equivalent,
  title={Equivalent strain failure criterion for multiaxially loaded incompressible hyperelastic elastomers},
  author={Rosendahl, PL and Drass, M and Felger, J and Schneider, J and Becker, W},
  journal={International Journal of Solids and Structures},
  volume={166},
  pages={32--46},
  year={2019},
  publisher={Elsevier}
}

@article{abeyaratne1980discontinuous,
  title={Discontinuous deformation gradients in plane finite elastostatics of incompressible materials},
  author={Abeyaratne, Rohan C},
  journal={Journal of Elasticity},
  volume={10},
  number={3},
  pages={255--293},
  year={1980},
  publisher={Springer}
}

@article{de2022nucleation,
  title={Nucleation under multi-axial loading in variational phase-field models of brittle fracture},
  author={De Lorenzis, Laura and Maurini, Corrado},
  journal={International Journal of Fracture},
  volume={237},
  number={1},
  pages={61--81},
  year={2022},
  publisher={Springer}
}

@article{lopez2025classical,
  title={Classical variational phase-field models cannot predict fracture nucleation},
  author={Lopez-Pamies, Oscar and Dolbow, John E and Francfort, Gilles A and Larsen, Christopher J},
  journal={Computer Methods in Applied Mechanics and Engineering},
  volume={433},
  pages={117520},
  year={2025},
  publisher={Elsevier}
}

@article{avril2026state,
  title={State-of-the-art and tomorrow’s challenges and opportunities in constitutive modeling of soft biological tissues with a focus on arterial, cardiac and brain biomechanics},
  author={Avril, St{\'e}phane and Goriely, Alain and Holzapfel, Gerhard A and Kuhl, Ellen and Nordsletten, David and Ogden, Ray W},
  journal={Acta Biomaterialia},
  year={2026},
  publisher={Elsevier}
}

@article{yin2024rate,
  title={Rate-dependent fracture behaviors of agar-based hybrid double-network hydrogels},
  author={Yin, Haiyan and You, Min and Yu, Hui and Si, Xinlei and Zheng, Yong and Cui, Wei and Zhu, Lin and Chen, Qiang},
  journal={Macromolecules},
  volume={57},
  number={9},
  pages={4024--4033},
  year={2024},
  publisher={ACS Publications}
}

@article{kwon2010compressive,
  title={Compressive strain rate sensitivity of ballistic gelatin},
  author={Kwon, Jiwoon and Subhash, Ghatu},
  journal={Journal of biomechanics},
  volume={43},
  number={3},
  pages={420--425},
  year={2010},
  publisher={Elsevier}
}

@article{falope2025experiments,
  title={Experiments on the finite torsion of nearly incompressible rubber-like materials: Nonlinear effects, analytic modeling and rubber characterization},
  author={Falope, Federico Oyedeji and Lanzoni, Luca and Tarantino, Angelo Marcello},
  journal={International Journal of Engineering Science},
  volume={211},
  pages={104254},
  year={2025},
  publisher={Elsevier}
}

@article{gao2010constitutive,
  title={Constitutive modeling of liver tissue: experiment and theory},
  author={Gao, Zhan and Lister, Kevin and Desai, Jaydev P},
  journal={Annals of biomedical engineering},
  volume={38},
  number={2},
  pages={505--516},
  year={2010},
  publisher={Springer}
}

@article{budday2020fifty,
  title={Fifty Shades of Brain: A Review on the Mechanical Testing and Modeling of Brain Tissue: S. Budday et al.},
  author={Budday, Silvia and Ovaert, Timothy C and Holzapfel, Gerhard A and Steinmann, Paul and Kuhl, Ellen},
  journal={Archives of Computational Methods in Engineering},
  volume={27},
  number={4},
  pages={1187--1230},
  year={2020},
  publisher={Springer}
}

@incollection{volokh2020new,
  title={New approaches to modeling failure and fracture of rubberlike materials},
  author={Volokh, KY},
  booktitle={Fatigue Crack Growth in Rubber Materials: Experiments and Modelling},
  pages={131--151},
  year={2020},
  publisher={Springer}
}

@book{volokh2024modeling,
  title={Modeling failure and fracture of soft solids and fluids},
  author={Volokh, Konstantin},
  year={2024},
  publisher={Springer}
}

@article{balakhovsky2012inflation,
  title={Inflation and rupture of rubber membrane},
  author={Balakhovsky, K and Volokh, KY},
  journal={International journal of fracture},
  volume={177},
  number={2},
  pages={179--190},
  year={2012},
  publisher={Springer}
}

@article{xiao2020granular,
  title={Granular hyperelasticity with inherent and stress-induced anisotropy},
  author={Xiao, Yang and Zhang, Zhichao and Wang, Jingkai},
  journal={Acta Geotechnica},
  volume={15},
  number={3},
  pages={671--680},
  year={2020},
  publisher={Springer}
}

@book{hadamard1903leccons,
  title={Le{\c{c}}ons sur la propagation des ondes et les {\'e}quations de l'hydrodynamique},
  author={Hadamard, Jacques},
  year={1903},
  publisher={A. Hermann}
}

@article{hill1962acceleration,
  title={Acceleration waves in solids},
  author={Hill, Rodney},
  journal={Journal of the Mechanics and Physics of Solids},
  volume={10},
  number={1},
  pages={1--16},
  year={1962},
  publisher={Elsevier}
}

@article{d2016localization,
  title={Localization of deformation and loss of macroscopic ellipticity in microstructured solids},
  author={d'Avila, MP Santisi and Triantafyllidis, Nicolas and Wen, Guangyang},
  journal={Journal of the Mechanics and Physics of Solids},
  volume={97},
  pages={275--298},
  year={2016},
  publisher={Elsevier}
}





\end{document}


\maketitle

\begin{center}
\large \textbf{Supplementary Material}
\end{center}
\vspace{1em}

\section*{S.1\quad Influence of softening sharpness parameters on the stress–reduction behavior}
\setcounter{figure}{0}
\renewcommand{\thefigure}{S.\arabic{figure}}

This section provides supplementary context for the evolution of the stress–reduction factor $\mathrm{Red}(\lambda)$ with softening sharpness parameters $m^{\pm})$, discussed in Section~6.1 of the main manuscript. Both the failure parameters $(\Phi^{\pm}, m^{\pm})$ govern the decay of the intact hyperelastic response via the mode-specific reduction factor prior to rupture. To illustrate the qualitative role of $m^{\pm}$, we examine the stress--reduction factor $\mathrm{Red}(\lambda)
    = \exp\!\left[-\left(\frac{W(\lambda)}{\Phi^{\pm}}\right)^{m^{\pm}}\right]$, for representative tensile $(m^{+})$ and compressive $(m^{-})$ softening sharpness parameters while holding the corresponding pseudo–failure energies $(\Phi^{\pm})$ fixed.

For this illustration, we consider the mode-specific pseudo–failure energies $(\Phi^{\pm})$ from the calibrated values of a representative 3\%~w/v agarose specimen using the proposed asymmetric failure model reported in the main manuscript: $\Phi^{+}=3.97~\text{kPa}$ and $\Phi^{-}=14.49~\text{kPa}$. With these values fixed, the softening sharpness parameters are varied as follows: $m^{+} \in \{2,\;5,\;50\}$ and $m^{-} \in \{0.2,\;0.4,\;0.8\}$.

Figure~S.1 shows the resulting stress–reduction trajectories $\mathrm{Red}(\lambda)$ in tension 
($\lambda>1$) and compression ($\lambda<1$).  
These curves illustrate how the softening sharpness parameters $m^{\pm}$ govern the steepness of 
the reduction profile as the stored energy $W(\lambda)$ approaches the corresponding 
mode-specific pseudo–failure thresholds $\Phi^{\pm}$.  
For fixed $\Phi^{\pm}$, variations in $m^{\pm}$ determine whether the onset and progression of 
softening occur gradually or abruptly, thereby highlighting the distinct tensile and compressive 
softening behaviors encoded within the proposed asymmetric–failure formulation.

\begin{figure}[t]
    \centering
    \includegraphics[width=3.555in]{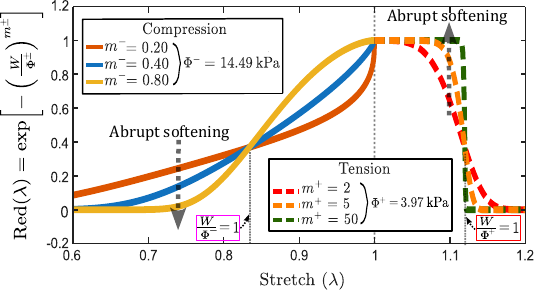}
    \caption{Representative stress--reduction curves in tension and compression for selected values
    of the softening sharpness parameters $m^{+}\in\{2,5,50\}$ and $m^{-}\in\{0.2,0.4,0.8\}$, using 
    fixed mode-specific pseudo–failure energies $\Phi^{+}=3.97$~kPa and $\Phi^{-}=14.49$~kPa corresponding to 
    the calibrated parameters of a 3\%~w/v agarose specimen using the proposed asymmetric failure model.  
    These curves provide a conceptual illustration of the behavior of the stress--reduction factor $\mathrm{Red}(\lambda)$ discussed in Section~6.1 of the main manuscript.}
    \label{fig:sm1}
\end{figure}

The influence of the compressive softening parameter \( m^- \), when \( m^- < 1\), is more nuanced, since the $\mathrm{Red}(\lambda)$ depends significantly on both the normalized stored energy \( W(\lambda)/\Phi^- \) and \( m^- \).  In the small-stretch regime, where \( W(\lambda)/\Phi^- < 1 \), a smaller \( m^- \) yields a larger value of \( \bigl(W(\lambda)/\Phi^-\bigr)^{m^{-}} \), causing \( \mathrm{Red}(\lambda) \) to deviate from unity earlier in the loading history. Consequently, softening activates at relatively smaller compressive stretches.

As compression progresses and \( W(\lambda)/\Phi^- \) exceeds unity (see the figure), this ordering reverses: a larger \( m^- \) amplifies the term \( \bigl(W(\lambda)/\Phi^-\bigr)^{m^{-}} \), leading to a steeper decline in the $\mathrm{Red}(\lambda)$ as damage propagates.  
This crossover behavior, evident in Fig.~S.1 near \( \lambda \approx 0.83 \) (where \( W(\lambda)/\Phi^- = 1 \)), explains why the curve with the smallest \( m^- \) departs earliest from unity, while the curve with the largest \( m^- \) exhibits the most abrupt decay at larger compressive strains.

In contrast, in the tensile branch, where $m^{+}$ $>>$1, the abruptness of failure exhibits a monotonic dependence on $m^{+}$ regardless of the applied stretch value: 
increasing $m^{+}$ keeps $\mathrm{Red}(\lambda)$ close to unity over most of 
the deformation path and then produces an increasingly abrupt drop to zero as it approaches the 
tensile failure threshold.  
Together, these behaviors mirror the asymmetric softening patterns discussed 
in Section~6.1 of the main manuscript and highlight the distinct mechanisms 
governing tensile and compressive failure in agarose gels.


\section*{S.2\quad Comparative analysis of tension–compression asymmetry in the elastic regime}

\begin{figure}[htbp]
    \centering
    \includegraphics[width=3.478in, height=1.80in]{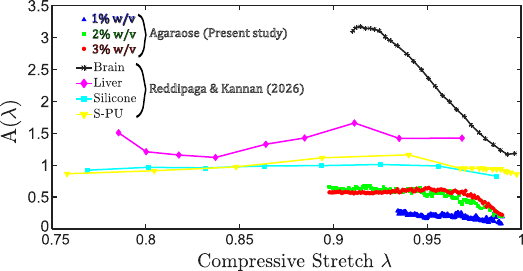}
    \caption{Mooney's asymmetry measure, \(A(\lambda)\), comparing the experimental 1\%, 2\%, and 3\%~w/v agarose hydrogel data against literature values for human brain tissue, liver, silicone rubber, and soft polyurethane (S-PU) reported by Reddipaga and Kannan~[1]. Agarose hydrogels display mild tension-stiffening asymmetry at small deformations, in contrast to the pronounced compression-stiffening response of biological tissues such as brain tissue and liver.}
    \label{fig:sm2}
\end{figure}

To better contextualize the mechanical response of agarose hydrogels within the broader spectrum of soft materials, we quantify their tension–compression asymmetry in the small-strain regime. Following the Mooney-type asymmetry analysis used by Reddipaga and Kannan~[1] to examine deviations from the idealization of symmetric response postulated by Mooney~[2], we compute Mooney's asymmetry measure $A(\lambda)$ as 

\begin{equation}
    A(\lambda) =
    \left|
    \frac{T_{\mathrm{comp}}(\lambda)}
    {T_{\mathrm{tension}}(1/\lambda)}
    \right|,
    \qquad \lambda \leq 1,
    \label{eq:mooney_asymmetry}
\end{equation}
where $T_{\mathrm{tension}}$ is the tensile Cauchy stress considered as the reference, and $T_{\mathrm{comp}}$ is the corresponding compressive Cauchy stress evaluated at the reciprocal stretch. Deviations of \(A(\lambda)\) from unity indicate departures from tension–compression symmetry. This facilitates comparison of the intact tension-compression asymmetric responses of the tested agarose hydrogels relative to previously reported biological tissues and synthetic elastomers within the broader class of soft materials.

Figure~S.2 shows the experimentally evaluated $A(\lambda)$ for representative 1\%, 2\%, and 3\%~w/v agarose hydrogel samples, together with literature data reported by Reddipaga and Kannan~[1] for biological tissues, such as human brain tissue and liver, and synthetic elastomers, such as silicone rubber and soft polyurethane (S-PU). In the ideal tension–compression symmetric case, the compressive stress at a given stretch has the same magnitude as the tensile stress at the reciprocal stretch, giving $A(\lambda) = 1$. Accordingly, $A(\lambda)>1$ indicates compression-stiffening, whereas $A(\lambda)<1$ indicates tension-stiffening.

Within the small-deformation regime for which tensile data are available, the 1\%~w/v agarose specimen exhibits a clear tension-stiffening tendency. The 2\% and 3\%~w/v specimens also show tension-stiffening, although the effect is less pronounced. In contrast, the examined biological tissues, such as brain tissue and liver, display strong compression-stiffening behavior, while synthetic elastomers like silicone rubber and soft polyurethane (S-PU) largely maintain tension–compression symmetry.

This comparison also clarifies the role of the Lode angle-dependent modulus function \(\mathcal{G}(K_3)\) within the selected Prasad--Kannan intact potential \(W_{\mathrm{PK}}\). The function \(\mathcal{G}(K_3)\) was originally formulated to represent compression-stiffening materials such as soft biological tissues. In contrast, the present Mooney-type analysis shows that agarose hydrogels exhibit tension-stiffening within the small-deformation range. Therefore, in the present work, \(\mathcal{G}(K_3)\) should be interpreted as a phenomenological, mode-dependent approximation of the elastic distortional response of agarose hydrogels over this accessible small-strain regime.

It is important to note that the Mooney-type comparison is limited to the deformation range for which tensile data are available. At larger finite deformations under compression, the idealized intact response is expected to transition toward compression-stiffening. In this regime, the chosen function \(\mathcal{G}(K_3)\) provides an appropriate mode-dependent representation of the emergent compressive response. Thus, although agarose hydrogels show tension-stiffening in the measured small-strain regime, the inclusion of \(\mathcal{G}(K_3)\) retains the flexibility needed to represent possible compression-stiffening behavior at larger deformations.



\section*{S.3\quad Theoretical failure-envelope construction using the Hessian-based criterion}

\begin{figure}[htbp]
    \centering
    \includegraphics[width=3.776in, height=3.272in]{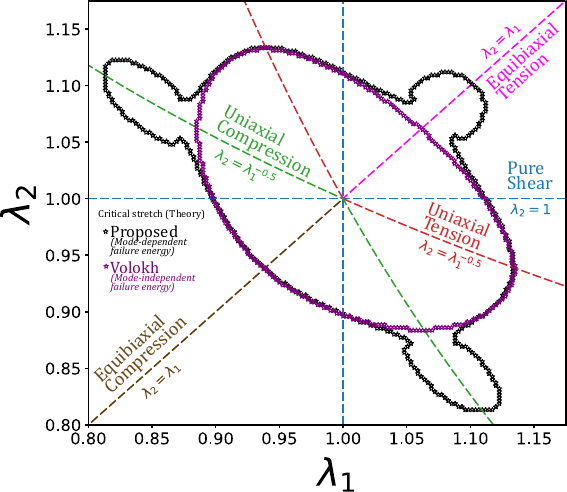}
    \caption{Theoretical failure envelopes in the principal-stretch space computed using the vanishing Hessian determinant criterion. Results are shown for the proposed asymmetric-failure model (black) and Volokh's classical mode-independent framework (purple), with critical stretches denoted using star markers. Dashed lines indicate characteristic deformation paths: uniaxial tension ($\lambda_2=\lambda_1^{-0.5}<1$), uniaxial compression ($\lambda_2=\lambda_1^{-0.5}>1$), pure shear ($\lambda_2=1$), equibiaxial tension ($\lambda_2=\lambda_1>1$), and equibiaxial compression ($\lambda_2=\lambda_1<1$).}
    \label{fig:sm3}
\end{figure}

To complement the strong-ellipticity-based failure envelope presented in Appendix E, we also examined a Hessian-based failure-envelope construction. In this approach, the onset of failure is identified by the vanishing determinant of the Hessian matrix of the reduced strain energy density \(\hat{\psi}_{\mathrm{prop}}(\lambda_1,\lambda_2)\) within a prescribed principal-stretch space, following the energy-based failure criterion illustrated by Volokh~[3,4,5]:

\begin{equation}
    H(\lambda_1,\lambda_2)
    =
    \det
    \begin{bmatrix}
    \dfrac{\partial^2 \hat{\psi}_{\mathrm{prop}}}{\partial \lambda_1^2}
    &
    \dfrac{\partial^2 \hat{\psi}_{\mathrm{prop}}}{\partial \lambda_1 \partial \lambda_2}
    \\[8pt]
    \dfrac{\partial^2 \hat{\psi}_{\mathrm{prop}}}{\partial \lambda_1 \partial \lambda_2}
    &
    \dfrac{\partial^2 \hat{\psi}_{\mathrm{prop}}}{\partial \lambda_2^2}
    \end{bmatrix}
    =0 .
    \label{eq:supp_hessian_failure}
\end{equation}

The resulting failure envelope predicts the critical stretches associated with the onset of static instability and local failure within a homogeneous deformation subspace. In this energy-limiter setting, the onset of failure is identified with the loss of positive definiteness of the Hessian, which signals a loss of local convexity in the strain energy density and provides a critical strength criterion~[6]. Derived directly from the underlying constitutive framework, this approach offers a rigorous, physics-based alternative to traditional phenomenological failure criteria.

Figure~\ref{fig:sm3} compares the theoretical failure envelopes derived from the Hessian determinant criterion for two distinct formulations: the proposed mode-dependent asymmetric model and the classical, distortional-mode-independent Volokh framework. To ensure a consistent baseline, both formulations employ the same intact elastic backbone, the Prasad--Kannan model. The analysis uses calibrated material parameters for a representative 3\%~w/v agarose gel. In the classical model, the failure energy is restricted to a single constant value, $\psi_f$, with a fixed sharpness of transition to material instability corresponding to the tensile limit. Consequently, the predicted critical stretches are predominantly governed by this fixed lower threshold. In contrast, the proposed formulation introduces a mode-dependent failure energy, $\psi_{\mathrm{failure}}^{(\mathrm{prop})}(K_3)$, via a Lode-weighted framework calibrated using the combined uniaxial tension--compression response. This allows the failure threshold and, in turn, the gradualness of the transition to material instability to evolve continuously with deformation mode, rather than being constrained by a constant scalar bound. This distinction is particularly important for capturing failure in compression-dominated regimes and intermediate loading modes not explicitly included in the calibration. In such cases, the proposed formulation smoothly interpolates failure/instability thresholds across principal-stretch space to capture the mode-dependent onset of softening.

The Hessian-based envelope is qualitatively similar to the strong-ellipticity-based failure envelope presented in Appendix~E for the calibrated case considered here, although the two criteria are not generally equivalent. While the Hessian-based criterion probes loss of local convexity within a prescribed homogeneous deformation subspace, the Legendre--Hadamard condition tests stability with respect to all rank-one perturbations through the acoustic tensor. Importantly, both criteria identify theoretical onset conditions for failure or constitutive instability; neither criterion describes subsequent localization, crack initiation, or propagation processes. These criteria provide critical-stretch predictions across multiaxial deformation states and align with expected trends reported in the literature, as well as with the primary modes evaluated in this work, namely uniaxial deformation and pure shear. However, rigorous quantitative validation against experimental measurements, especially for intermediate loading regimes beyond uniaxial loading and pure shear, remains an important direction for future work.

\vspace{3em}
\noindent \textbf{References}

\noindent [1] M. Reddipaga and K. Kannan, On the construction of a viscoelastic constitutive model for brain tissue maximizing tension--compression asymmetry, \textit{International Journal of Engineering Science} 218 (2026) 104415.

\noindent [2] M. Mooney, A theory of large elastic deformation, \textit{Journal of Applied Physics} 11(9) (1940) 582--592.

\noindent [3] K. Y. Volokh, Review of the energy limiters approach to modeling failure of rubber, \textit{Rubber Chemistry and Technology } 86, no. 3 (2013): 470-487. 

\noindent [4] K. Y. Volokh,  Constitutive model of human artery adventitia enhanced with a failure description, \textit{ Mechanics of Soft Materials} 1, no. 1 (2019): 8. 

\noindent [5] K. Balakhovsky, K. Volokh, Inflation and rupture of rubber membrane, \textit{International journal of fracture} 177 (2) (2012) 179–190.

\noindent [6] Y. Xiao , Z. Zhang , J. Wang,  Granular hyperelasticity with inherent and stress-induced anisotropy. \textit{Acta Geotechnica} (2020) 15(3):671-80.
